\documentstyle[12pt,prb,aps,psfig]{revtex}
\begin{document}
 \draft
 \title{ Stability of the Ground State of a Harmonic Oscillator \\
in  a Monochromatic Wave}
 \author{Gennady P. Berman, Daniel F.V. James, and Dimitry I. Kamenev
 \footnote{On leave from Nizhny Novgorod State University, Nizhny Novgorod,
 603600, Russia}}
 \address{Theoretical Division and CNLS, Los Alamos National Laboratory,
 Los Alamos, NM 87545}
\maketitle
\vspace{20mm}

\begin{abstract}

Classical and quantum dynamics of a harmonic oscillator 
in a monochromatic wave is studied 
in the exact resonance and near resonance cases.  This 
model describes, in particular, a dynamics of a cold 
ion trapped in a linear ion trap and interacting with 
two lasers fields with close frequencies.
Analytically and numerically a stability of the
``classical ground state'' (CGS) --  the vicinity
of the point ($x=0, p=0$) -- is analyzed. In the quantum 
case, the method for studying a stability of the quantum 
ground state (QGS) is suggested, based on the quasienergy representation. 
The dynamics depends on four parameters: the detuning from the resonance, 
$\delta=\ell-\Omega/\omega$, where
$\Omega$ and $\omega$ are, respectively, the wave and the 
oscillator's frequencies;
the positive integer (resonance) number, $\ell$;
the dimensionless Planck constant, $h$, 
and the dimensionless wave amplitude, $\epsilon$. For $\delta=0$,
the CGS and the QGS are unstable for resonance numbers $\ell=1,\,2$.
For small $\epsilon$, the QGS becomes more stable
with increasing $\delta$ and decreasing $h$.
When $\epsilon$ increases, the influence
of chaos on the stability of the QGS is analyzed for different parameters of
the model, $\ell$, $\delta$ and $h$.
\end{abstract}

\section{Introduction}
One of the major difficulties in developing quantum technologies,
such as quantum computers, \cite{fortsch,bdmt} are different kinds of
specifically quantum dynamical instabilities that can occur due
to interactions between different degrees of freedom and resonant 
interaction with the external fields. Instabilities in quantum systems 
have different nature than instabilities in classical systems, in which
such kind of phenomenon as dynamical
chaos occurs as a result of exponential divergence of initially close
trajectories. In quantum systems, the notion of a trajectory is
not well defined. This is one of the  main reasons why most of
well-developed methods for stability analysis can not be directly applied to
quantum systems.

In this paper the dynamics of a harmonic oscillator in a monochromatic
wave field is studied.
This system describes, in particular, a quantum dynamics of a cold ion
trapped in a linear ion trap and interacting with two laser fields with
close frequencies.\cite{BD}
Our attention is focused mainly on classical and quantum  behavior in the
region of parameters close to the quantum ground state (QGS) of the harmonic
oscillator. This case corresponds to classical dynamics in the vicinity of
the point ($x=0,\,p=0$) in the phase space. We shall call the corresponding
region of parameters for the classical system a ``classical ground state''
(CGS) of a harmonic oscillator.
The stability of the CGS and the QGS at stable and chaotic
classical regimes is analyzed for different parameters of the model.
In Sec. II, we consider classical dynamics in the vicinity of the
CGS in the exact resonance case, $\delta=0$.
In this situation the system we study is degenerate,\cite{Z} and an
infinitely small perturbation generates in the classical phase space an
infinite number of the ``resonance cells'' separated by the infinite
stochastic web. A classical dynamics inside the resonance cells at small
$\epsilon$ is described by using the resonance perturbation theory.\cite{L}
It is shown, that in the vicinity of the CGS (the ``central
cell''), and for small enough $\epsilon$, the dynamics in the case
$\delta=0$ is unstable for
resonance numbers $\ell=1,2$ and stable for $\ell\ge 3$.

We show that in spite of the  ``resonant Hamiltonian'' describes many
features of the dynamics inside the resonance cells, it can not be used
for describing the stability of the system in the central cell, at
$\ell\ge 3$. The classical dynamics in this region is determined,
by the Mathieu equation. It is shown that at small enough value of
$\epsilon$ the area of the central cell
increases with increasing the resonance number, $\ell$, and increasing
the perturbation amplitude, $\epsilon$.

The dynamics of the system near the CGS
in the near resonance case, when $\delta\ne 0$,
is considered in Sec. III. It is shown that in this case
at small $\epsilon$ the classical
dynamics near the CGS is stable at any value of
the resonance number $\ell$. The cases $\ell=1$ and $\ell=2$ are
considered in detail. It is shown that the CGS in these two 
cases becomes unstable at much less values of $\epsilon$,
than for large $\ell$ ($\ell=4,5$...).  

A stability of the quantum system is considered in Sec. 4. In this case, an 
additional parameter, a dimensionless Planck constant, $h$, influences
significantly the behavior of the system.
Because the Hamiltonian is time-periodic, with the period $2\pi/\Omega$, 
we use the Floquet theory to study localization properties of quantum 
system in the region of the QGS. For $\epsilon\ll 1$, and small enough 
$h$, the quasienergy (QE) states can be conventionally divided into the
following groups. The first group includes the QE states which belong to
some particular ``resonance cells'' in the Hilbert space.
These QE states are well-localized inside a given resonance cell. The second
group includes the delocalized ``separatrix'' QE states which correspond to
the classical stochastic web.
Usually, these ``separtrix'' QE states are responsible for tunneling effects
between different ``resonance cells'',  even for small $\epsilon$.\cite{1}

We show that the QGS is stable when the following conditions are
satisfied: (a) an existence of the QE state mainly localized in 
the QGS of the harmonic oscillator, 
(b) when $\epsilon\ll 1$ and chaos is weak, (c) small enough $h$, when 
one can neglect the tunneling phenomenon. 
When $h$ is larger than the size of the 
central cell, no QE state localized in the QGS of the
harmonic oscillator was found. 
We show, that at small enough values of $\epsilon$,
the stability of the QGS can be improved by choosing the non-resonant
frequency of the wave, so that $\delta=\ell-\Omega/\omega\ne 0$.

For small enough $\epsilon$, a stability of the QGS is mainly determined
by the structure of the QE state mostly localized at the QGS 
of the harmonic oscillator (QGS QE state). In the exact resonance case 
this  particular QE state has a complicated structure.  
On the one hand, it is mostly localized on the QGS,
but, on the other hand, it has the  ``separatrix'' structure\cite{3} 
and provides a tunneling from the region of the QGS to other 
resonance cells. In order to improve stability of the QGS, the separatrix 
structure should be destroyed by choosing $\delta\ne 0$.  

In the exact and near resonance cases
we study numerically the probability, $P_0$, for a particle to remain 
in the QGS, depending on $\epsilon$ and for different values of $h$, 
$\ell$ and $\delta$. 
When $\delta=0$ and $\epsilon$ is small,
the dynamics depends on the value of the 
quantum parameter, $h$, which controls degree of delocalization
of the QGS QE state. When $h$ is small,
the probability of tunneling to other cells is small too, and
$P_0$ decreases with $\epsilon$ increasing, because chaos makes the 
QGS QE state more delocalized.
For large enough values of $h$, 
the dependence of $P_0$ on $\epsilon$ 
becomes more complicated because in this case we have 
two superimposing effects: 
the influence of chaos on the dynamics, and
the effect of tunneling to the classically unacceptable cells. 
These two features make the dependence $P_0=P_0(\epsilon)$ non-monotonic. 
In the region of small $\epsilon$, $P_0$ increases 
with $\epsilon$ increasing.  This behavior of $P_0$ can be explained 
as a result of 
destroying the delocalized QE states with increasing $\epsilon$. Further 
increase of $\epsilon$ leads to decreasing $P_0$. This behavior of 
$P_0$ is connected with delocalization of QE states due to increasing
of the chaotic component. At large enough values of $\epsilon$, the
dependence of $P_0$ on $\epsilon$ becomes more complicated, and generally
must be considered using many QE states. It is shown that the value of
$\delta$ influence significantly the stability of the QGS only at small
$\epsilon\le 1$.
The stability of the QGS at $\ell=1,\,2$
and $\delta\ne 0$ is explored. It is shown that the QGS becomes unstable
at much less values of the wave amplitude than
in the case when $\ell$ is large. The derived in this paper results
are important for understanding a stability of quantum dynamics in the
vicinity of the ground state.

\section{Classical dynamics near the CGS in the case of exact resonance}
The Hamiltonian of the harmonic oscillator in a monochromatic
wave is,
\begin{equation}
\label{cl_H}
H=\frac{p^2}{2M}+\frac{M\omega^2x^2}2
+v_0\cos(kx-\Omega t)=H_0+V(x,t),
\end{equation}
where $M$ is the mass of the
particle, $p$ is the momentum, $k$ is the wave vector,
$v_0$ is the amplitude of the perturbation, and
$H_0$ is the Hamiltonian of the harmonic oscillator.
In the first part of this section we discuss the case of the resonance,
when $\Omega=\ell\omega$. It is known
(see for example Ref.~\cite{Z}) that under the resonance condition 
the infinitely small perturbation, $v_0$, is enough to generate 
in the classical phase space the infinite stochastic web. 
The web is inhomogeneous, and its width decays with decreasing 
the perturbation amplitude, $v_0$, and increasing $|p|$ and $|x|$.             
Inside the cells of the web a particle moves along stable 
closed trajectories.

To analyze the 
dynamics of the harmonic oscillator in a monochromatic wave,
described by the Hamiltonian (\ref{cl_H}), 
it is convenient to use
the resonance perturbation theory \cite{L}
discussed below.   
Let us perform a transformation  from the variables ($p,x$)
to the canonically
conjugated variables ($\bar J_\varphi,\,\varphi$),
\begin{equation}
\label{xtheta}
x=(2\bar J_\varphi/M\omega)^{1/2}\sin\varphi=
r(\bar J_\varphi)\sin\varphi,
\end{equation}
\begin{equation}
\label{ptheta}
p_x=(2\bar J_\varphi M\omega)^{1/2}\cos\varphi=
M\omega r(\bar J_\varphi)\cos\varphi,
\end{equation}
where $r(\bar J_\varphi)=(2\bar J_\varphi/M\omega)^{1/2}$
is the amplitude of oscillations.
It is more convenient to work with
the dimensionless coordinate,
$X=kx$, and the dimensionless momentum, $P=kp/M\omega$, which are related to
 the variables ($\bar J_\varphi,\,\varphi$)
by the formulas,
\begin{equation}
\label{Xtheta}
X=\rho(\bar J_\varphi)\sin\varphi,
\end{equation}
\begin{equation}
\label{Ptheta}
P=\rho(\bar J_\varphi)\cos\varphi,
\end{equation}
where $\rho(J_\varphi)=\sqrt{X^2+P^2}=kr(J_\varphi)$. 
In order to treat the time on the same ground as the phase 
$\varphi$ let us introduce the new pair of canonically conjugated 
variables, $(\bar J_\beta,\,\beta)$, where $\beta=\Omega t$.  
The initial Hamiltonian (\ref{cl_H}) expressed through the new variables
takes the form,  
\begin{equation}
\label{cl_H0}
H=\bar J_\varphi\omega+\bar J_\beta\Omega+
v_0\cos\left(\rho\sin\varphi-\beta\right).
\end{equation}
It is independent of time, but describes the motion in 
the two-dimensional space.  
The nonlinear perturbation in Eq. (\ref{cl_H0})
can be expressed in a series,
\begin{equation}
\label{wave_decomposition}
v_0\cos\left(\rho\sin\varphi-\beta\right)=v_0\sum_{m=-\infty}^\infty
J_m(\rho)\cos\left(m\varphi-\beta\right),
\end{equation}
where $J_m(\rho)$ is the Bessel function.  
Under the resonance condition, 
$\Omega=\ell\omega$, all terms
in the sum (\ref{wave_decomposition}) quickly oscillate
and can be averaged out, except for one term with $m=\ell$.
In this approximation, the
Hamiltonian (\ref{cl_H0}) is reduced to,
\begin{equation}
\label{cl_H1}
H=\bar J_\varphi\omega+\bar J_\beta\Omega+v_0J_\ell(\rho)
\cos\left(\ell\varphi-\beta\right).
\end{equation}
It is convenient to introduce new, resonance, variables,
($\tilde I,\,\theta$), ($\tilde J,\,\tilde\beta$), 
by using the generating function,
$$
F=\tilde I\left(\ell\varphi-\beta\right)+\tilde J\beta,
$$
 
The new Hamiltonian,
\begin{equation}
\label{cl_H3}
H=\tilde I(\ell\omega-\Omega)+\tilde J\omega+v_0J_\ell(\rho)\cos\theta,
\end{equation}
where $\theta=\ell\varphi-\beta$,
is independent of the variable $\tilde\beta$. Hence,
$\tilde J=const$. The resonance Hamiltonian,
\begin{equation}
\label{cl_HR}
H_\ell(\rho,\theta)=H-\tilde J\omega=v_0J_\ell(\rho)\cos\theta,
\end{equation}
(where we used the resonance condition $\ell\omega=\Omega$)
is independent of time, unlike the initial Hamiltonian (\ref{cl_H}).

The Poincar\`e surfaces of section of the system
described by the 
Hamiltonian (\ref{cl_H}) in variables ($X,\,P$) 
are shown in Figs. 1 (a) - 1 (e), for the cases 
$\ell=1,2,3,4,5$. The phase 

 \begin{figure}[tb]
 \begin{center}
 \mbox{\psfig{file=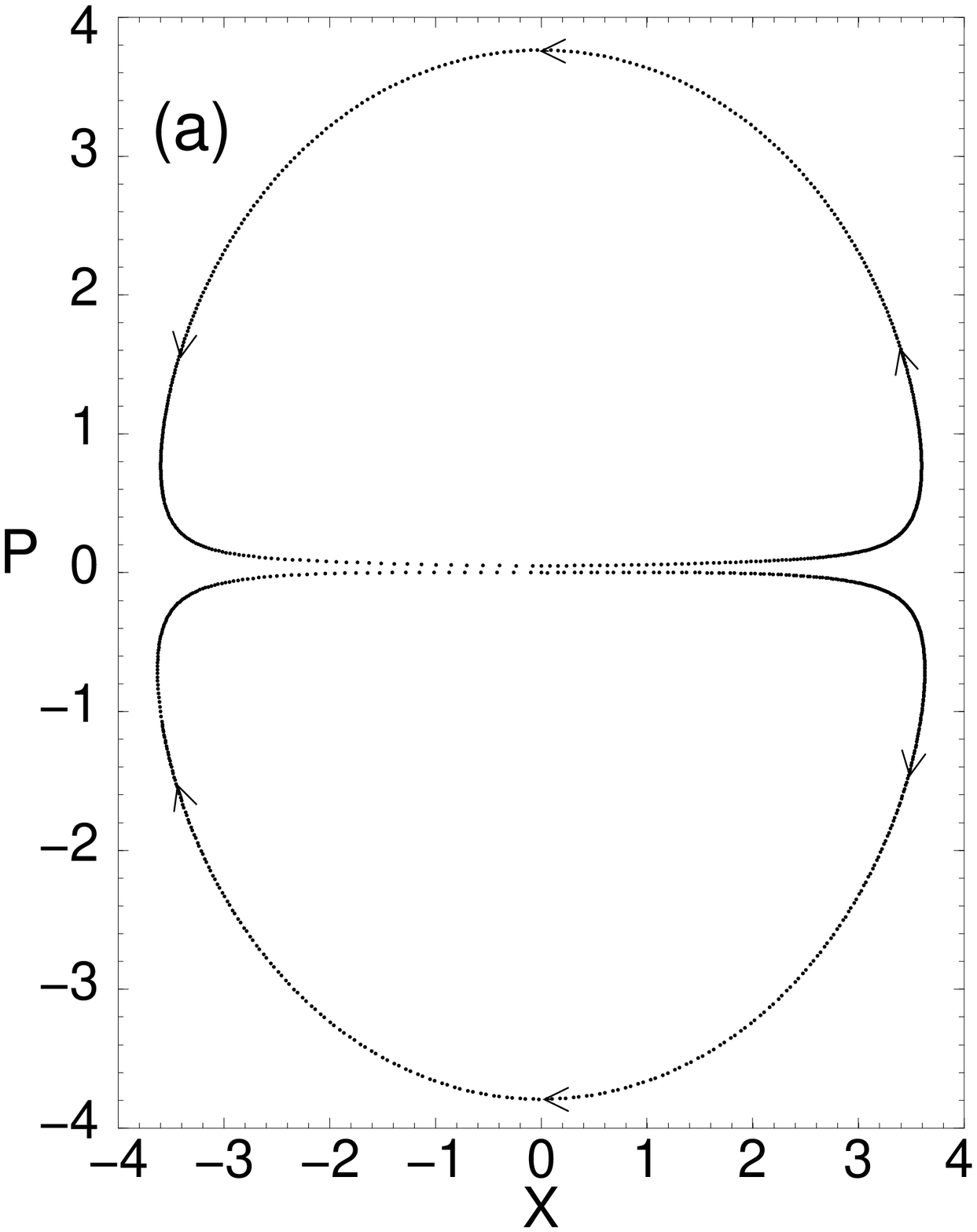,width=7cm,height=7cm}
       \psfig{file=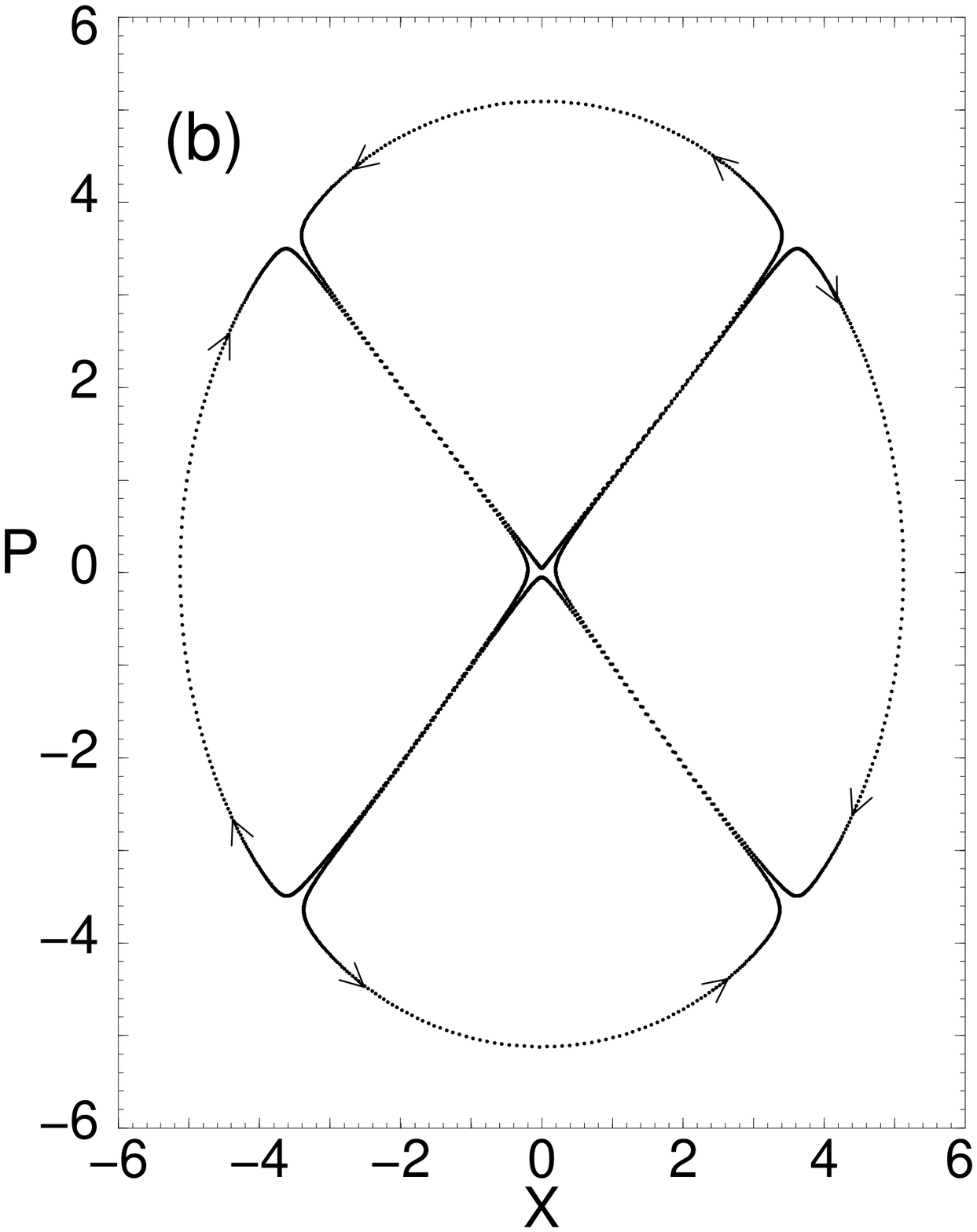,width=7cm,height=7cm}}
 \mbox{\psfig{file=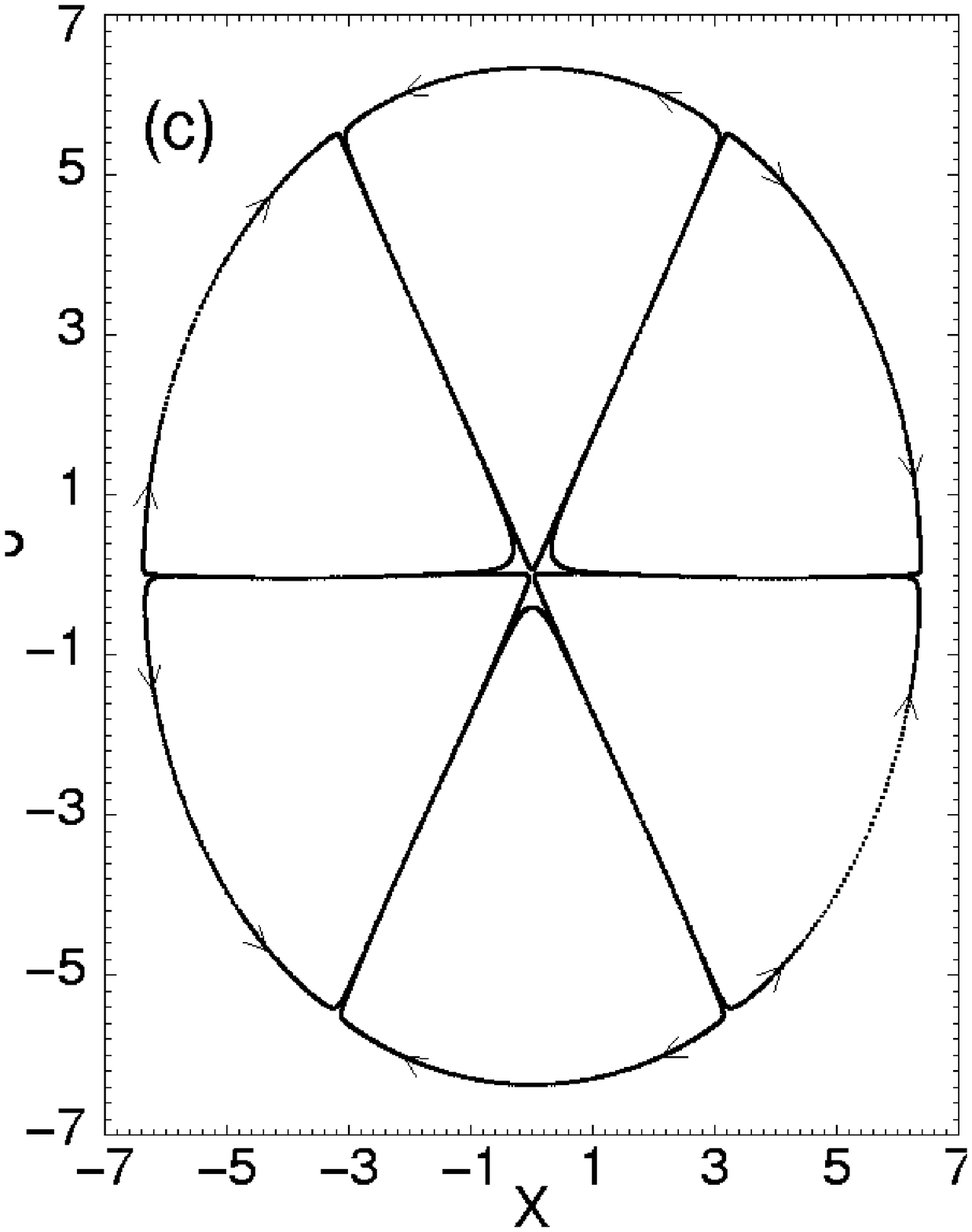,width=7cm,height=7cm}
       \psfig{file=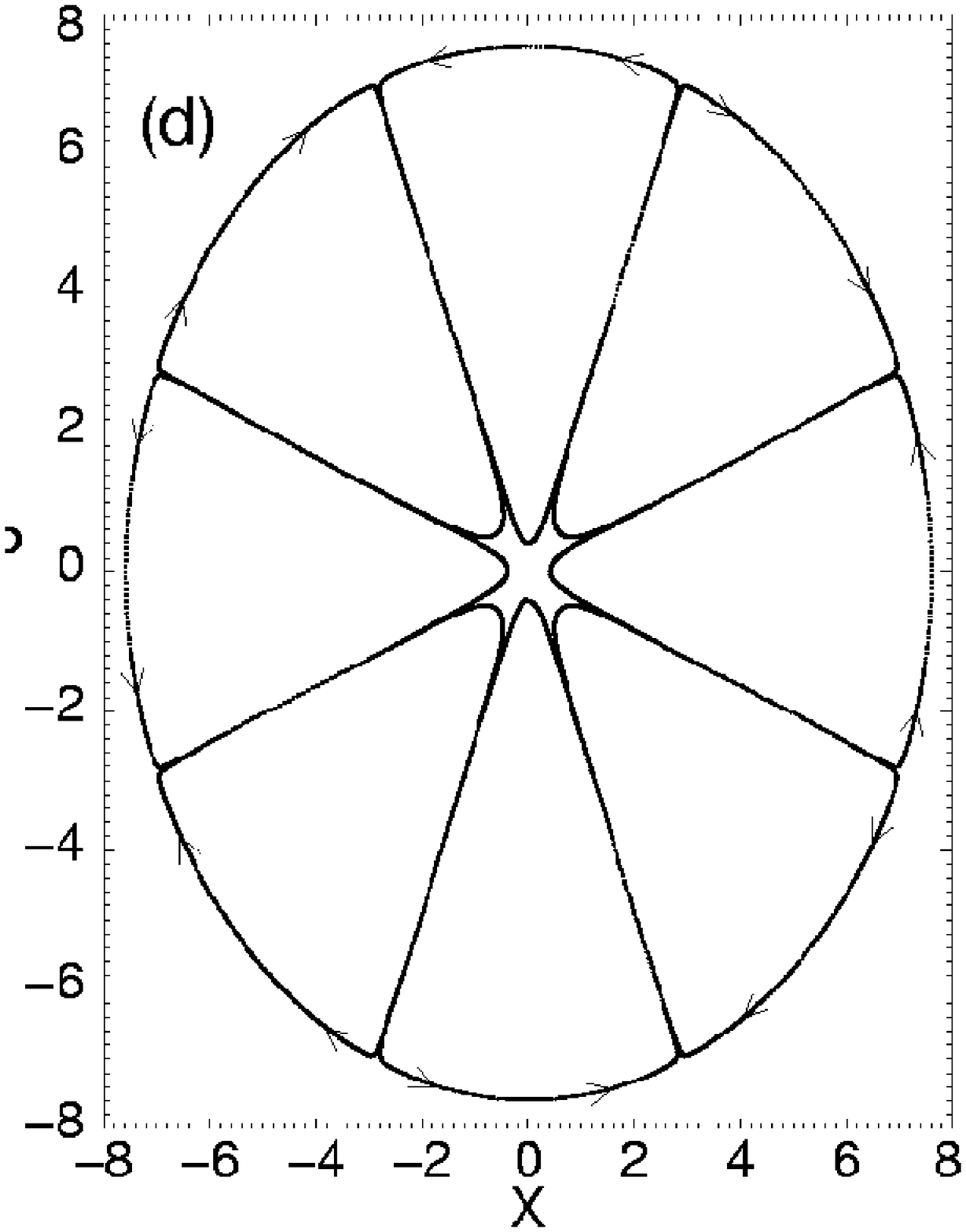,width=7cm,height=7cm}}
 \psfig{file=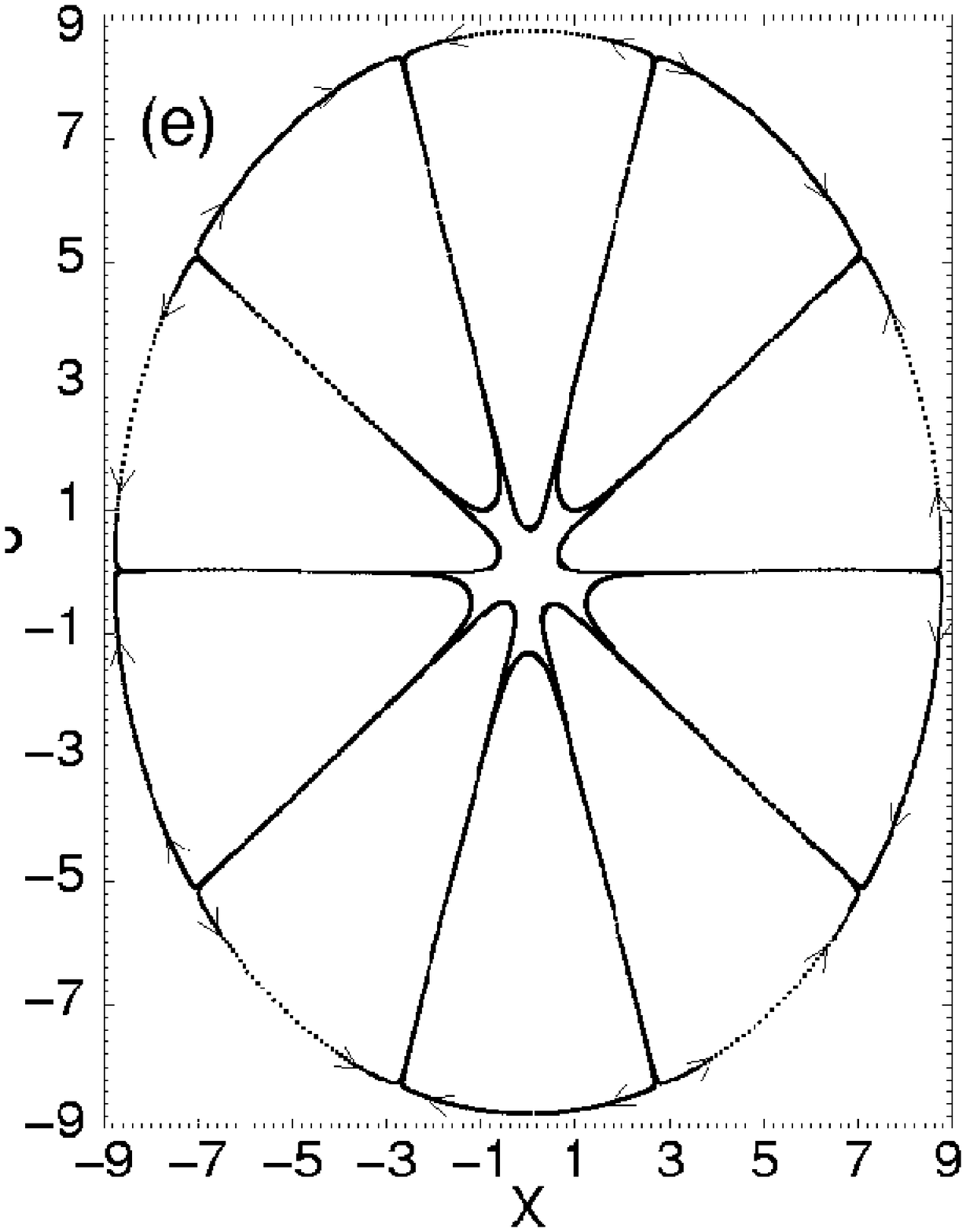,width=7cm,height=7cm}
 \vspace{0.5cm}
 \caption{The resonance cells in the phase space for 
 $\epsilon=0.05$, $\delta=0$ and: (a) $\ell=1$, (b) $\ell=2$,
 (c) $\ell=3$, (d) $\ell=4$, (e) $\ell=5$.}
 \end{center}
 \label{fig:1}
 \end{figure}

\noindent
points are plotted at the moments
$t_j=2\pi j/\Omega$, where $j=0,\,1,\,2,\dots$.  
One can see that the phase space  
has an axial symmetry of the order $\ell$. The phase 
space is divided into the cells. 
A particle moves along closed trajectories inside the cells. 
(In Figs. 1 (a) - 1 (e) only the boundaries of the cells are demonstrated.) 
At small values of $v_0$, the motion inside the resonant cells, 
illustrated in Figs. 1 (a) - 1 (e), 
can be considered in the resonance approximation. 
The next order approximation 
is required only for analysis of the motion inside the exponentially 
small chaotic regions near the separatrices. It is shown below that 
the resonance approximation also fails to describe the motion 
in the region near the point $(X=0,\,P=0)$.   

The symmetry of the phase space in Figs. 1 (a) - 1 (e)
follows from the form of the resonance
Hamiltonian (\ref{cl_HR}), which is invariant under the transformation,
\begin{equation}
\label{transform1}
\varphi\rightarrow\varphi + 2\pi/\ell. 
\end{equation}
In the phase space there are $\ell$ cells connected
by the transformation (\ref{transform1}), and $\ell$ cells  
connected by the same transformation but described by the 
Hamiltonian $H_\ell(\rho,\theta+\pi)$ in 
Eq. (\ref{cl_HR}). The total number of cells at given $\rho$
is  $2\ell$.  
Thus, in Fig. 1 (a)
the upper cell corresponds to the Hamiltonian $H_1(\rho,\theta)$, while
the lower cell corresponds to the Hamiltonian $H_1(\rho,\theta+\pi)$ in 
Eq. (\ref{cl_HR}); the upper and lower cells in Fig. 1 (b) 
correspond to $H_2(\rho,\theta)$, and the right and the left cells 
correspond to $H_2(\rho,\theta+\pi)$, and so on. 

It is easy to see, the resonance Hamiltonian    
(\ref{cl_HR}) yields unstable solution near the CGS
(the point ($X=0,\,P=0$)). 
To show this, we present the Hamiltonian (\ref{cl_HR}) in the form,
\begin{equation}
\label{cl_HR1}
H_\ell=v_0\frac{\rho^\ell}{2^\ell\ell !}\cos\ell\varphi=E_\ell,
\end{equation}
where $E_\ell=const$ (because the resonance Hamiltonian is independent 
of time). Also, we took into account that at $\rho\ll 1$ the Bessel 
function can be expressed in the form, 
\begin{equation}
\label{Bessel1}
J_\ell(\rho)\approx \rho^\ell/2^\ell\ell !,   
\end{equation}
and the fact that in the Poincar\`e surfaces of section the position
of the particle is taken at the moments 
$\Omega t_k=2\pi k$, $k=0,\,1,\,2,\dots$. 
It follows from Eq. (\ref{cl_HR1}) that at the 
angles $\varphi_k=\frac\pi{2\ell}(2k-1)$, $k=1,\,2,\dots,\,2\ell$,
the radius $\rho$ should sharply increase or decrease. 
It is seen from Figs. 1 (a) - 1 (e) that
at angles $\varphi_k$ the particle moves in the radial 
direction. The grows of $\rho$ is limited due to nonlinear effects. 

However, the resonance perturbation theory does not adequately describe
the motion of a particle in the vicinity of the point ($X=0,\,P=0$)
at large values of the resonance number $\ell$, because the amplitude 
of the resonance term due to Eq. (\ref{Bessel1}) quickly decays when 
the radius, $\rho$, decreases. Indeed, at $\ell=4$ and $\rho=0.1$ the 
amplitude 
of the resonance term with $m=\ell=4$ in Eq. (\ref{wave_decomposition}) 
is $80$ times less than the amplitude of the non-resonant term with
$m=\ell-1=3$. In order to describe the motion in the region near
the CGS,   we 
consider the initial Hamiltonian (\ref{cl_H}) under the condition $X\ll 1$. 
The exact classical equation of motion reads,
\begin{equation}
\label{e_motion1}
\frac{d^2}{dt^2}X+\omega^2 X=\frac{v_0k^2}{M}\sin(X-\Omega t).   
\end{equation}

 \begin{figure}[tb]
 \begin{center}
 \mbox{\psfig{file=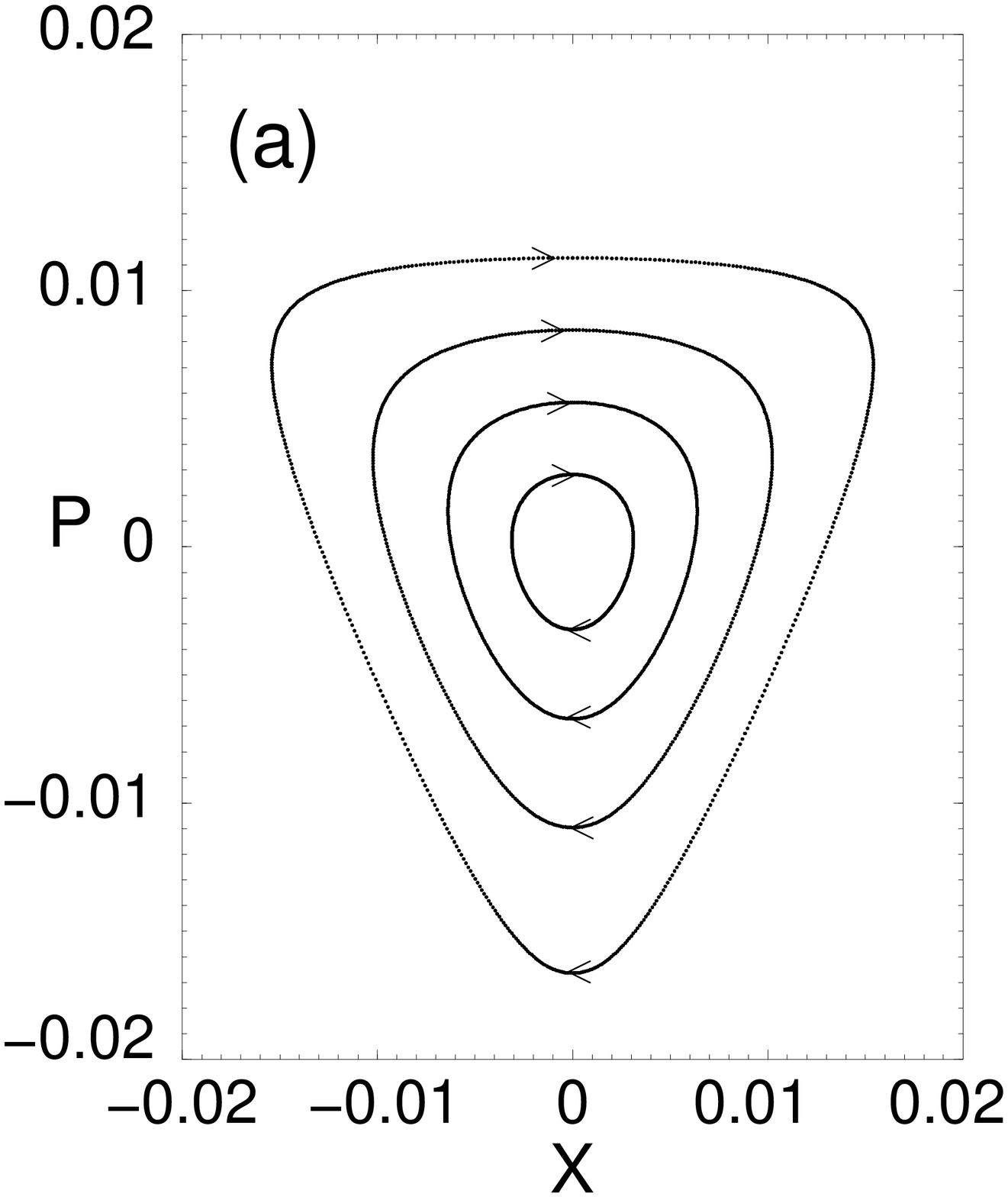,width=5cm,height=5cm}
       \psfig{file=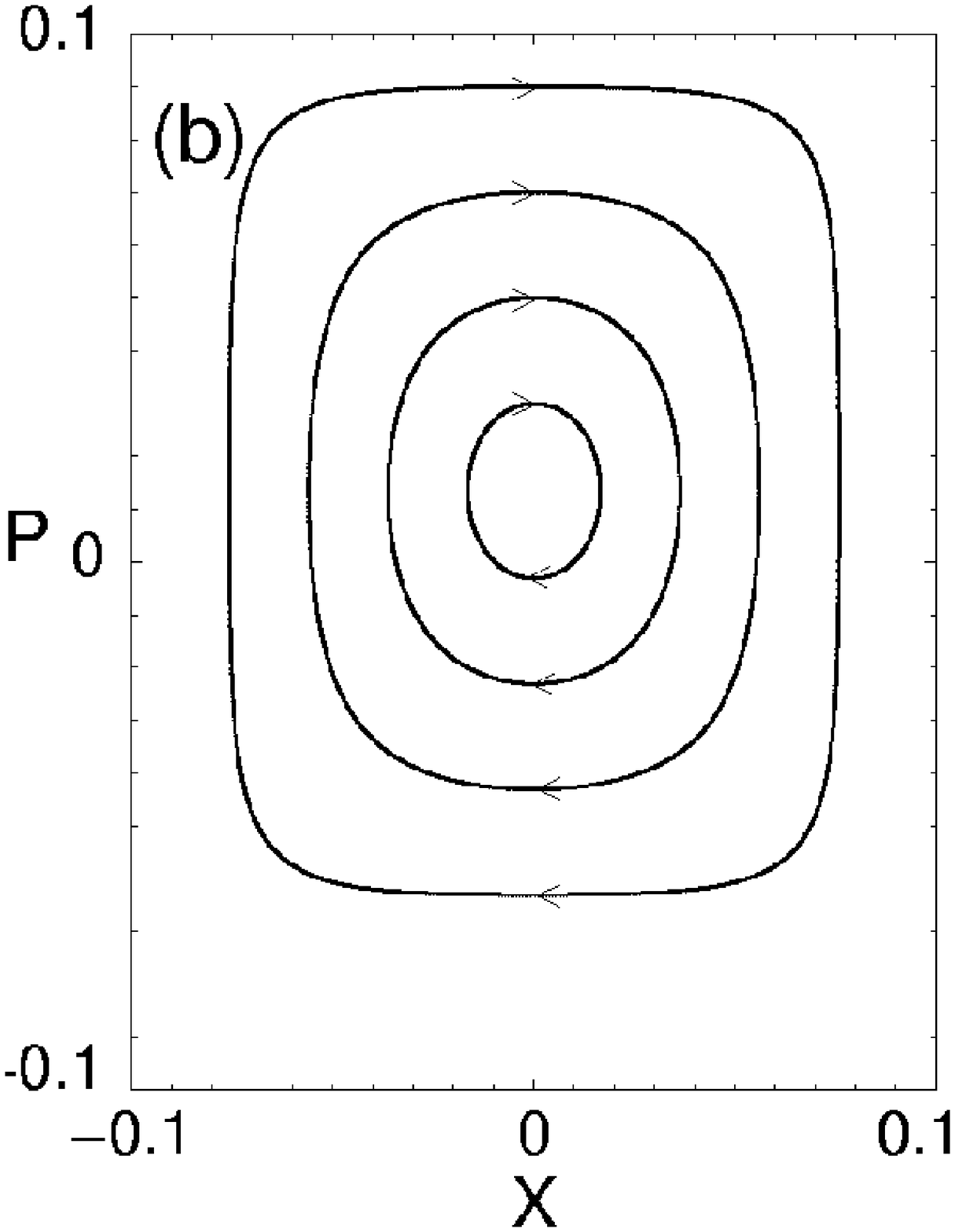,width=5cm,height=5cm}
       \psfig{file=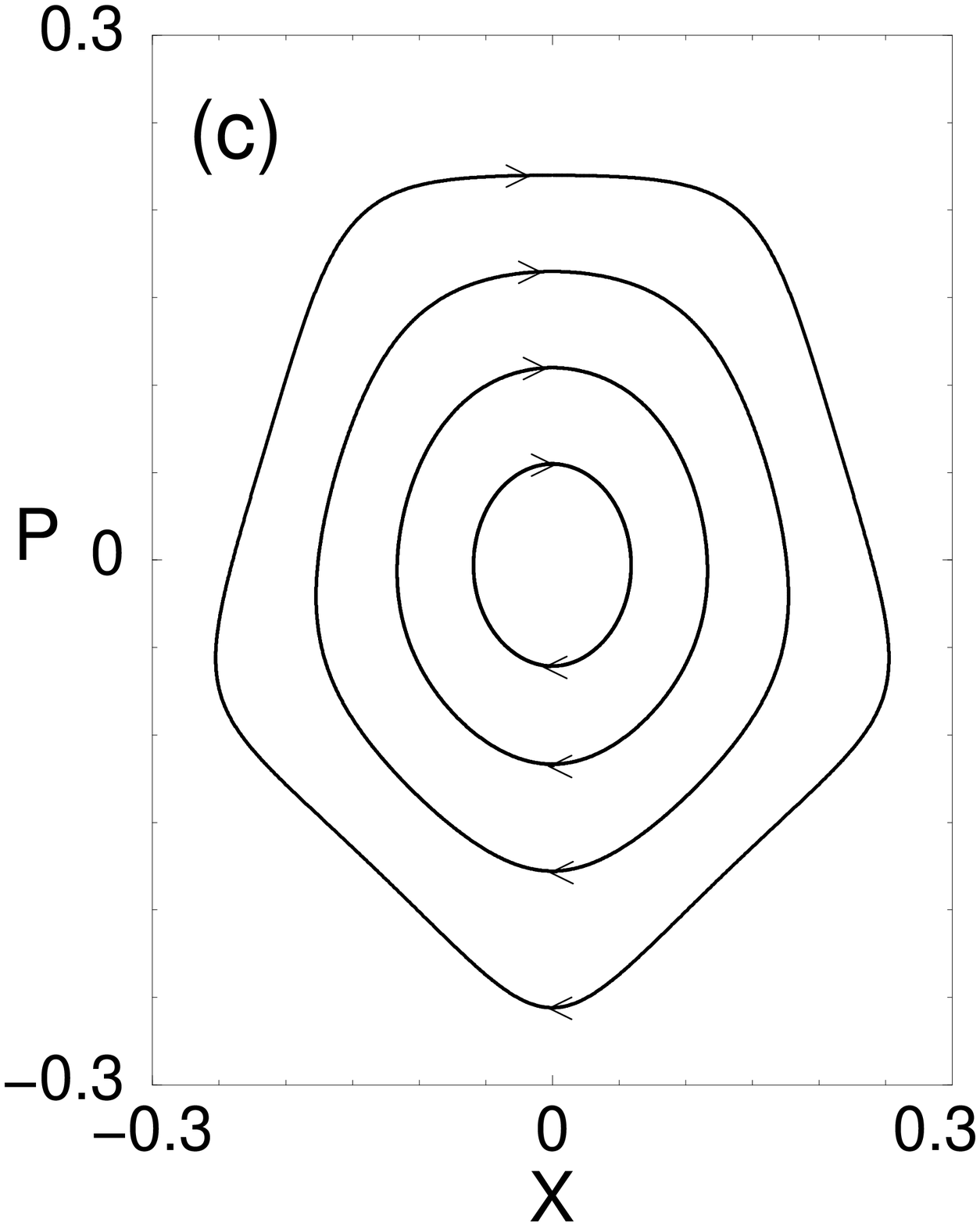,width=5cm,height=5cm}}
 \vspace{0.5cm}
 \caption{The trajectories in the central resonance
 cell in the phase space for $\delta=0$ and (a) $\ell=3$,
 $\epsilon=5\times 10^{-4}$, (b) $\ell=4$, $\epsilon=0.05$,
 (c) $\ell=5$, $\epsilon=0.05$.}
 \end{center}
 \label{fig:2}
 \end{figure}

\noindent
Up to the first order in $X$ Eq. (\ref{e_motion1}) is,
\begin{equation}
\label{e_motion2}
\frac{d^2}{dt^2}X+\omega^2(1-\epsilon\cos(\Omega t))X=
\frac{v_0k^2}{M}\sin(\Omega t),
\end{equation}
where $\epsilon=v_0k^2/M\omega^2$ is the dimensionless
perturbation amplitude. 
If we introduce a new dimensionless time, $2\tau=\Omega t$,
then from Eq. (\ref{e_motion2}) we obtain the Mathieu equation 
with the additional right-hand side term  
in the form, 
\begin{equation}
\label{Mathieu}
\frac{d^2}{d\tau^2}X+
a_\ell(1-\epsilon\cos(2\tau))X=a_\ell\epsilon\sin(2\tau),
\end{equation}
where $a_\ell=\left(2/\ell\right)^2$. 
From the theory of Mathieu 
functions \cite{Mathieu} it is known 
that for small $\epsilon$,
Eq.~(\ref{Mathieu}) has unstable general solutions  
at $a_\ell=1$ and $a_\ell=4$ which correspond to the resonance numbers, 
$\ell=2$ and $\ell=1$. The additional term in the 
right-hand side of Eq. (\ref{Mathieu}) does not influence
the stability of trajectories (see Ref. \cite{Mathieu}, \S 6.22).
At $a_\ell<1$ and small enough values
of $\epsilon$, the Mathieu equation has periodic solutions which
correspond to the stable dynamics for resonance numbers $\ell>2$.
In Fig. 2 stable trajectories in the system described by the
Hamiltonian (\ref{cl_H}) are shown for $\ell=3,\,4,\,5$.
Stable region in the vicinity of the CGS can be
considered as additional ``central cells'' to those resonance cells shown
in Figs. 1~(c) - 1~(e). The motion in the central cell is characterized
by the following features: i) the size of the cell increases
with increasing the resonance number, $\ell$;
ii) trajectories in the cell have the
axial symmetry of the order $\ell$;  iii) as follows from  
numerical calculations, the period of the motion in the central 
cell along all trajectories in Figs. 2 (a) - 2 (c) is 
 \vspace{0.3cm}

 \begin{figure}[tb]
 \begin{center}
 \mbox{\psfig{file=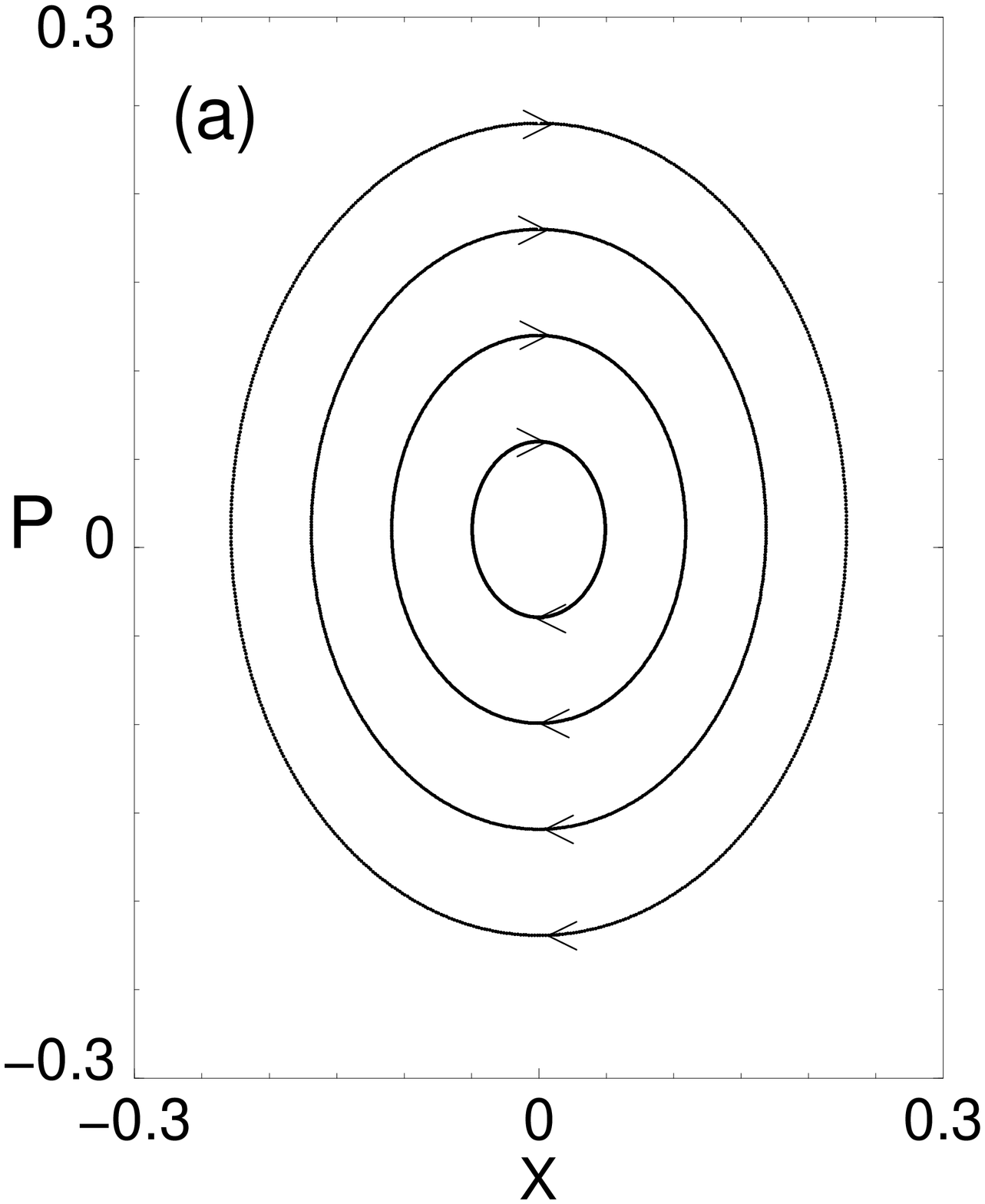,width=8cm,height=8cm}
       \psfig{file=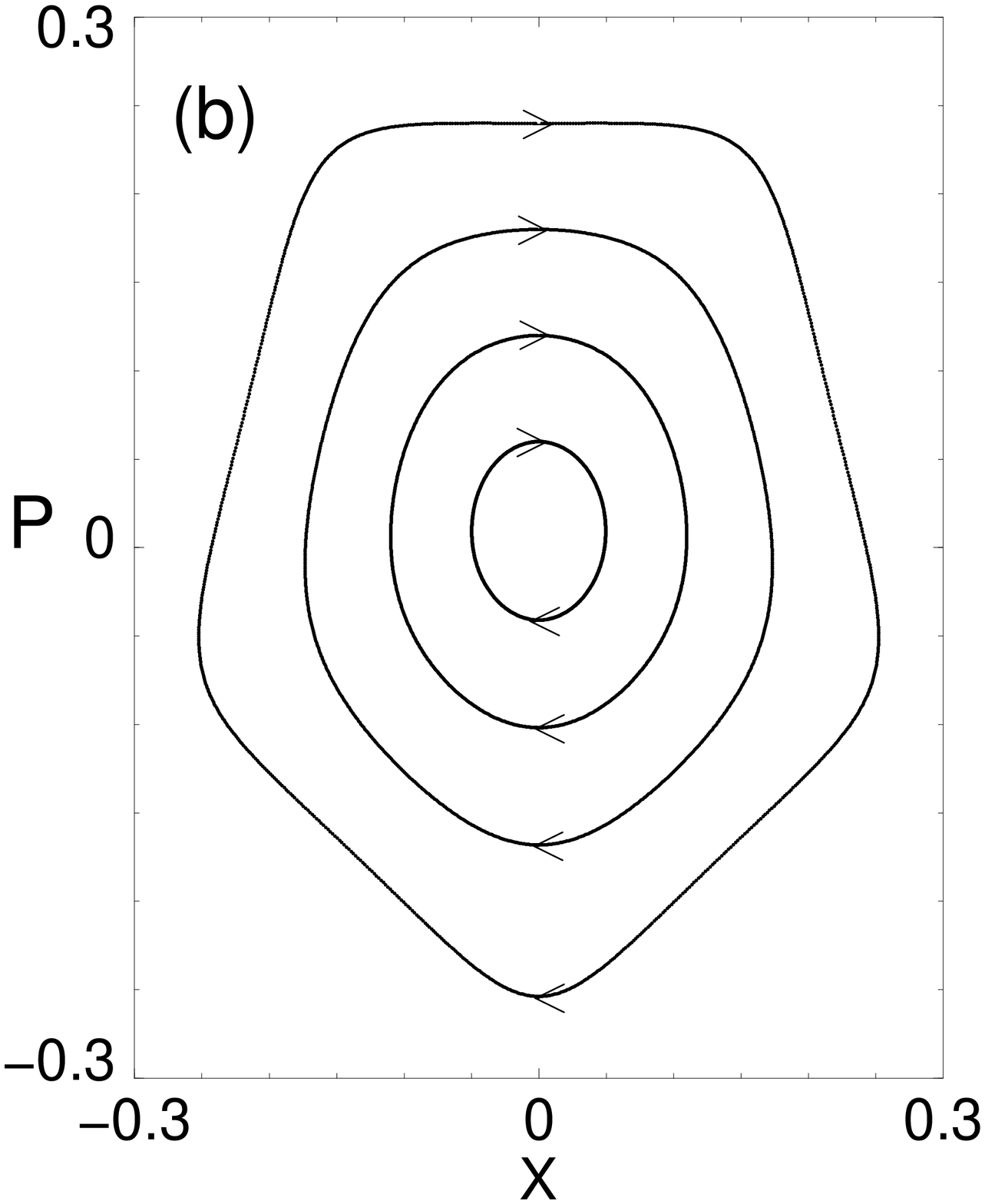,width=8cm,height=8cm}}
 \end{center}
 \label{fig:3}
 \end{figure}

 FIG.3. The trajectories obtained by solution of Eq.
 (\ref{e_motion1}) when only the terms up to (a) the
 first order (Eq. (\ref{Mathieu})) and (b) the fourth
 order in $X$ are taken into 
 account; $\ell=5$, $\delta=0$,
 $\epsilon=0.05$.

\noindent
approximately the same (in each figure), and very large.
For example, the period of motion along the trajectories in Fig. 2 (b)
for $\ell=4$ is $(1-1.2)\times 10^5T$, where $T=2\pi/\Omega$,
while the period of motion along the
trajectories in Fig. 1 (d) is of the order $\sim 5\times 10^3T$.
The property i) can be described by the fact
that the resonance term, which leads to unstable solution near the
point ($X=0,\,P=0$), for
larger values of $\ell$ has less influence on the dynamics
because its amplitude at $\rho\ll 1$ quickly decreases with
$\ell$ increasing, due to Eq. (\ref{Bessel1}).

In order to treat the properties ii), iii), we have considered
the influence of the terms of higher order in $X$ in Eq. (\ref{e_motion1})
on the dynamics described by Mathieu equation (\ref{Mathieu}). The results
for the case $\ell=5$ are shown in Figs. 3 (a), 3 (b).
From Figs. 3 (a), 3 (b) it is seen that
the terms of high order in $X$
change the shape of trajectories, and make
them symmetrical with the axial symmetry of the order $\ell$.    
This fact follows also from comparison of different trajectories
in Fig. 2 (c). Indeed, internal trajectories with small values
of $\rho$ have the round form, unlike external trajectories with 
larger values of $\rho$, which have the form of the pentagon.  
From comparison of Figs. 2 (c) and 3 (b) it is clear that the 
terms of high order in $X$ (the terms of the order $X^i$ where
$i>\ell-1$) in Eq.(\ref{e_motion1}) do not influence significantly
the dynamics,
because the form of trajectories in Fig. 3 (b), computed for
the approximate model, is practically the same
as the form of trajectories
in the exact model shown  in Fig. 2 (c).

It was established numerically that slowness of the motion
in the vicinity of
the CGS is also result of influence of the terms of high order in $X$. Thus,
the period of motion along the external trajectory described by the
approximate equation (\ref{Mathieu}) in Fig. 3 (a) is $\sim 4\times 10^4T$.
Including into consideration the second order term in $X$ increases the
period of motion up to $\sim 3.3\times 10^5T$. If we include,
as well, the third order term, the period becomes $\sim 3.6\times 10^5T$.
Finally, including the 
resonance term leads to increase of the period up to    
the value $\sim 3.8\times 10^5T$ shown in Figs. 2 (c) and 3 (b).

Next, we shall analyze the dynamics in the 
central cell depending on $\epsilon$. Modification of trajectories 
at increasing the wave amplitude, $\epsilon$, for the resonance
number $\ell=4$ is shown in Figs. 4 (a) - 4 (d). 
Two features in the structure of the trajectories of the central cell 
can be observed. i) Increase in $\epsilon$ shifts the central cell up. 
ii) The size of the cell increases considerably in 
Figs. 4 (a) - 4 (c) with increasing $\epsilon$ in comparison with 
the case shown in Fig. 2 (b), when the value of $\epsilon$ is very small.
The cell shrinks when $\epsilon$ increases, which is shown in Fig. 4 (d)
for $\epsilon=10$.
Further increase of $\epsilon$ destroys the central cell entirely.

 \vspace{0.5cm}
 \begin{figure}[tb]
\setcounter{figure}{3}
\centerline{\mbox{\psfig{file=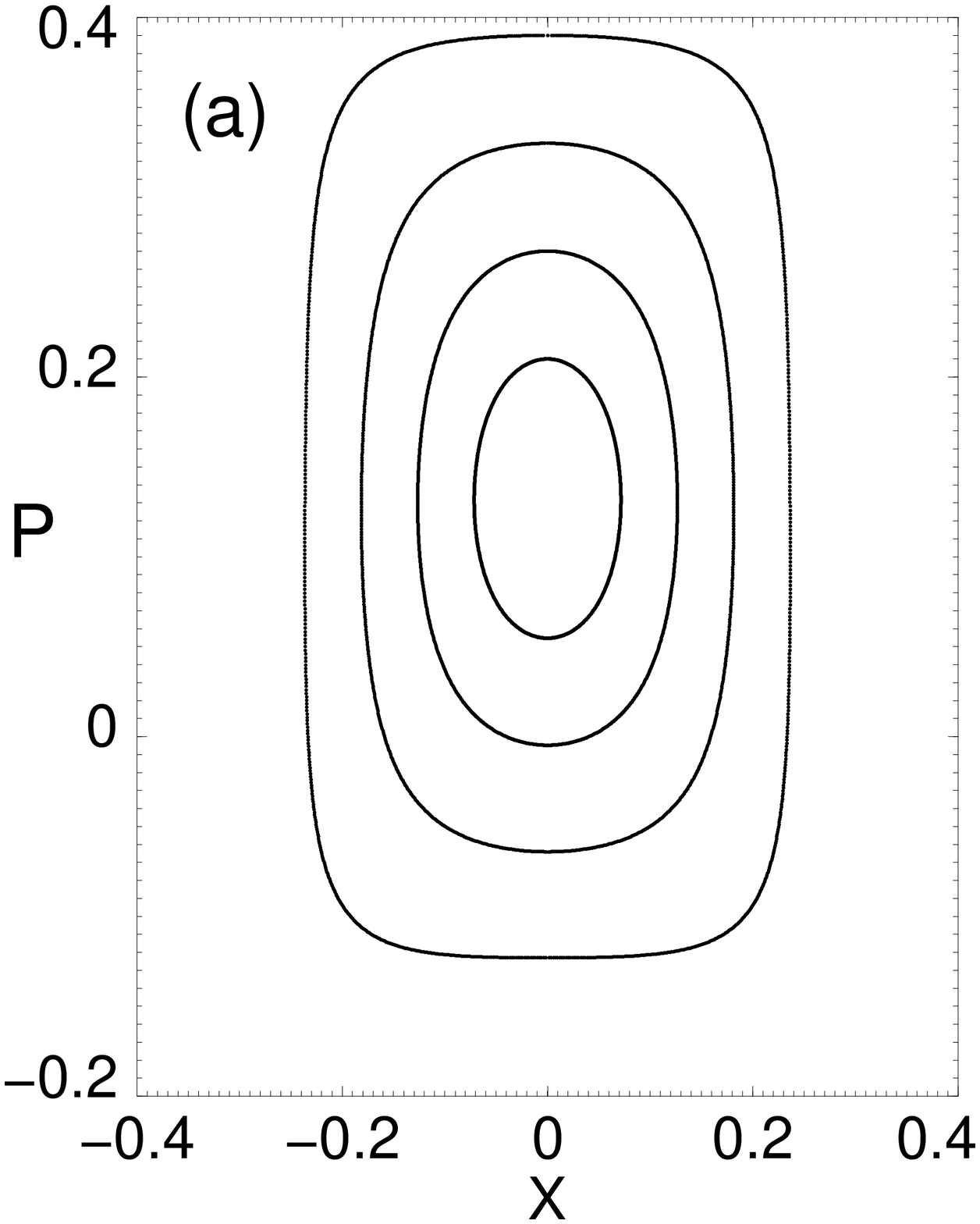,width=8cm,height=8cm}
       \psfig{file=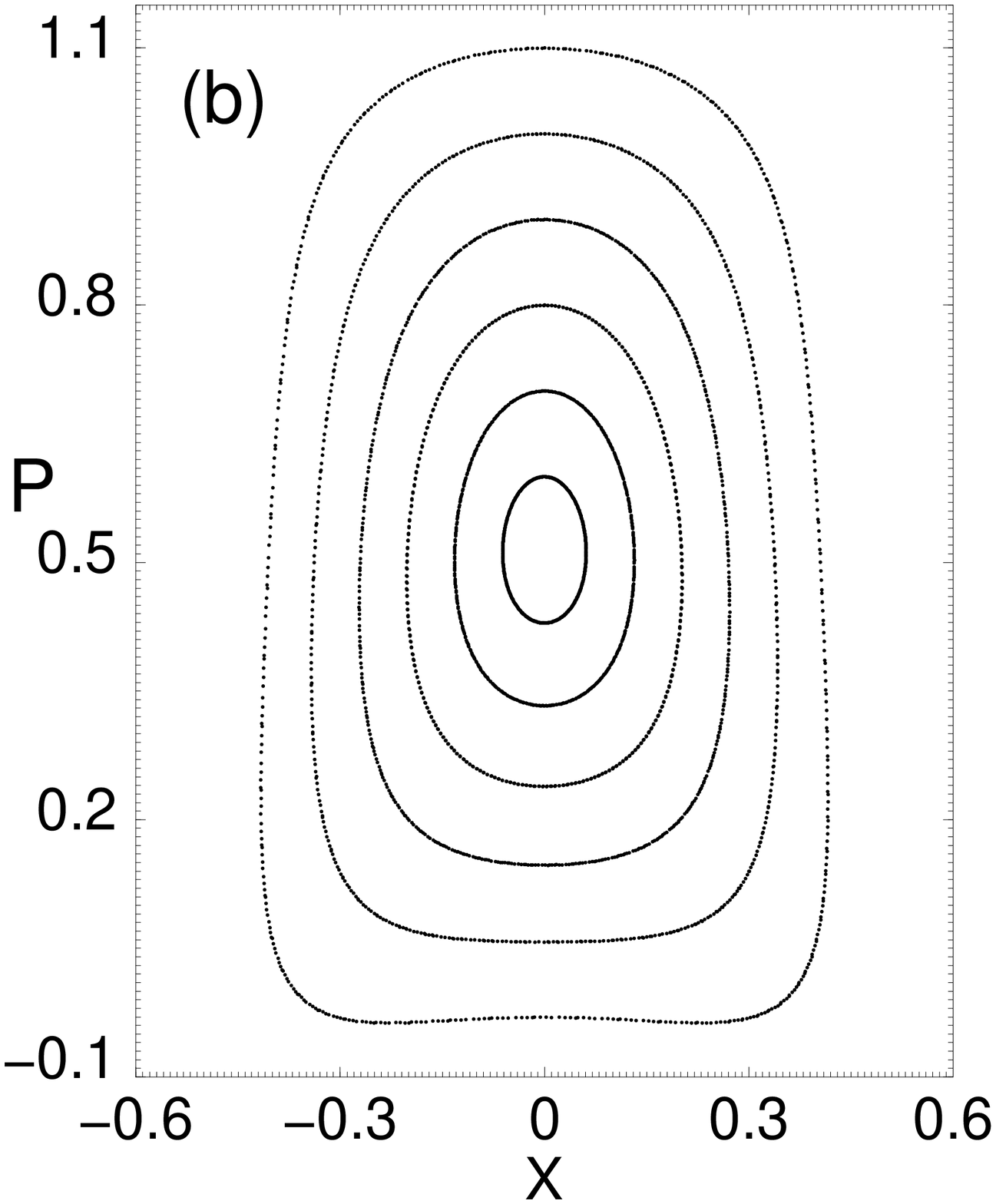,width=8cm,height=8cm}}}
 \vspace{0.4cm}
\centerline{\mbox{\psfig{file=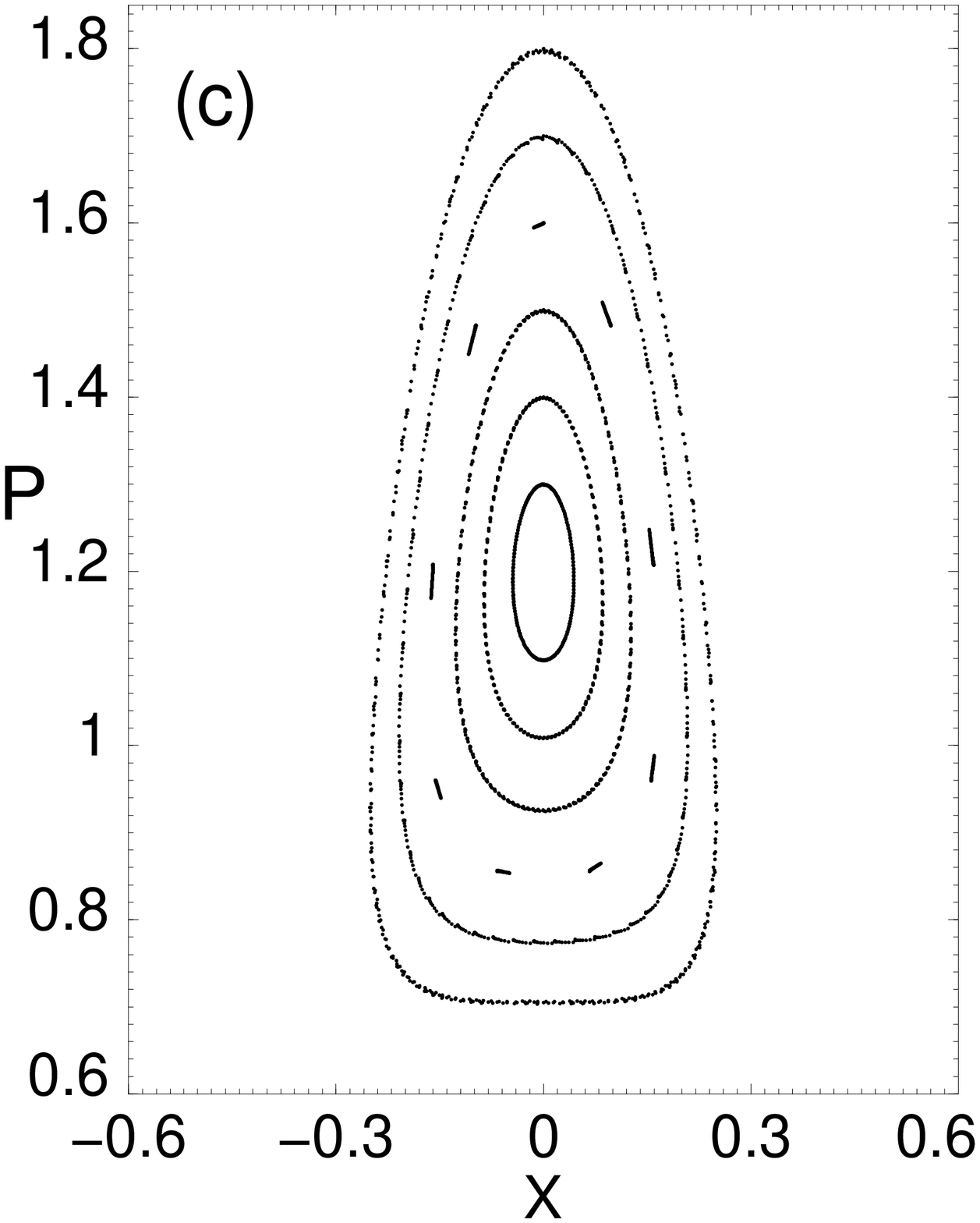,width=8cm,height=8cm}
       \psfig{file=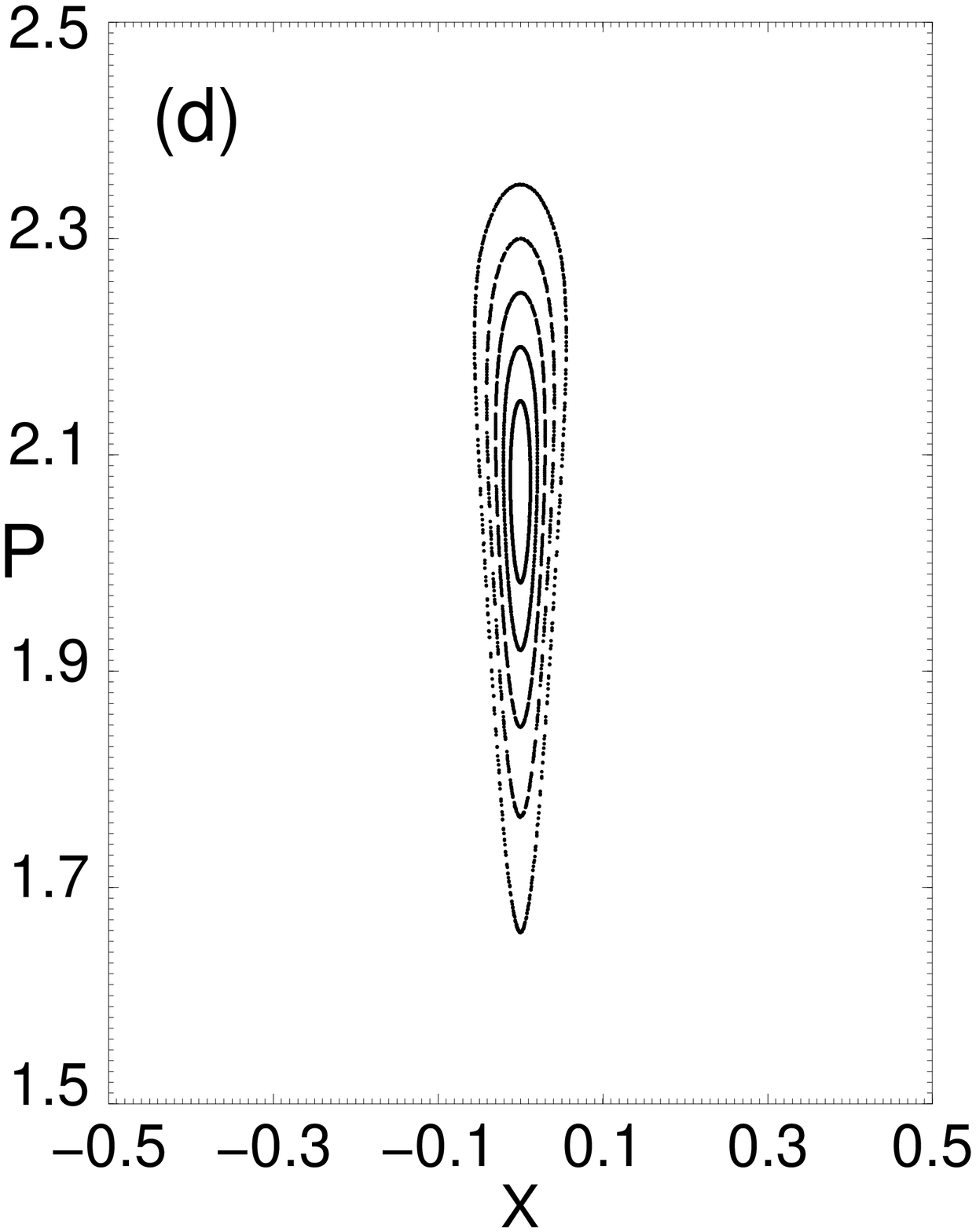,width=8cm,height=8cm}}}
 \vspace{0.4cm}
 \caption{Trajectories in the central resonance
 cell at $\ell=4$, $\delta=0$ under influence of the perturbation
with the  amplitude, $\epsilon$: a) $\epsilon=0.5$, b) $\epsilon=2$,
 c) $\epsilon=5$, d) $\epsilon=10$.}
 \label{fig:4}
 \end{figure}

Similar features were observed in dependence of the dynamics
in the central cell
on $\epsilon$ for the case $\ell=5$, shown in Figs. 5 (a) - 5 (d). 
Comparison of the data for $\ell=5$ in Figs. 5 (a) - 5 (d)
with those for $\ell=4$ in Figs. 4 (a) - 4 (d) allows us to 
conclude that the area of the 
central cell increases with increasing $\ell$, and chaotization 
of the motion in the central cell at 

 \vspace{0.9cm}
 \begin{figure}[tb]
 \begin{center}
 \mbox{\psfig{file=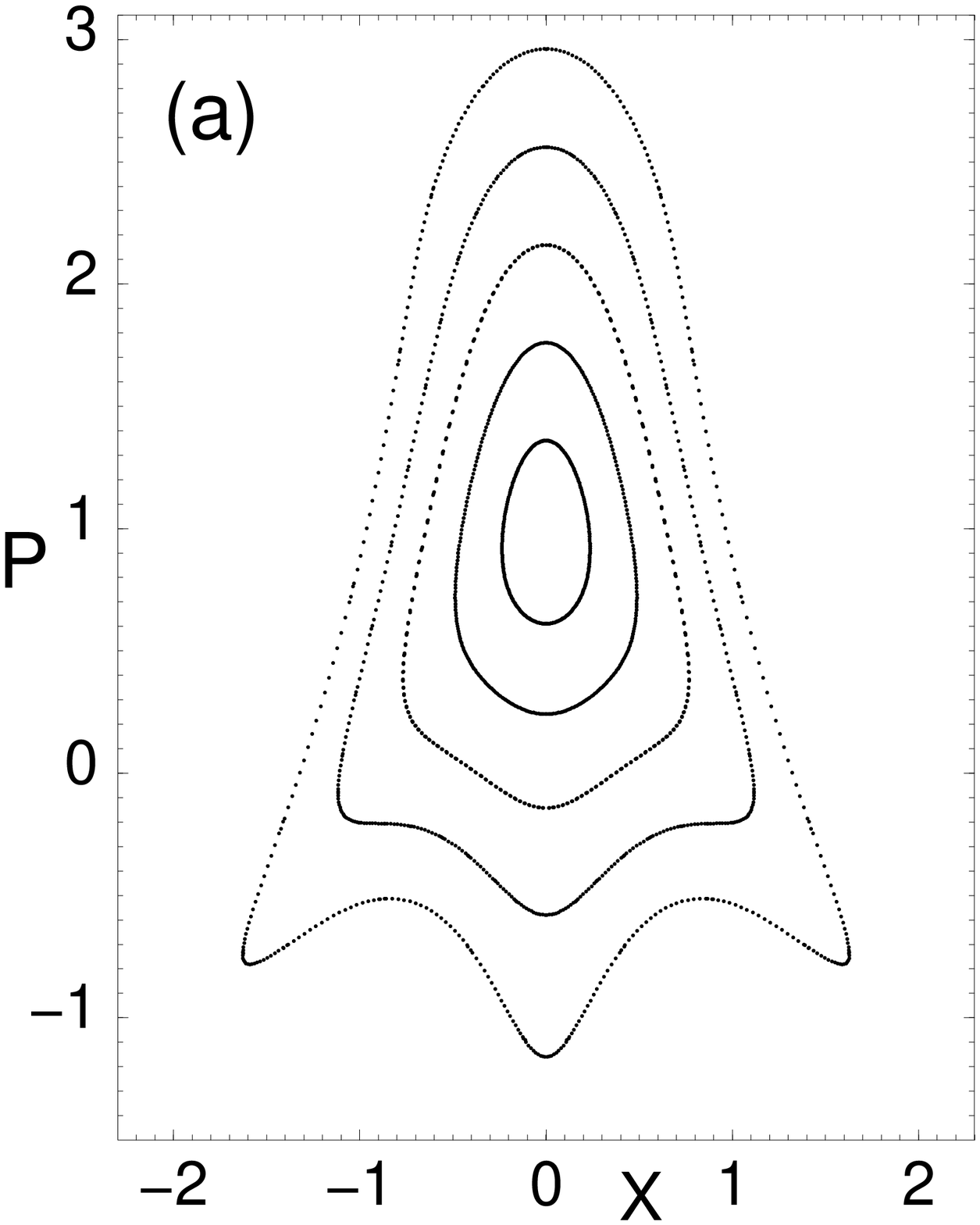,width=8cm,height=8cm}
       \psfig{file=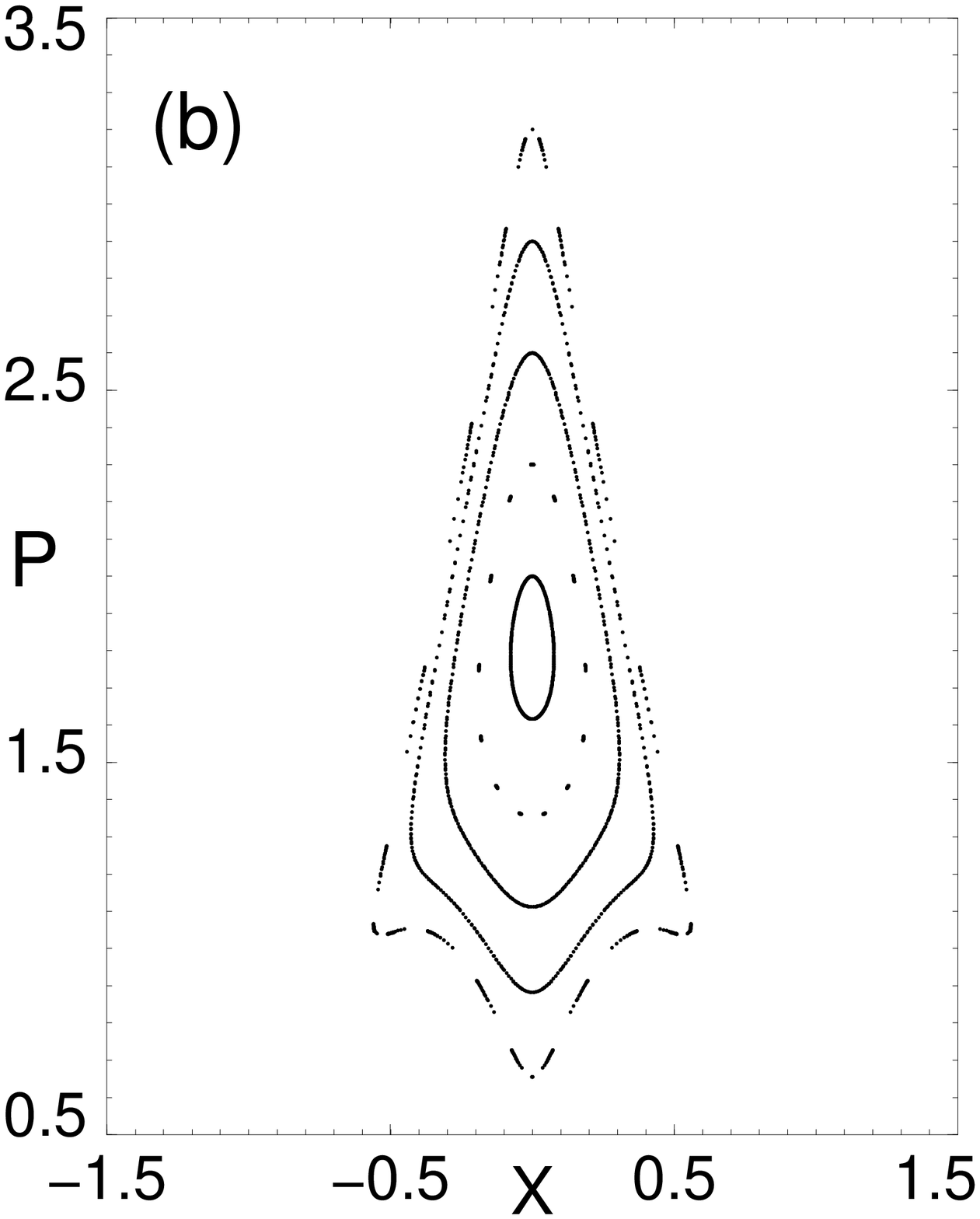,width=8cm,height=8cm}}
 \vspace{0.7cm}
 \mbox{\psfig{file=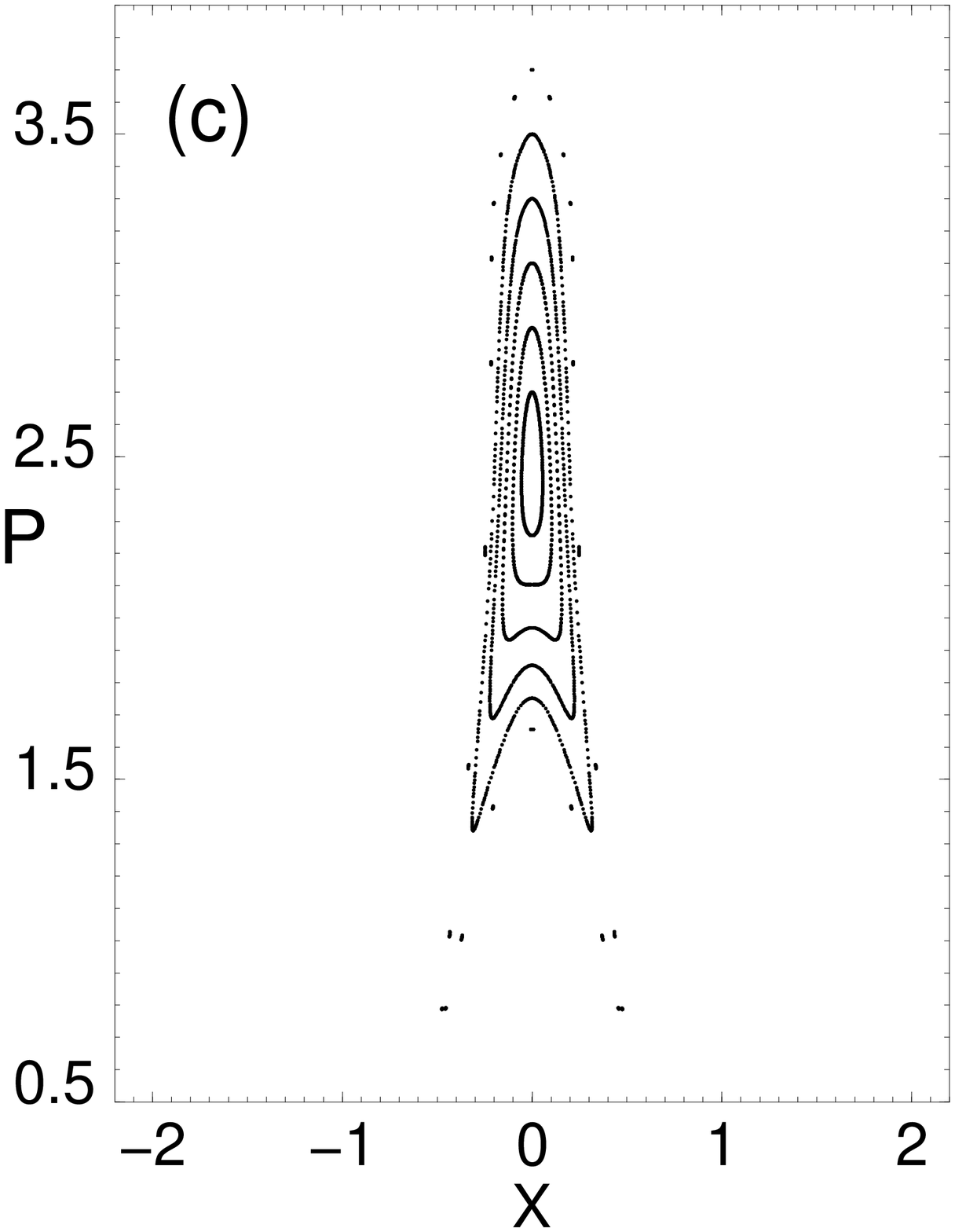,width=8cm,height=8cm}
       \psfig{file=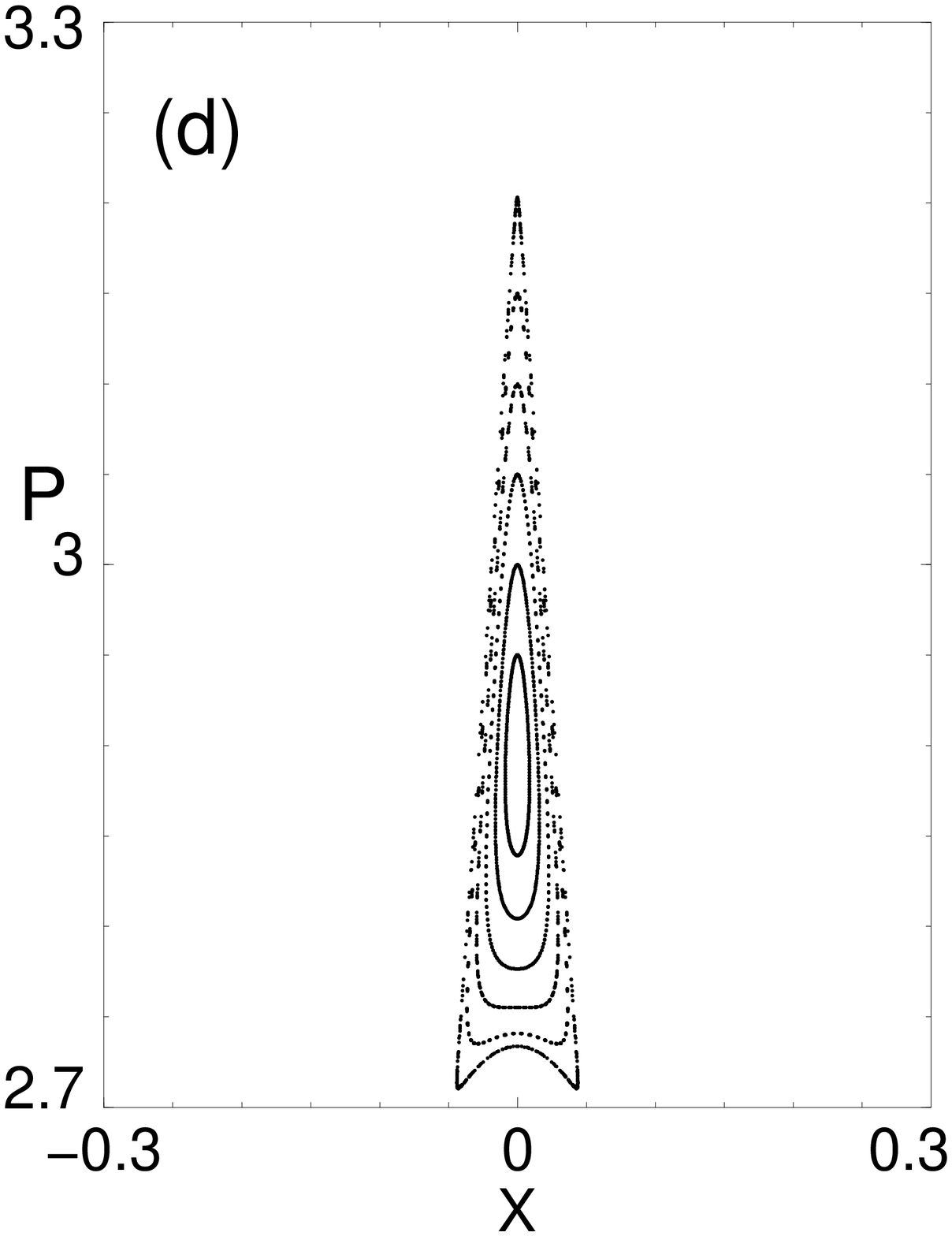,width=8cm,height=8cm}}
 \caption{Trajectories in the central resonance
 cell at $\ell=5$, $\delta=0$ under influence of the perturbation
 with the amplitude, $\epsilon$: a) $\epsilon=5$, b) $\epsilon=10$,
 c) $\epsilon=15$, d) $\epsilon=19$.}
 \end{center}
 \label{fig:5}
 \end{figure}

\noindent
larger values of $\ell$ requires
larger values of $\epsilon$. In other words, the motion in the 
central cell becomes more stable with increasing the resonance 
number, $\ell$. 
In order to understand the observed dynamics, we 
again included into consideration only the 
first order terms in $X$ in Eq. (\ref{e_motion1}),
and computed the dynamics for large values of $\epsilon$. 
The trajectories described by the approximate equation (\ref{Mathieu})
for $\epsilon=5$, $\ell=4$ and for $\epsilon=9.5$, $\ell=5$ are
shown in Figs. 6 (a), 6 (b). The following features can be observed.
i) As follows from our calculations, 

 \begin{figure}[tb]
 \begin{center}
 \mbox{\psfig{file=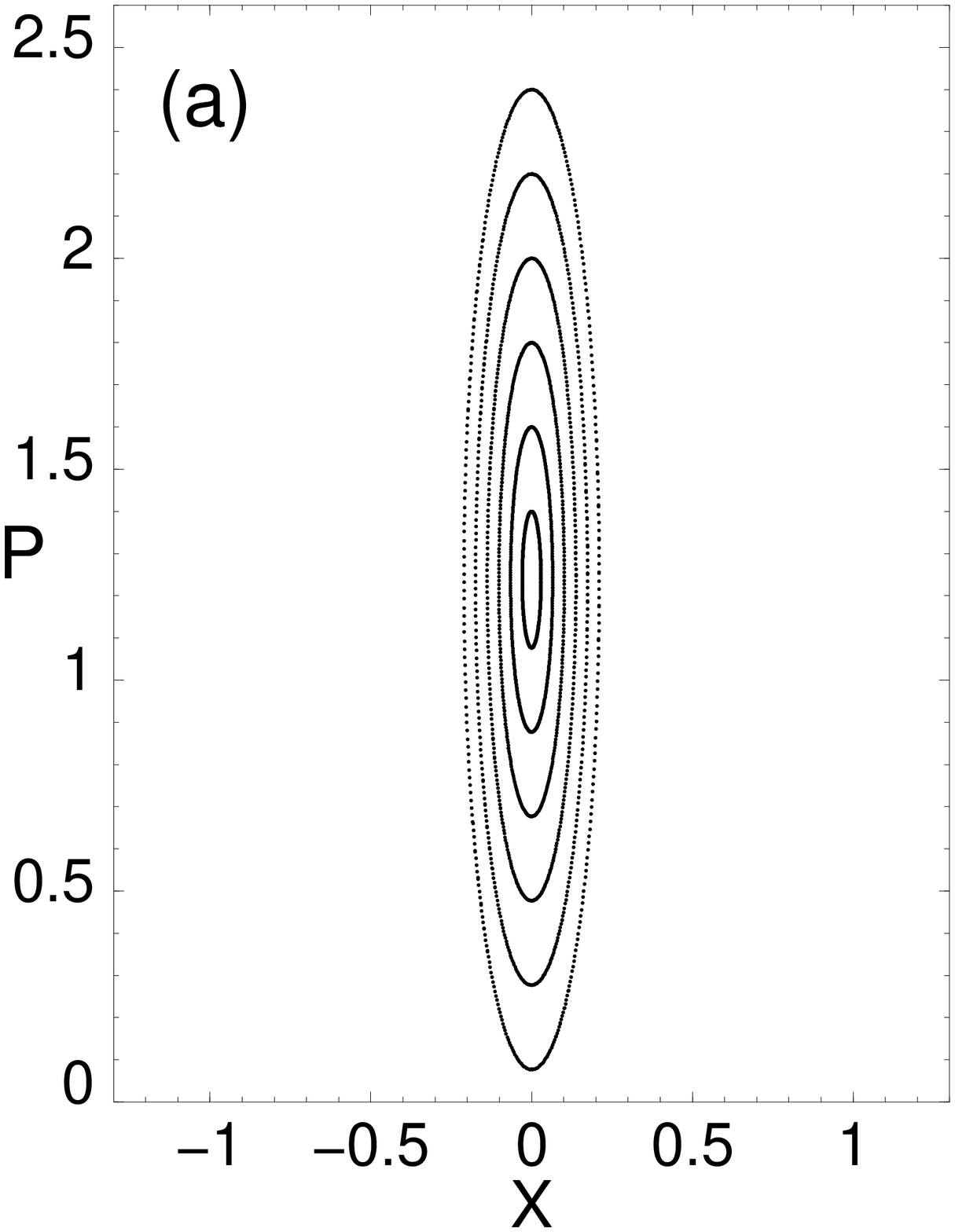,width=8cm,height=8cm}
       \psfig{file=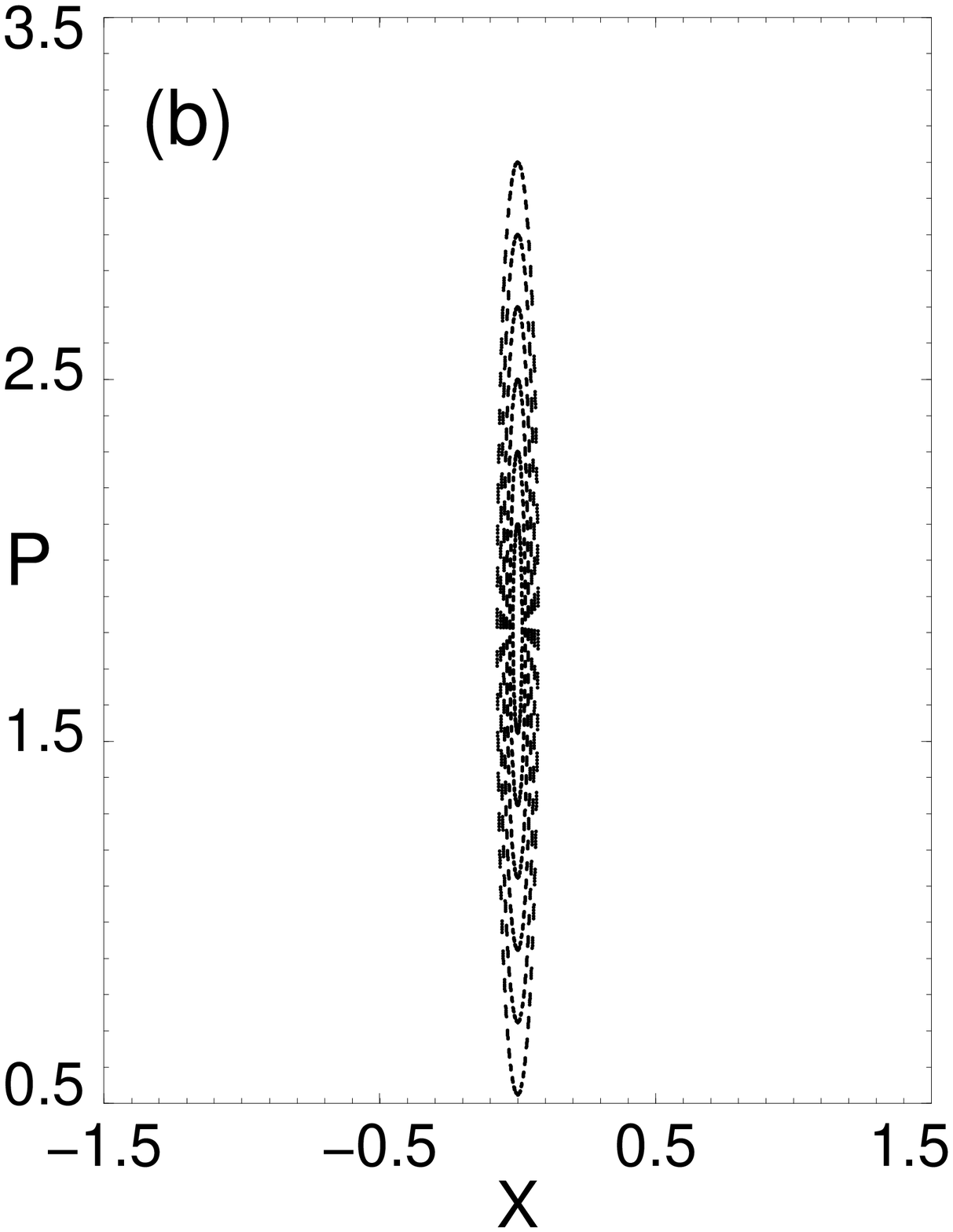,width=8cm,height=8cm}}
 \end{center}
 \label{fig:6}
 \end{figure}

 FIG. 6. The phase trajectories described by the approximate 
 equation (\ref{Mathieu}) for $\delta=0$ and (a) $\ell=4$, $\epsilon=5$,
 (b) $\ell=5$, $\epsilon=9.5$.
\vspace{0.5cm}

\noindent
the phase portrait shifts up 
from the point $(X=0,\,P=0)$
under the
influence of the term in the right-hand side of the approximate
equation (\ref{Mathieu}). ii) Comparison of Fig. 6 (a) with Fig. 4 (c)
and Fig. 6 (b) with Fig. 5~(b) allows us to conclude that the terms 
of the high order in $X$ in Eq. (\ref{e_motion1}) change the 
shape of trajectories and restrict the region of stable motion. 
iii) The motion described by the approximate equation (\ref{Mathieu})
becomes unstable 
at values of $\epsilon>\epsilon_\ell$ where $\epsilon_\ell$ lies
in the interval $2.3<\epsilon_3<2.4$ for $\ell=3$;
$5.5<\epsilon_4<5.6$ for 
$\ell=4$; and $9.6<\epsilon_5<9.7$ for $\ell=5$. One can see from Fig. 4 (d) 
and Figs. 5 (c), 5 (d) that the motion described by 
exact equation (\ref{e_motion1}) 
in the region $\epsilon>\epsilon_\ell$ remains stable.
Thus, the high order terms in $X$ in Eq. (\ref{e_motion1})    
stabilize the dynamics at large values of the perturbation amplitude, 
$\epsilon$.    

 \begin{figure}[tb]
 \begin{center}
\setcounter{figure}{6}
 \mbox{\psfig{file=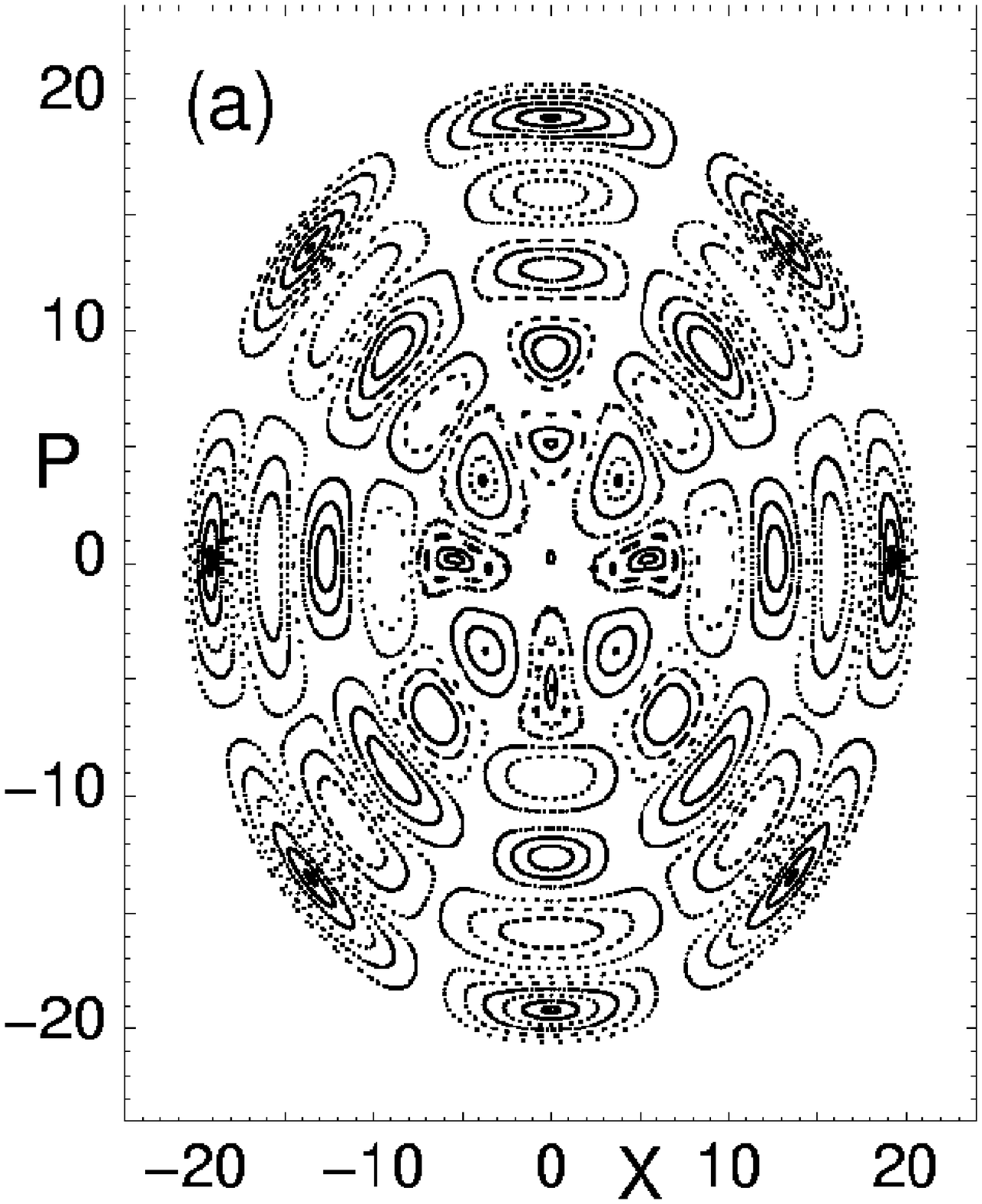,width=8cm,height=8cm}
       \psfig{file=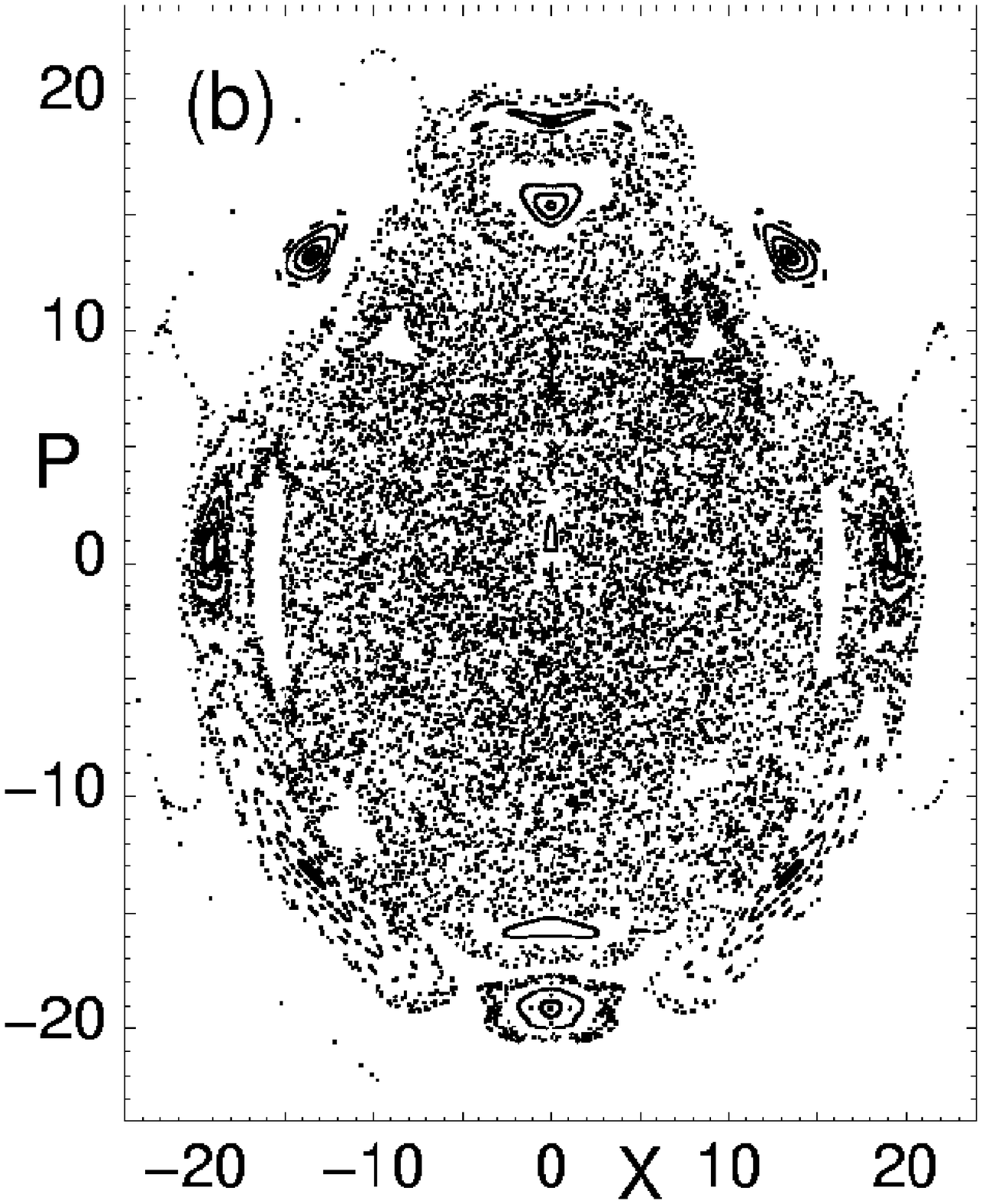,width=8cm,height=8cm}}
 \mbox{\psfig{file=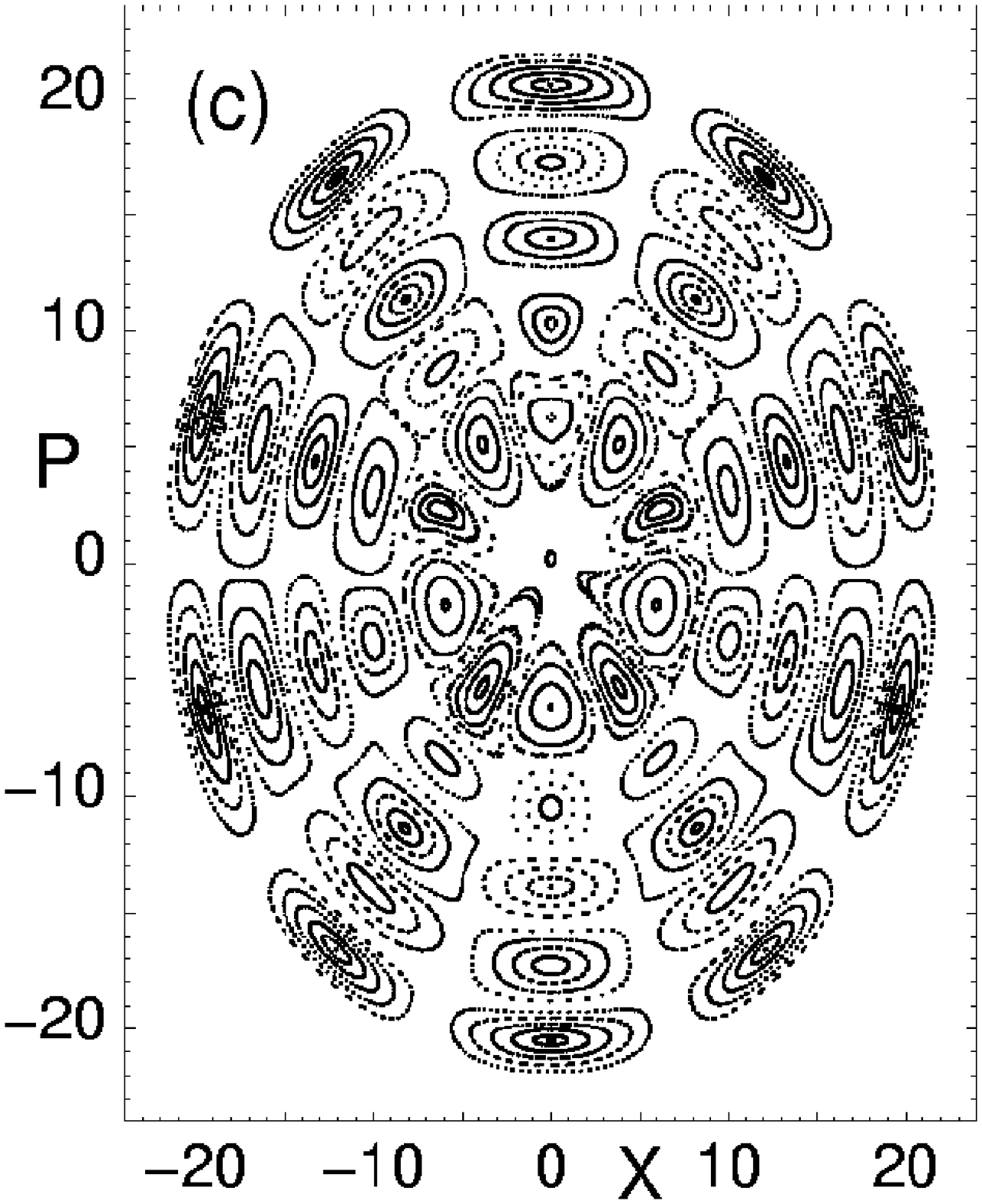,width=8cm,height=8cm}
       \psfig{file=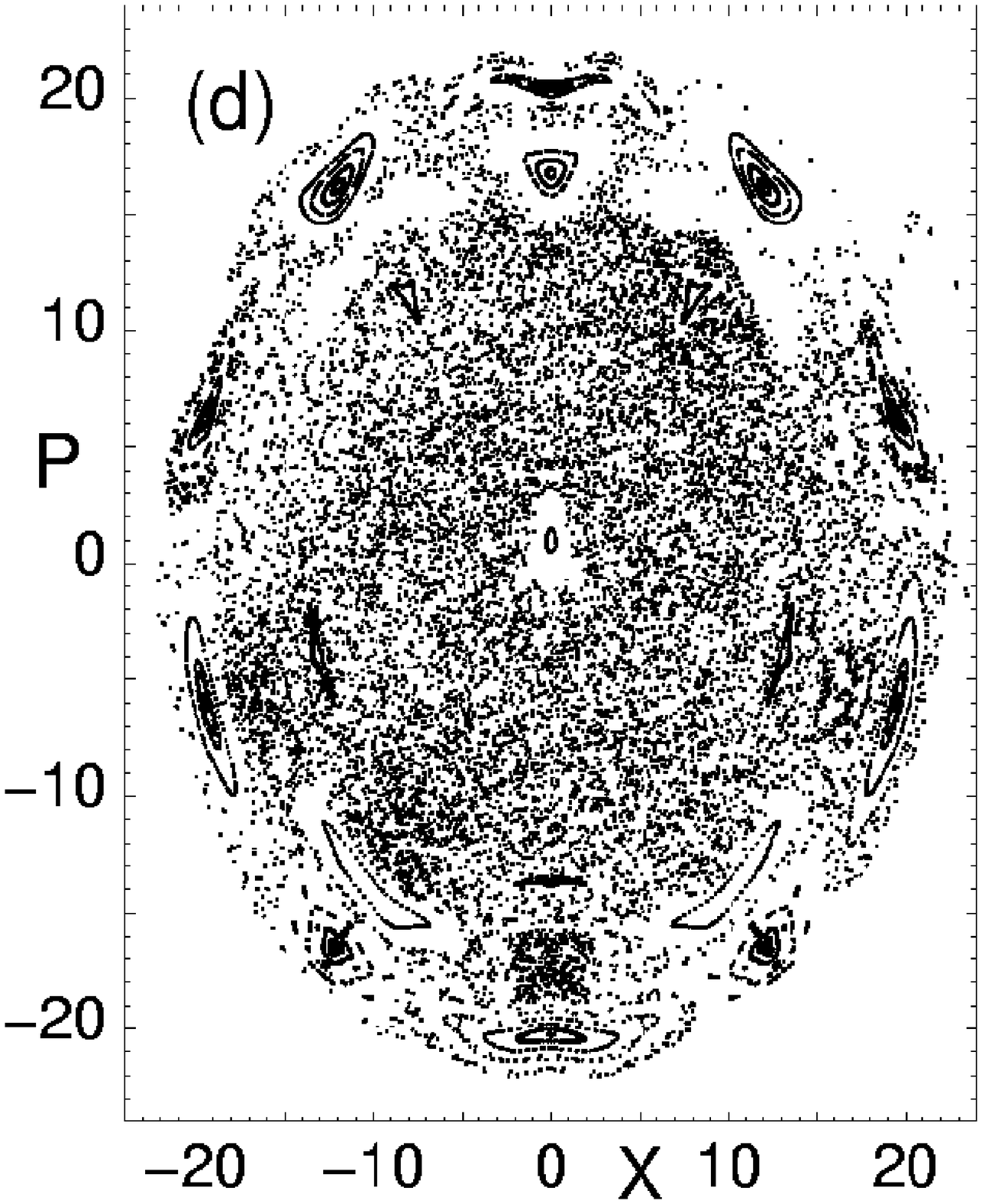,width=8cm,height=8cm}}
 \vspace{0.2cm}
 \caption{Influence of chaos on different resonant cells, $\delta=0$,
 (a) $\ell=4$, $\epsilon=1$; (b) $\ell=4$, $\epsilon=5$;
 (c) $\ell=5$, $\epsilon=1$; (d) $\ell=5$, $\epsilon=5$.}
 \end{center}
 \label{fig:7}
 \end{figure}

Let us compare the classical dynamics in the central 
cell with the dynamics in the other cells
when the perturbation parameter $\epsilon$ is not small. The results of 
calculation of the dynamics in several cells are shown in
Figs. 7 (a) - 7 (d).
From comparison Fig. 7 (b) with Figs. 4 (c), 4 (d) and Fig. 7 (d) with
Figs. 5 (a) - 5 (d) one can see that the trajectories in the central
cell remain stable, while other nearest cells are completely destroyed
by chaos. The extremely high stability of trajectories in the central
cell can be explained by relatively small influence on the
dynamics of the 
terms of high order in $X$, oscillating with different frequencies, because 
their amplitudes are small at small $X$ (or $\rho$).

\section{The classical dynamics near the CGS in the near resonance case}
Now, let us consider the CGS in the near resonance case, when $\delta\ne 0$. 
The resonant Hamiltonian (\ref{cl_H3}) takes form 
\begin{equation}
\label{dw_H}
H_\ell=\tilde I(\delta\omega)+v_0J_\ell(\rho)\cos\theta.
\end{equation}
The stationary points for the dynamics generated by the 
Hamiltonian (\ref{dw_H}) 
are defined by the conditions,
$$
\dot\theta=\partial H_\ell/\partial\tilde I=0,\qquad
\dot{\tilde I}=-\partial H_\ell/\partial\theta=0.
$$
Positions of the elliptic stationary points are given by the expressions,
\begin{equation}
\label{elliptic1}
v_0{\partial J_\ell[kr(\tilde I_e)]\over\partial\tilde I}=\mp \delta\omega,
\qquad \theta_e=0,\pi,
\end{equation}
where the sing $``-''$ corresponds to the stable point, with the angle 
$\theta_e=0$ with the $P$-axis, and the sing $``+''$ corresponds 
to the stable point with the angle $\theta_e=\pi$. 
In the dimensionless form Eq. (\ref{elliptic1}) is, 
\begin{equation}
\label{elliptic}
\frac 1{\rho_e}{\partial J_\ell(\rho_e)\over\partial\rho}=
\mp{\delta\over\ell\epsilon},
\end{equation}
where $\rho_e=kr(\tilde I_e)$. 
For the positions of the hyperbolic stationary points one has,
\begin{equation}
\label{hyperbolic}
J_\ell\left[ kr(\tilde I_h)\right]=0, \qquad \theta_h=\pm\frac\pi 2.
\end{equation}
As one can see from Eq. (\ref{elliptic}), the number of the elliptic stable
points in the near resonance case, when $\delta\ne 0$,
is finite because the right-hand side of Eq. (\ref{elliptic}) is constant
while the left-hand side oscillates, and decreases on average.
As a consequence, there is a finite number of the resonance cells. 

The motion near the CGS can be described by the approximate equation 
(\ref{Mathieu}) with the parameter $a_\ell$ equal to: 
$a_\ell=[2/(\ell-\delta)]^2$. It is known \cite{Mathieu} that
at small $\epsilon$ and $\delta\ne 0$ the Mathieu equations have the stable 
solutions at any $\ell$  including 
the cases $\ell=1$ and $\ell=2$. 

Let us consider the cases $\ell=1$ and $\ell=2$ at $\delta\ne 0$ in 
detail. In the case $\ell=1$, one may use the results of the resonance 
theory at arbitrary small $X$ and $P$, because the term of the lowest order 
in $X$ (proportional to $X$) in the Hamiltonian (\ref{cl_H}) is resonant.
Let us suppose that the dimensionless radius $\rho_e$ in Eq. (\ref{elliptic}) 
is small, $\rho_e\ll 1$. Then $J_1'(\rho_e)\approx 1/2$, and  
Eq. (\ref{elliptic}) yields, 
\begin{equation}
\label{roe}
\rho_e=\mp \epsilon/(2\delta\ell).
\end{equation}
Thus, shift of the stable elliptic point from the CGS is small,
$\rho_e\ll 1$, when the condition $\epsilon\ll 2|\delta|\ell$ is satisfied.
At small values of the wave amplitude,
$\epsilon$, shift of the elliptic stable point from the point $X=0,\,P=0$ is
proportional to $\epsilon$. One can see from Eq. (\ref{roe}) that at
$\ell=1$ one elliptic stable point exists at arbitrary small value of
$\epsilon$ (which follows also from the theory of Mathieu functions).

 \begin{figure}[tb]
 \begin{center}
 \mbox{\psfig{file=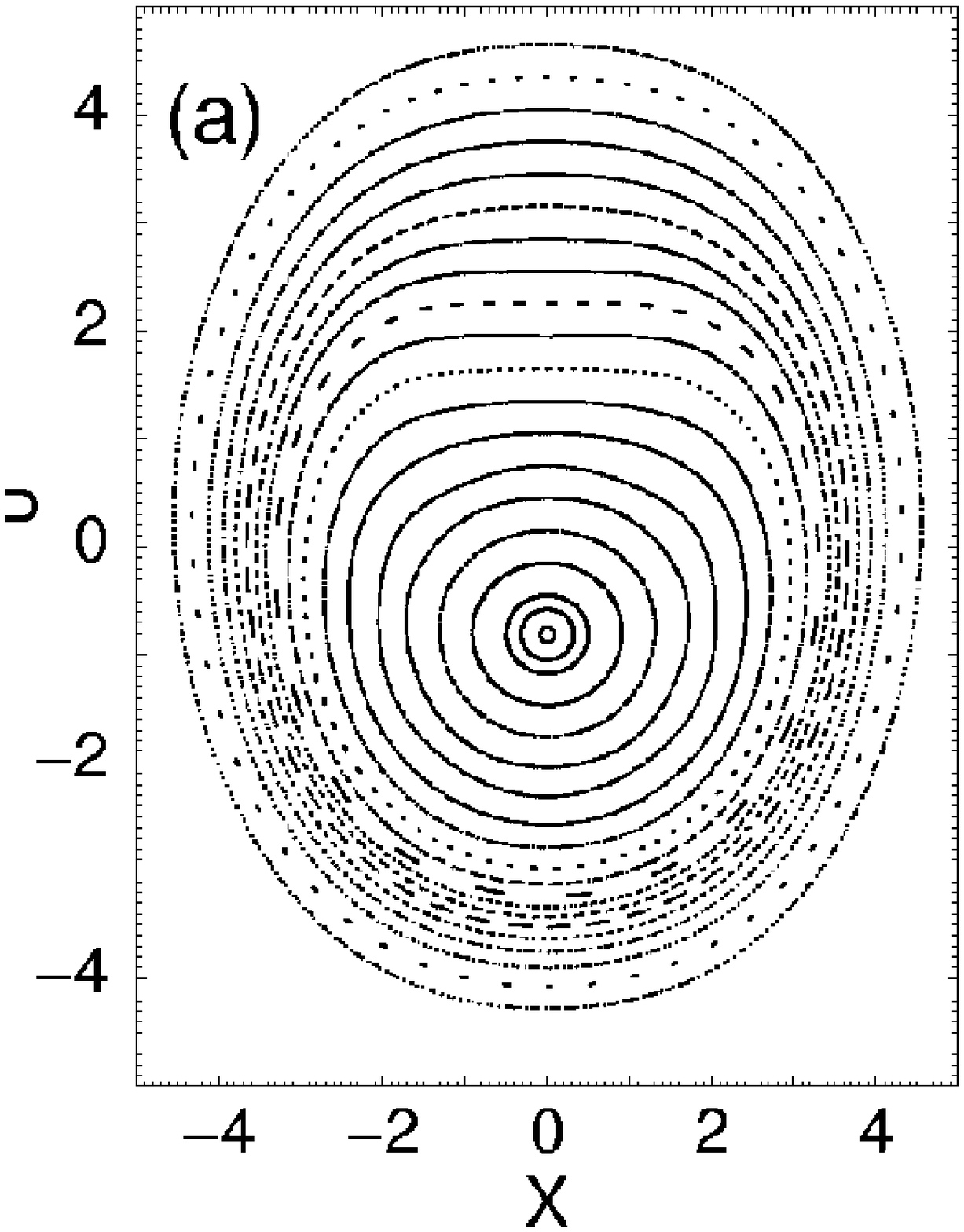,width=5.4cm,height=5.4cm}
       \psfig{file=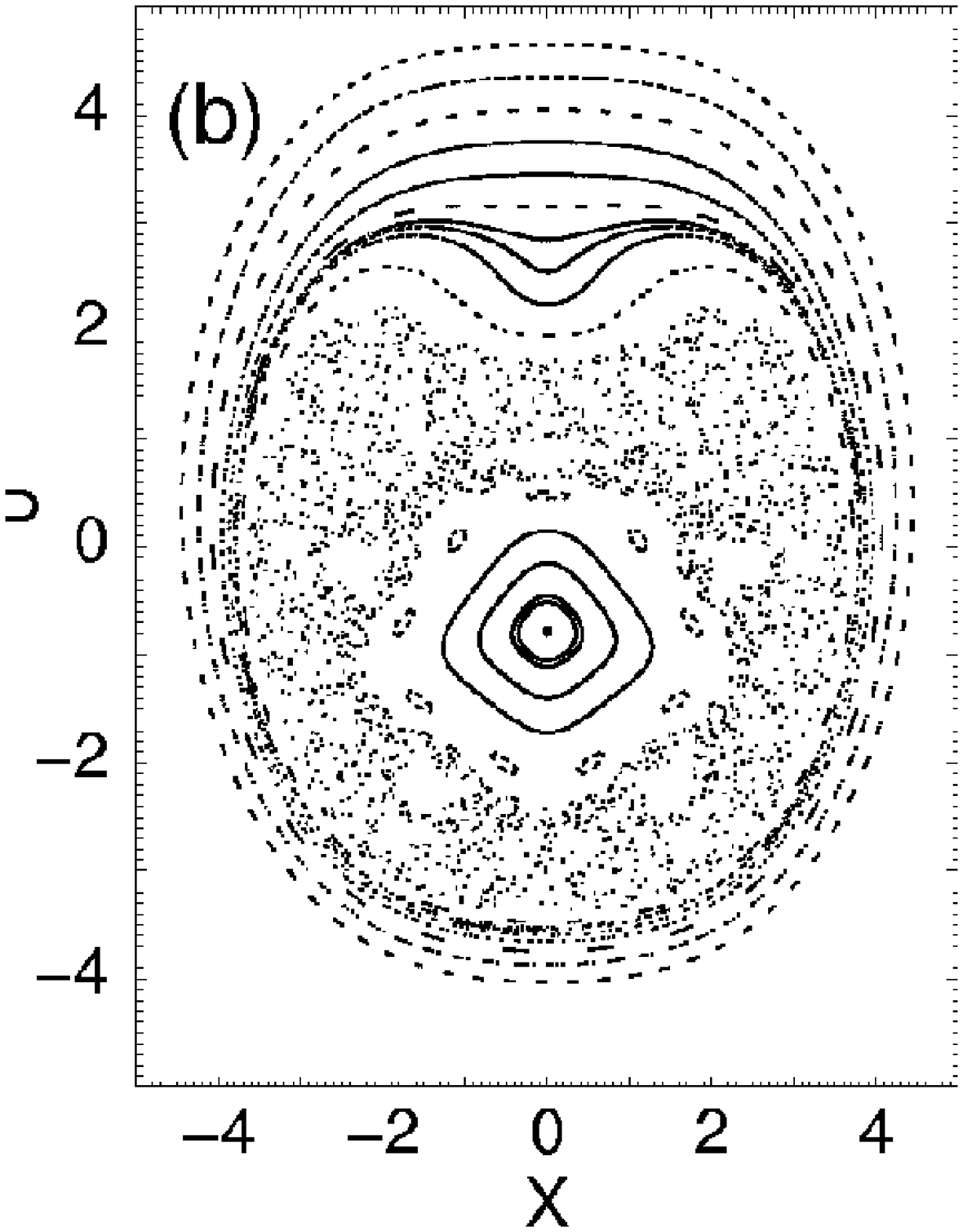,width=5.4cm,height=5.4cm}
       \psfig{file=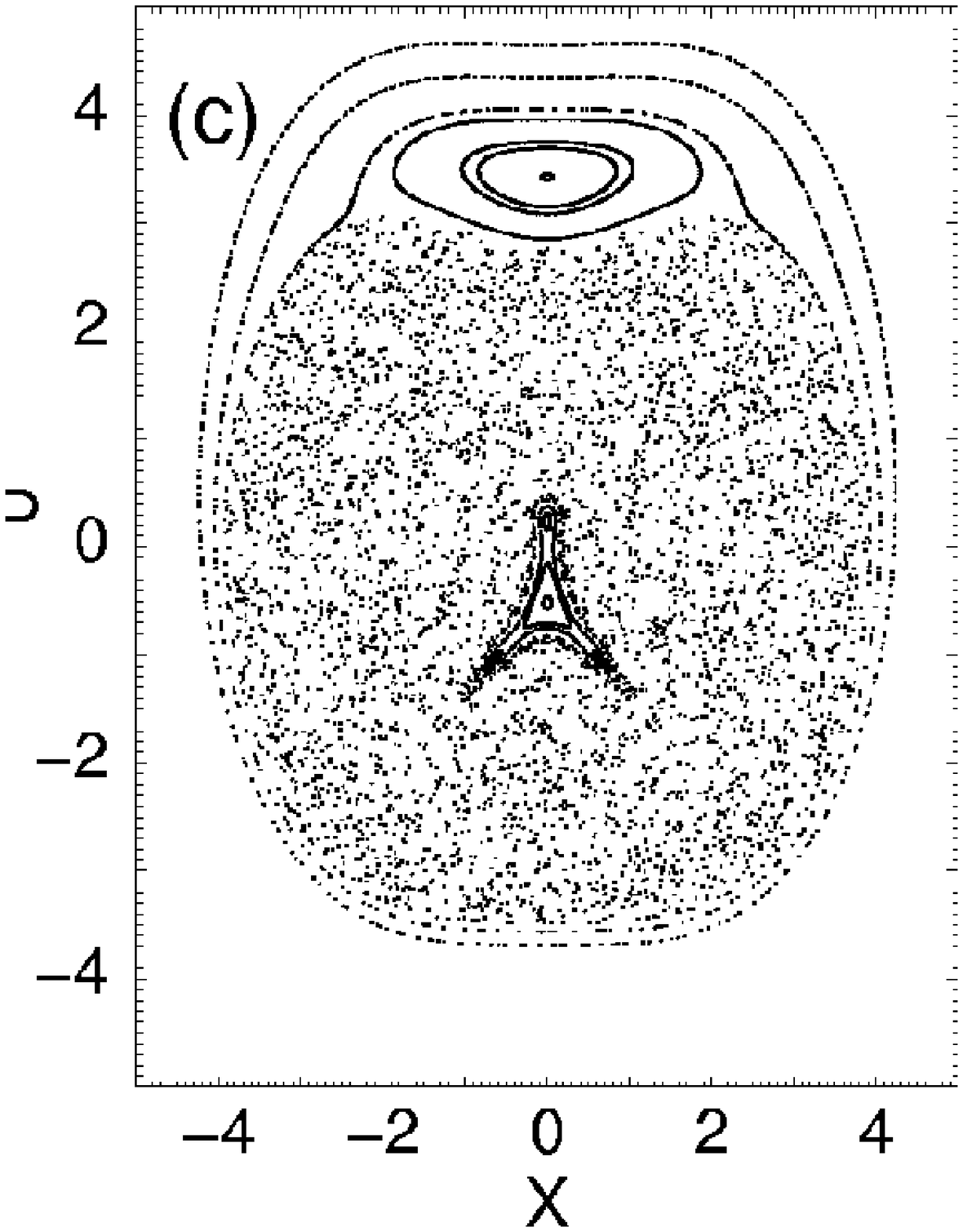,width=5.4cm,height=5.4cm}}
 \caption{The phase space for the near resonance case, $\delta=0.1$,
 $\ell=1$, and (a) $\epsilon=0.4$, (b) $\epsilon=0.7$, (c) $\epsilon=1.2$}
 \end{center}
 \label{fig:8}
 \end{figure}

When $\epsilon$ is small, the phase trajectories 
are the circles with the center located near the CGS.
Figuratively speaking, in the case $\delta\ne 0$  and  $\epsilon\ll 1$
there is only one resonant (central) cell with an infinite 
area, because at small enough value of $\epsilon$ the
equation (\ref{elliptic}) has no other solutions, except for 
Eq. (\ref{roe}), and in the phase space there are 
no other cells, except for the central one. 
When $\epsilon$ increases (we suppose $\epsilon>0$)
and $\delta>0$, the stable point shifts down, as shown in Fig. 8 (a),
because the left-hand side of Eq. (\ref{elliptic}) is positive
and we should take the sign ``$+$'' in the right-hand side,
which corresponds to the shift in the direction $\theta=\pi$.  

When $\epsilon$ increases (see Fig. 8 (b), 8 (c)), the dynamical
chaos appears, and the
area of the central cell decreases. As before, we have considered
the influence of the  high order terms in $X$ in the exact equation of motion
(\ref{e_motion1}) on the dynamics described by the approximate equation
(\ref{Mathieu}). The approximate equation
(\ref{Mathieu}) yields unstable solutions when $\epsilon>\epsilon_1$, where
$|\epsilon_1|=\sqrt{24|\delta|/5}$ if $\delta>0$, and
$\epsilon_1=\sqrt{24|\delta|}$ if $\delta<0$.\cite{Landau}
The parameter $\delta=0.1$ yields $\epsilon_1=0.69$.\footnote{We
also checked these criterion numerically.} As one can see from
Figs. 8 (b), 8 (c) the central cell remains undestroyed. Thus,
the nonlinear terms stabilize the dynamics in the near resonance case,
similar to the case of exact resonance.
At $\epsilon=1.2$ in Fig. 8 (c)
one more cell is generated, because the condition (\ref{elliptic}) is
satisfied for two values of $kr$.

 \begin{figure}[tb]
 \begin{center}
 \mbox{\psfig{file=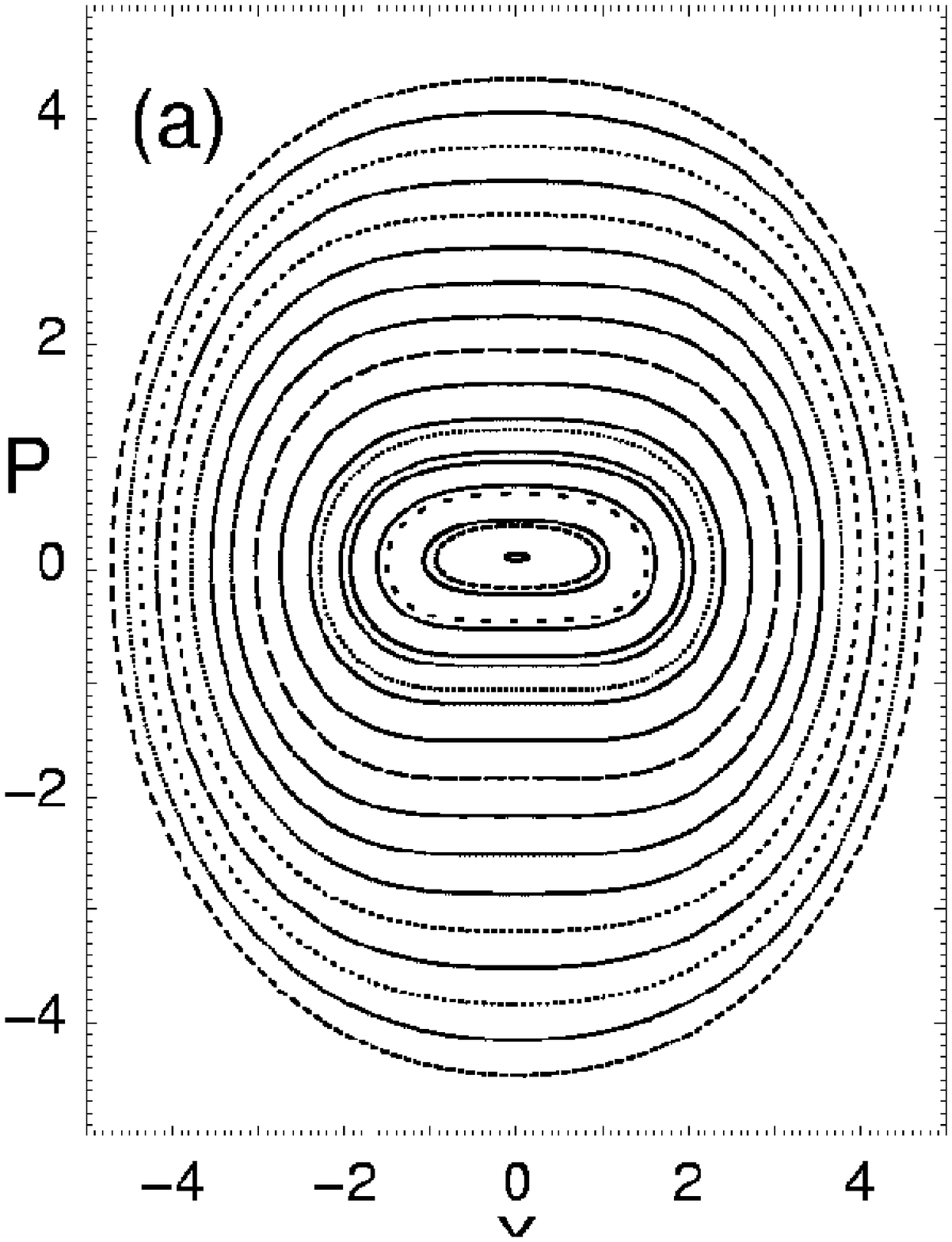,width=8cm,height=8cm}
       \psfig{file=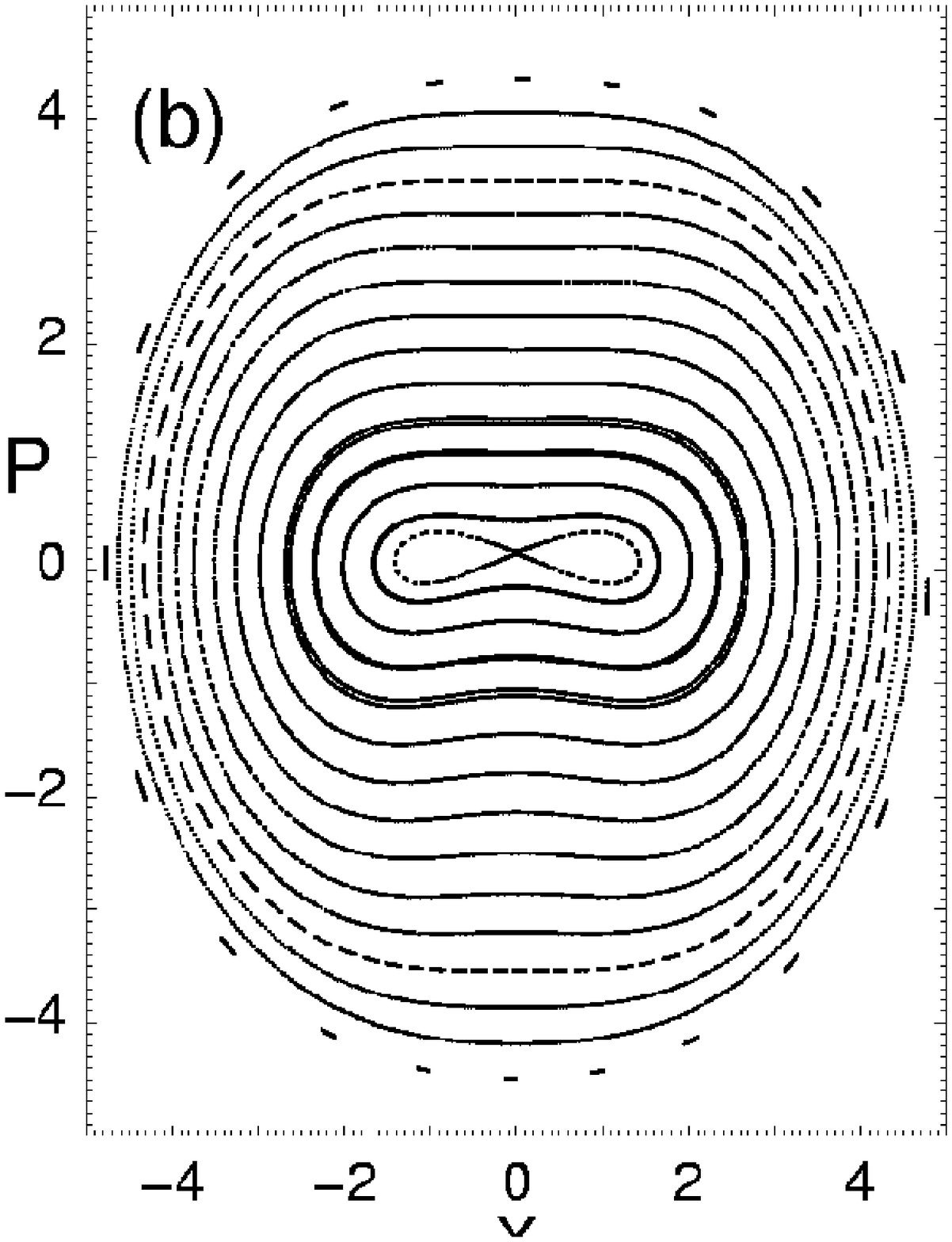,width=8cm,height=8cm}}
 \vspace{0.5cm}
 \caption{The phase space for the near resonance case, $\delta=0.1$,
 $\ell=2$, and (a) $\epsilon=0.17$, (b) $\epsilon=0.22$}
 \end{center}
 \label{fig:9}
 \end{figure}

Unlike the case $\ell=1$, 
when $\ell=2$ and $\epsilon$ is small enough (see Figs. 9 (a)), the 
stable point does not shift from the point $(X=0,\,P=0)$. Instead, 
in Fig. 9 (b) we
observe bifurcation at the value $\epsilon=\epsilon_2$, where 
$\epsilon_2$ can be estimated from the solution of the approximate
equation (\ref{Mathieu}). Namely, up to the second order in $\delta$,  
the dynamics becomes unstable at
$\epsilon_2=2\delta-\delta^2/2$.\cite{Landau} Our computed value
of $\epsilon_2$ lies in the interval $0.185<\epsilon_2<0.186$ which 
is slightly less than the estimated quantity due to the influence 
of nonlinear in $X$ terms, which are neglected in the approximate equation
(\ref{Mathieu}).        
    
\begin{figure}[tb]
\centerline{\psfig{file=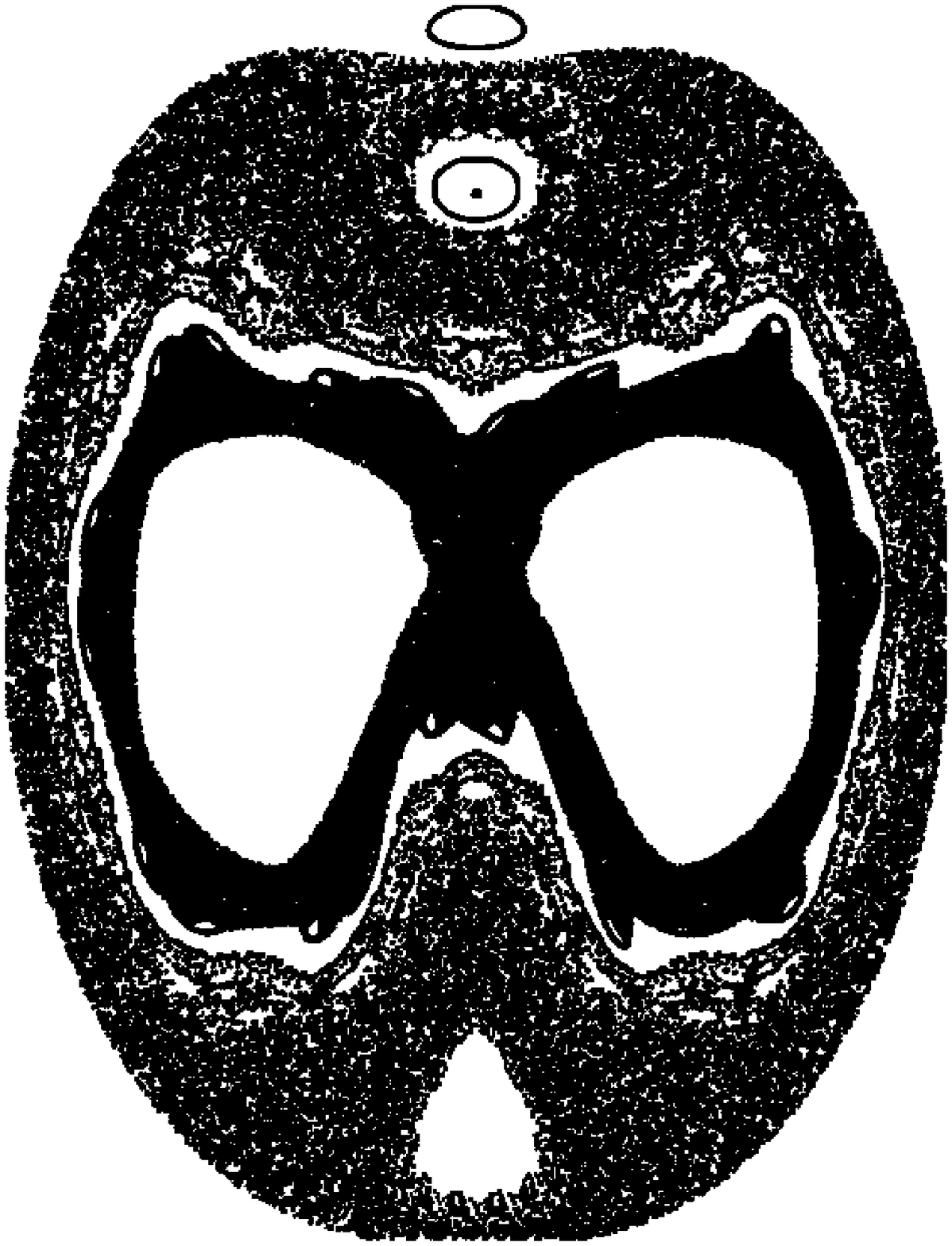,width=13cm,height=13cm}} 
\caption{The same as in Figs. 9 (a), 9 (b) but for $\epsilon=0.8$} 
\label{fig:10}
\end{figure}

As shown in Fig. 10, 
at further increase of $\epsilon$, two stable points, formed
after bifurcation, diverge 
at larger distance from each other, the chaotic area increases, 
and  additional cells appear because the condition (\ref{elliptic})
is satisfied for more number of points ($kr(\tilde I_e)$).      

\section{Stability of the quantum ground state}
Now we consider a stability of the ground state of the quantum
harmonic oscillator (QGS) under the same conditions as in the classical model.
The quantum Hamiltonian is,
\begin{equation}
\label{q_H}
\hat H=\frac{\hat p^2}{2M}+\frac{M\omega^2}2x^2
+v_0\cos(kx-\Omega t)=\hat H_0+\hat V(x,t),
\end{equation}
where $\hat p=-i\hbar\partial/\partial x$, and the same notation as in Eq. (\ref{cl_H}) where used. Since the 
Hamiltonian (\ref{q_H}) is periodic in time, we can use the Floquet 
theorem, and write the solution of the Schr\"odinger equation in the form, 
\begin{equation}
\label{qef}
\psi_q(x,t)=\exp(-iE_qt/\hbar)u_q(x,t),
\end{equation}
where $E_q$ is the quasienergy, $\psi_q(x,t)$ is the quasienergy (QE) 
eigenfunction, and the function $u_q(x,t)$ is periodic in time, 
$u_q(x,t)=u_q(x,t+T)$, where $T=2\pi/\Omega$. 
We expand $u_q(x,t)$ in the basis of the unperturbed harmonic 
oscillator,   
\begin{equation}
\label{qe_dec}
u_q(x,t)=\sum_{n=0}^\infty C_n^q(t)\psi_n(x),
\end{equation}
where coefficients, $C_n^q(t)$, are periodic in time, 
$C_n^q(t)=C_n^q(t+T)$. Due to periodicity of $C_n^q(t)$, the approach 
based on  Floquet states is very convenient for investigation 
of localization properties of the 
quantum system. Namely, if some initial state coincides with the
quasienergy function localized in some region of the Hilbert space,
$C_n(0)=C_n^q(0)$, then it will remain localized in this region for any time.   

We used the following numerical procedure to calculate the 
QE states. \cite{Ber1,Ber2,Reichl_a} The QE states are the 
eigenstates of the evolution operator for one period of the wave 
field, $\hat U(T)$. 
In order to build the matrix $U_{nm}$ of the operator  
$\hat U(T)$ we choose the 
representation of the Hamiltonian $\hat H_0$.   
Let us act with the
evolution operator on the wave function $\psi(x,0)$,
\begin{equation}
\label{qefind1}
\hat U(T)\psi(x,0)=\psi(x,T),
\end{equation}
and choose the initial state in the form $C_n(0)=\delta_{n,n_0}$.  
In this way we obtain a column in the evolution operator matrix,  

 \begin{figure}[tb]
 \begin{center}
 \mbox{\psfig{file=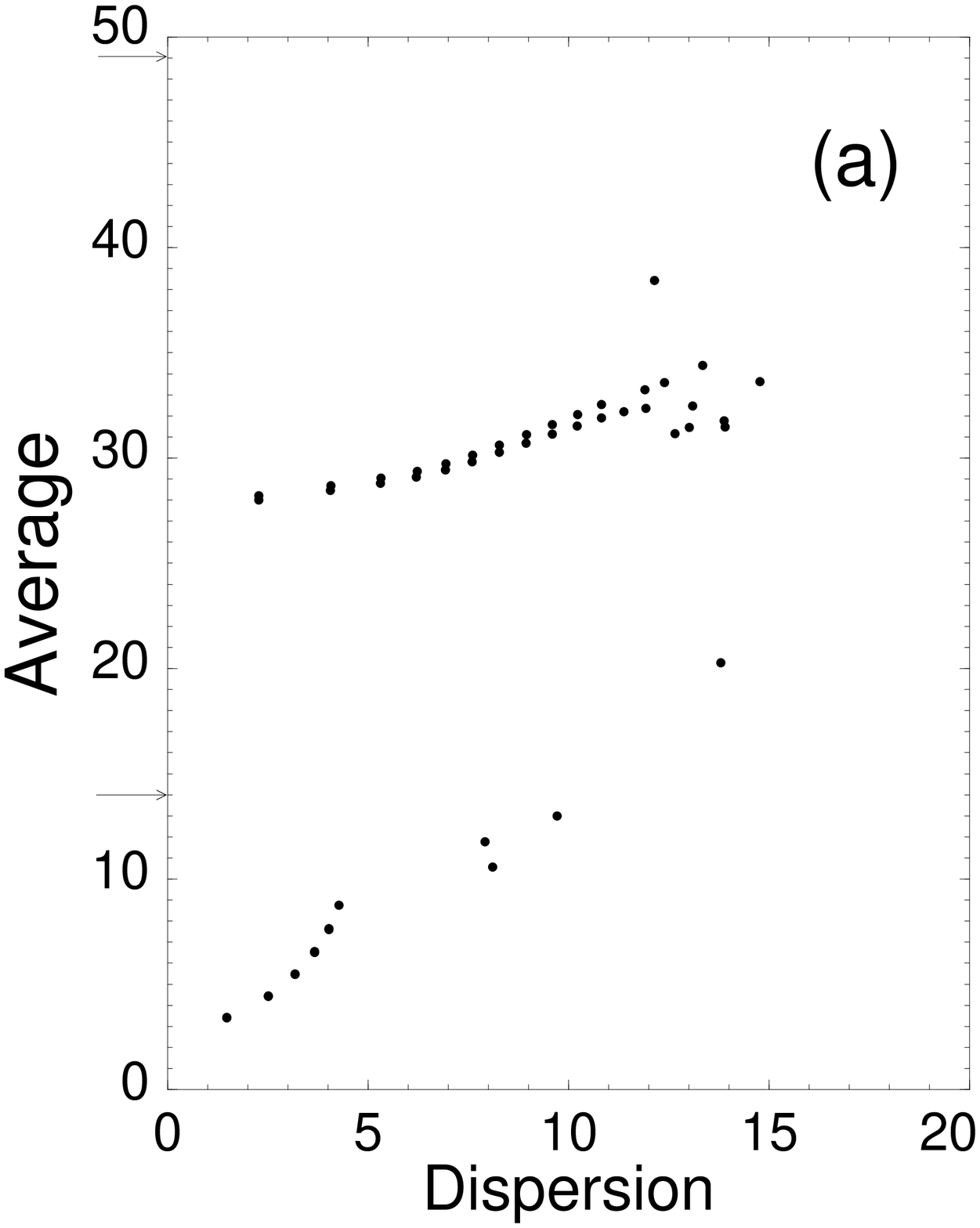,width=7cm,height=7cm}\hspace{0.5cm}
       \psfig{file=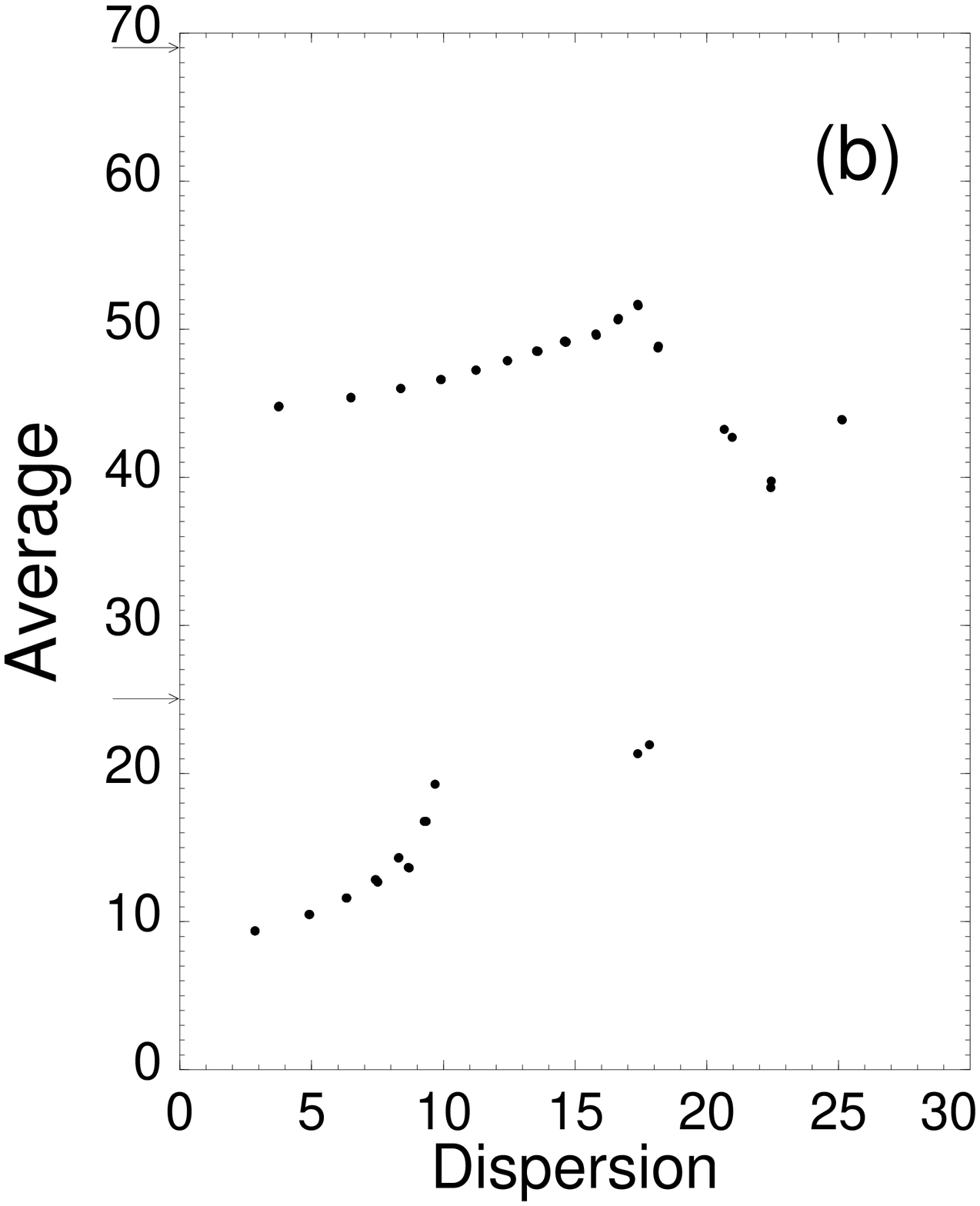,width=7cm,height=7cm}}
 \end{center}
 
 \begin{center}
 \mbox{\psfig{file=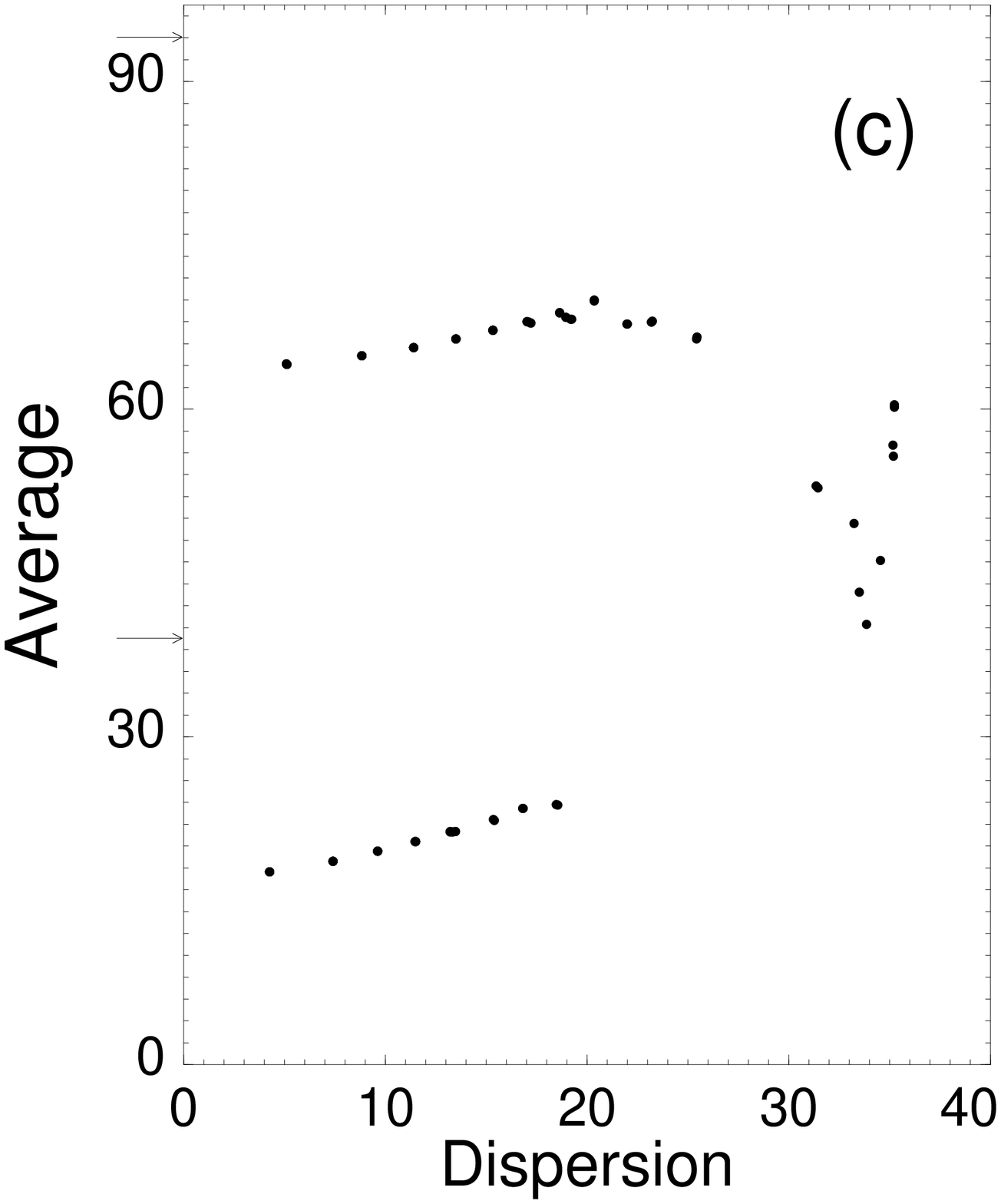,width=7cm,height=7cm}
       \hspace{0.5cm}\psfig{file=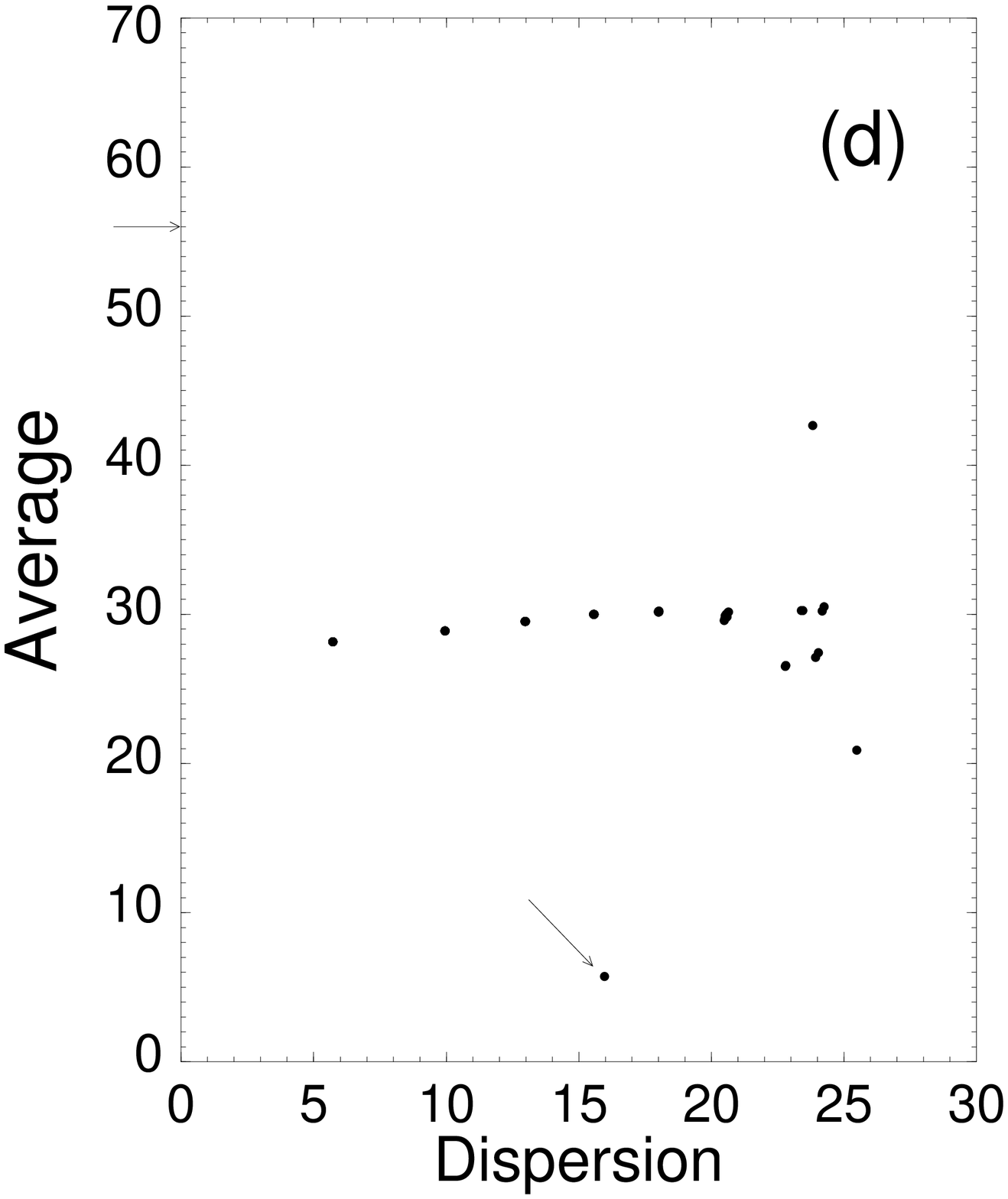,width=7cm,height=7cm}}
 \end{center}
 \centerline{\psfig{file=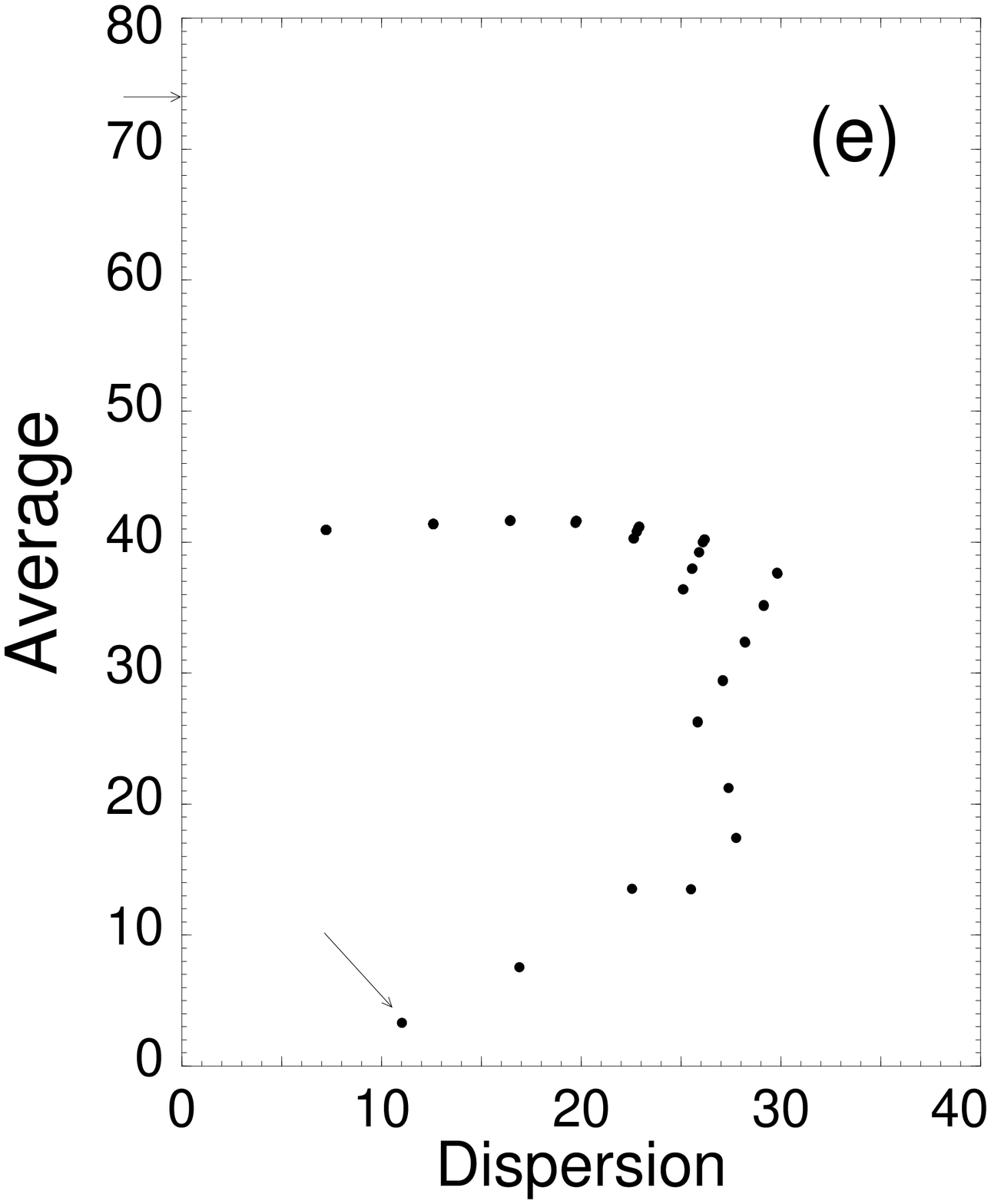,width=7cm,height=7cm}}
 \vspace{0.4cm}  
 \caption{ The plots of averages versus dispersions for 
  $h=0.5$, $\delta=0$ and
 (a) $\ell=1$, $\epsilon=0.05$, (b) $\ell=2$, $\epsilon=0.05$,
 (c) $\ell=3$, $\epsilon=5\times 10^{-4}$, (d) $\ell=4$, $\epsilon=0.05$,
 (e) $\ell=5$, $\epsilon=0.05$.}
 \label{fig:11}
 \end{figure}
\noindent
\begin{equation}
\label{qefind2}
U_{n,n_0}=C_n^{(n_0)}(T),
\end{equation}
where the coefficients, $C_n^{(n_0)}(T)$, can be obtained by numerical solution 
of the Schr\"odinger equation (for more detailed discussion see 
Ref. \cite{3}). After diagonalization  of $U_{n,m}$, we obtain the QE
functions, $C_n^q\equiv C_n^q(mT)$, $m=0,\,1,\,2,\dots$ and the
quasienergies, $E_q$. The values of matrix elements, $U_{n,m}$,
depend on three dimensionless parameters: the wave amplitude, 
$\epsilon$, the quantum parameter,
$h=k^2\hbar/M\omega$, which can be treated as a dimensionless Planck constant,
and from the ratio $\Omega/\omega=\ell-\delta$.
 \begin{figure}[tb]
 \begin{center}
 \mbox{\psfig{file=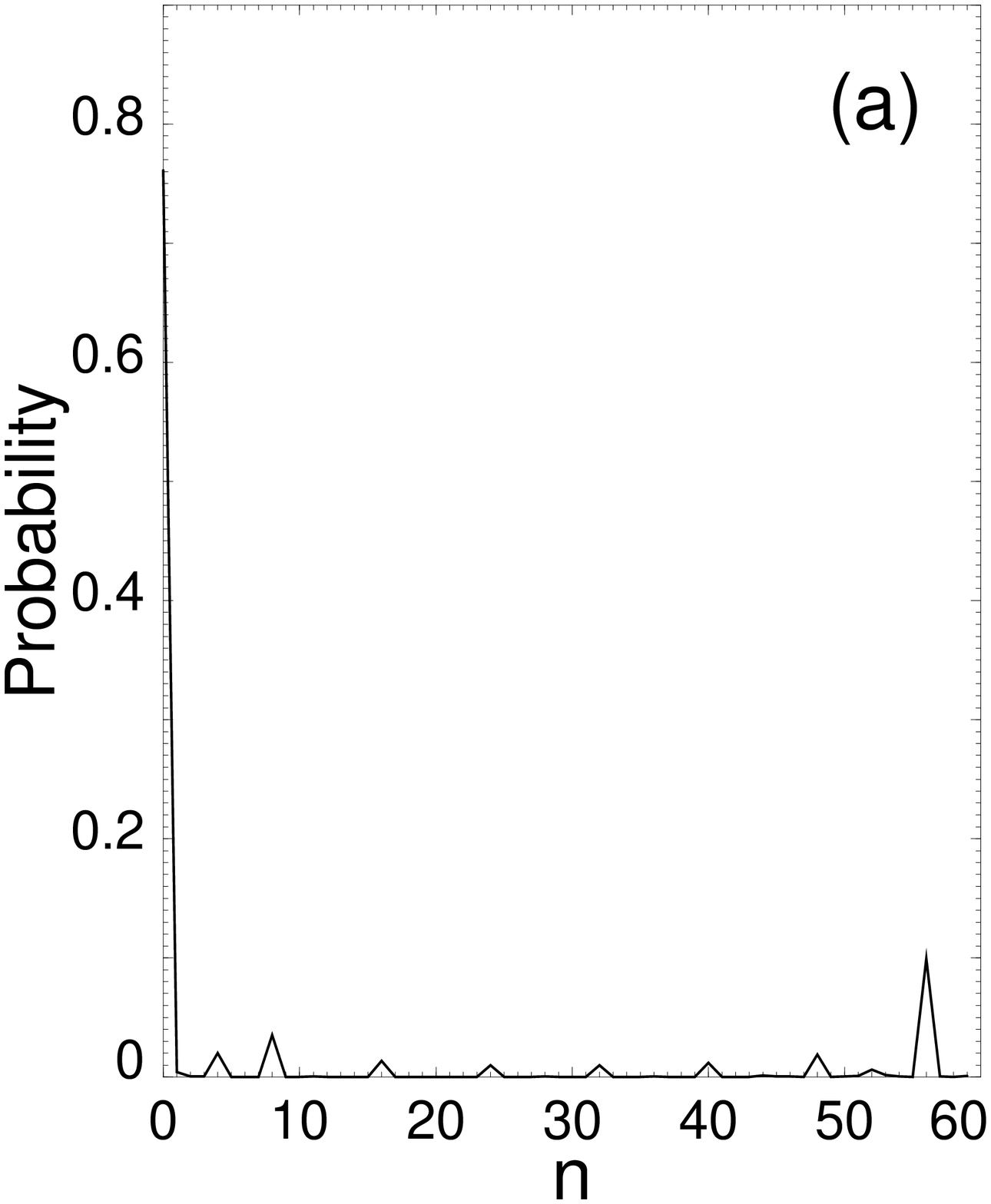,width=8cm,height=8cm}\hspace{0.3cm}
       \psfig{file=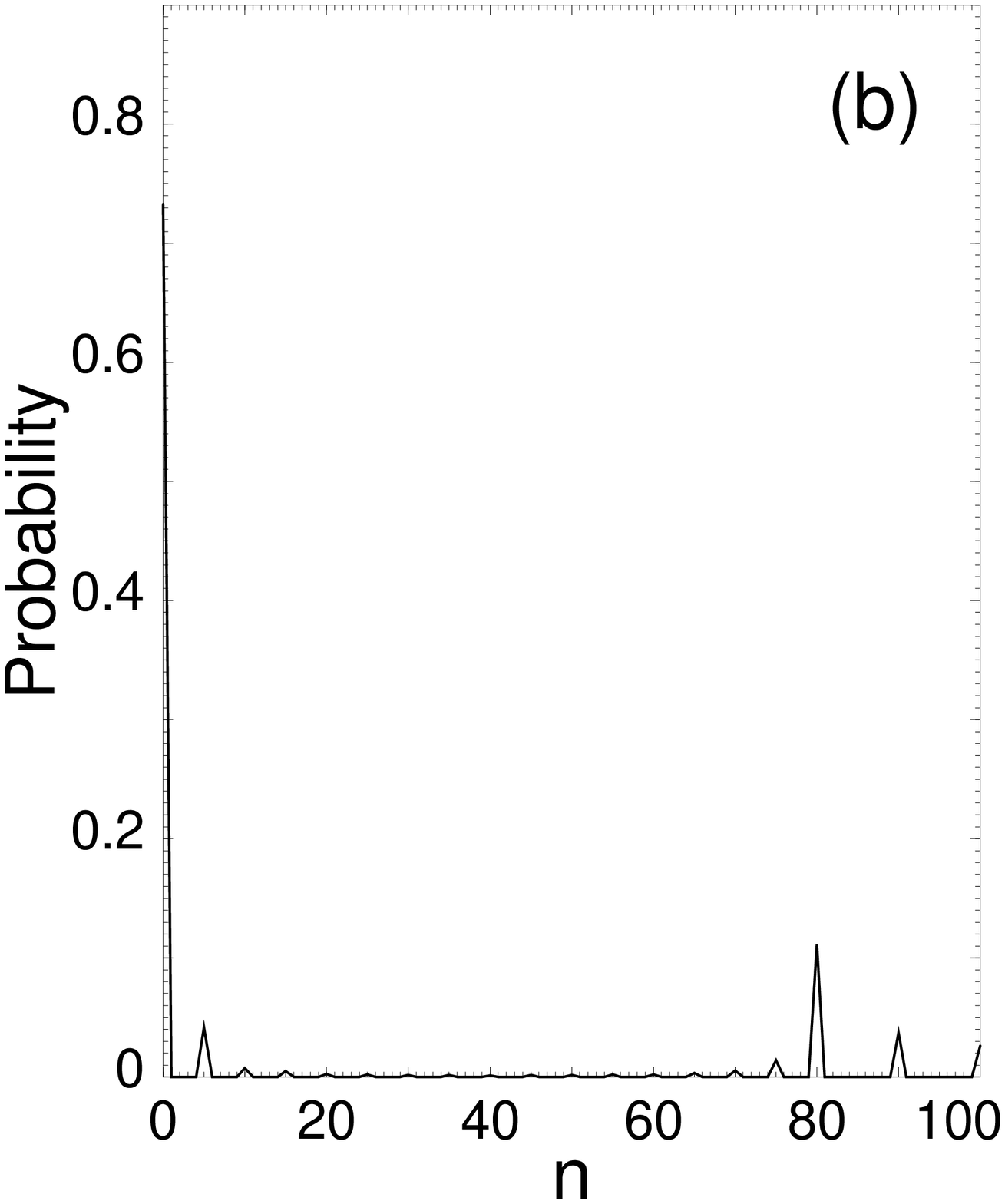,width=8cm,height=8cm}}
 \end{center}
 \caption{The probability distribution for the QGS QE states 
 with the smallest average, $n_q$, for the cases (a) $\ell=4$,
 (b) $\ell=5$; $\epsilon=0.05$, $h=0.5$, $\delta=0$.}
 \label{fig:12}
 \end{figure}
When $\delta=0$ and 
the amplitude of the wave is small, $\epsilon\ll 1$, most of the QE states 
are divided into almost independent groups, each located in one resonance cell 
of the Hilbert space. \cite{1} In order to show this let us characterize 
each QE state by its average, 
$n_q=\sum_nn|C_n^q|^2$, and a dispersion, 
$\sigma_q=\left[\sum_n(n-n_q)^2|C_n^q|^2\right]^{1/2}$ and plot 
$n_q$ versus $\sigma_q$ (see also Ref.\cite{bvi}). Such plots for 
different values of the resonance number, $\ell$,
and for small value of the wave amplitude, $\epsilon$, 
are shown in Figs. 11 (a) - 11 (e). The boundaries of the cells are 
marked by arrows. The radius of the external boundary of 
classical cells in Figs.  1~(a)~-~1~(e) corresponds to the position of the 
boundary of the first quantum cell, respectively,
in Figs. 11 (a) - 11 (e) with the quantized radius, $\rho_n=\sqrt{2nh}$.
One can see from Figs. 11 (a) - 11 (e) that the QE states
are mostly located within the quantum resonance cells, because their
averages, $n_q$, are situated inside the cells, and their widths, 
$\sigma_q$, do not exceed the width of the corresponding resonance
cell. Such states  form rows in Figs. 11 (a) - 11 (e). 
Each cell in the Hilbert space in the 
quasiclassical limit corresponds to $2\ell$ classical cells. \cite{4} 
There also
exist the QE states which do not belong to a particular resonance cell, 
but rather they belong to the stochastic web. These QE states have a 
``separatrix'' structure, i.e. they delocalized over several resonance cells
and have large dispersion, $\sigma$. Such states are represented by scattered 
points on the diagrams $n_q=n_q(\sigma_q)$ in Figs. 11 (a) - 11 (e).  
The structure of these QE states was in details discussed
in Refs. \cite{1,3,2}.
 \begin{figure}[tb]
 \begin{center}
 \mbox{\psfig{file=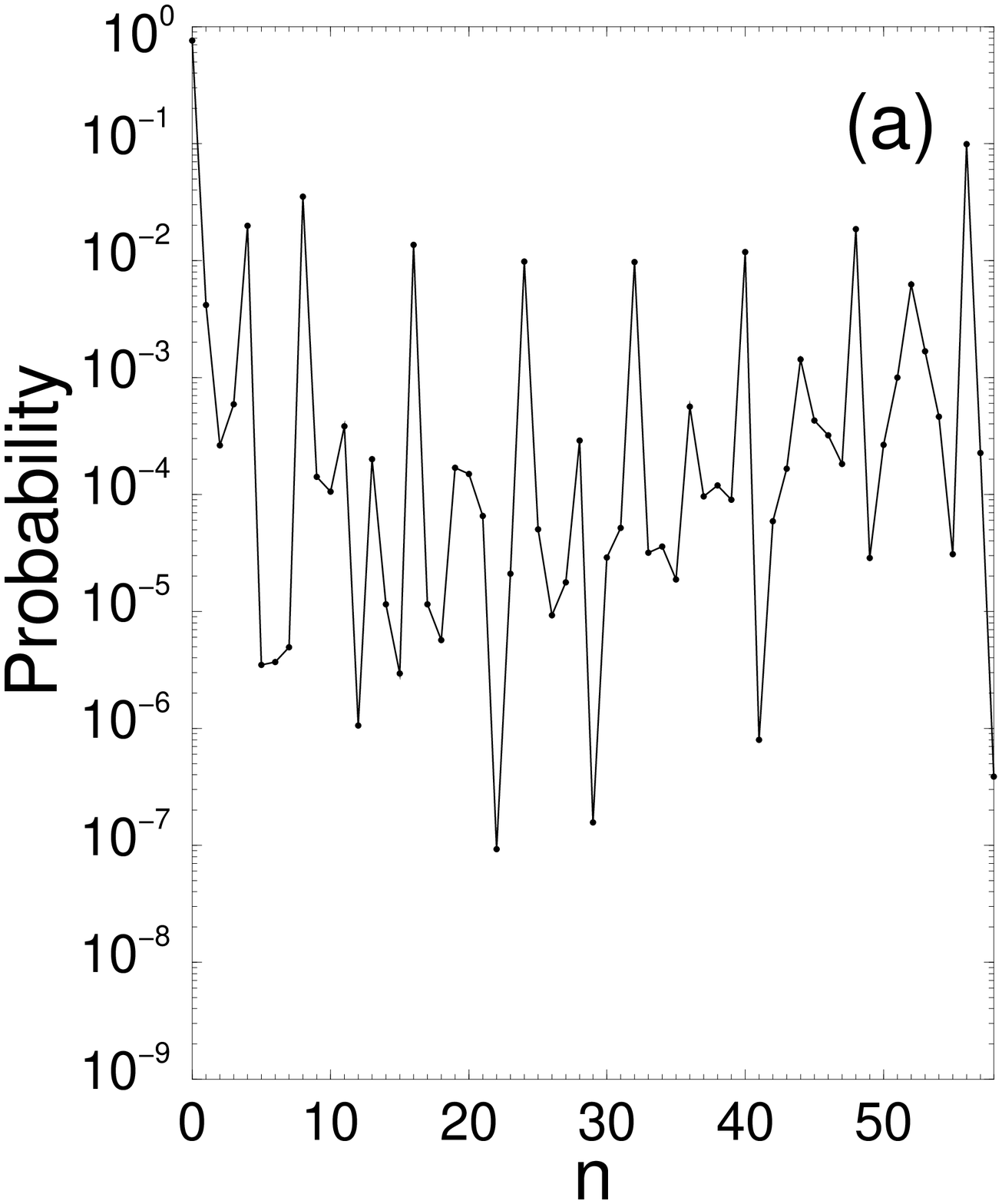,width=7.5cm,height=7.5cm}\hspace{0.5cm}
       \psfig{file=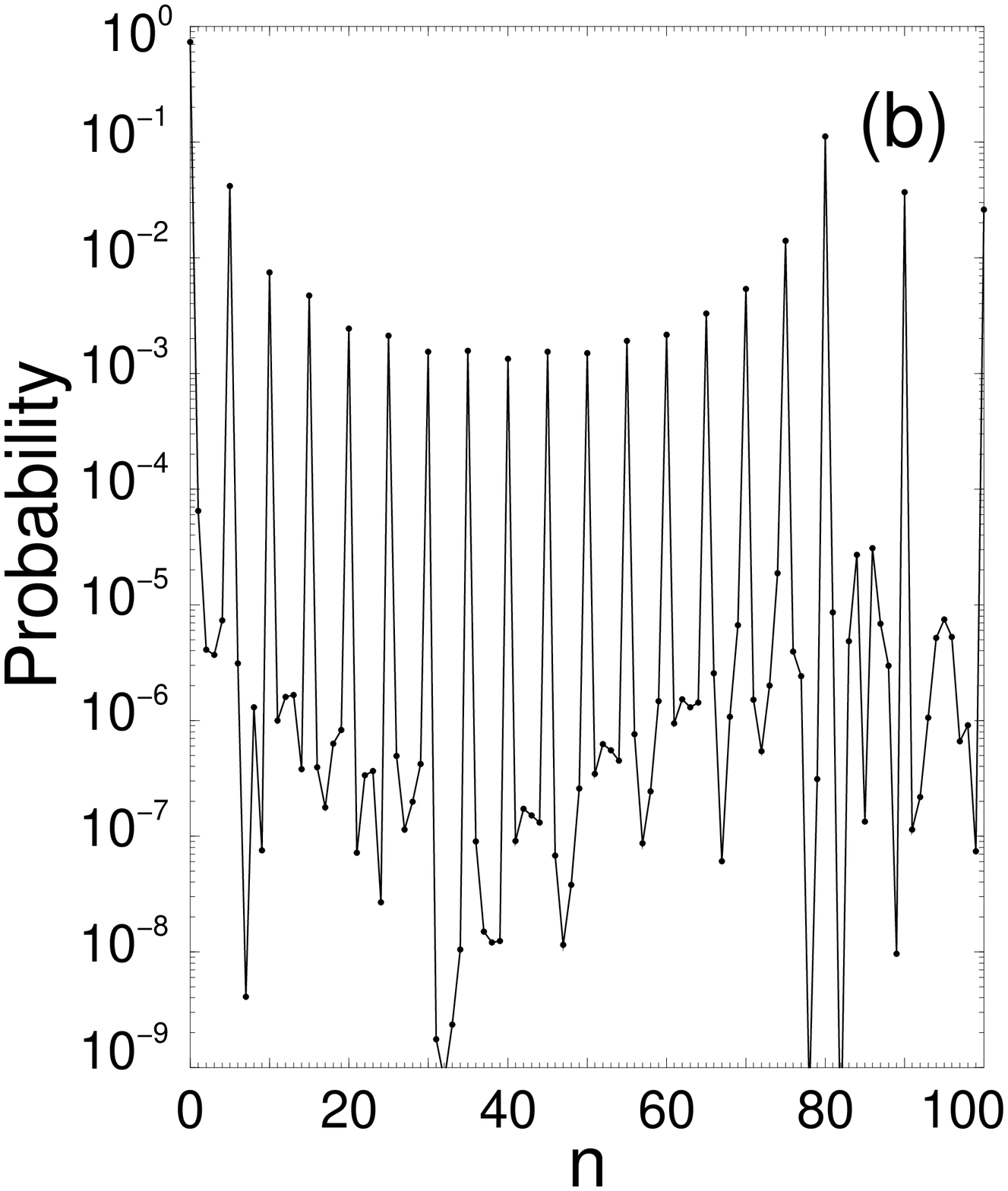,width=7.5cm,height=7.5cm}}
 \caption{ The same as in Figs. 12 (a), 12 (b) but in the
 logarithmic scale.}
 \end{center}
 \label{fig:13}
 \end{figure}
 In this paper, we focus on the QE states which belong to 
the central resonance cell (QGS QE states), because these 
states are mainly responsible for stability properties of the QGS. 
Such QGS QE states are characterized by 
small average, $n_q$, and are located in the low
part of the plots, $n_q(\sigma_q)$, in Figs. 11 (d), 11 (e). In the 
cases $\ell=1,2$ in Figs. 11 (a) - 11 (b) these states are absent,  
which corresponds to unstable dynamics near the CGS in 
the phase classical space, shown, respectively, in Figs. 1 (a) - 1 (b). 
In the case $\ell=3$, the area of the stable island in Fig. 2 (a) is much 
 less than the value of the dimensionless Planck constant $h$ ($h=0.5$).
 So, in this case, the QGS QE state is absent, too.      

 The plots of the probability distribution
for the QGS QE states $q'$ with 
the smallest average, $n_{q'}$, marked in Figs. 11 (d), 11 (e) by arrows, are 
shown in Figs. 12~(a), 12~(b), for the cases $\ell=4$ and $\ell=5$.  
In the logarithmic scale, these states  are illustrated in Figs.
13 (a), 13 (b). The Husimi functions of these states are 
shown in Figs. 14 (a) ($\ell=4$) and 14~(b) ($\ell=5$).      
As one can see from Figs. 12 (a), 12 (b) the QGS QE state is mainly localized 
in the CGS of the harmonic oscillator.
One can see from Figs. 12 (a), 12 (b) and 13 (a), 13 (b)  that the small part 
of the probability distribution is located at the levels with the 
numbers $n=\ell m$, where $m=1,\,2,\dots$. This 
can be explained by influence of the resonance terms in the
quantum equations of motion.\cite{1} In the resonance approximation 
the QE states can be defined from the set of algebraic equations
which in the dimensionless form can be written as,
\begin{equation}
\label{al_eq}
\left({E_q\over \hbar\omega}-\delta n\right)C_n^q=
\frac\epsilon h(V_{n,n+\ell}C_{n+\ell}^q+V_{n,n-\ell}C_{n-\ell}^q).
\end{equation}

 \begin{figure}[tb]
 \begin{center}
 \mbox{\psfig{file=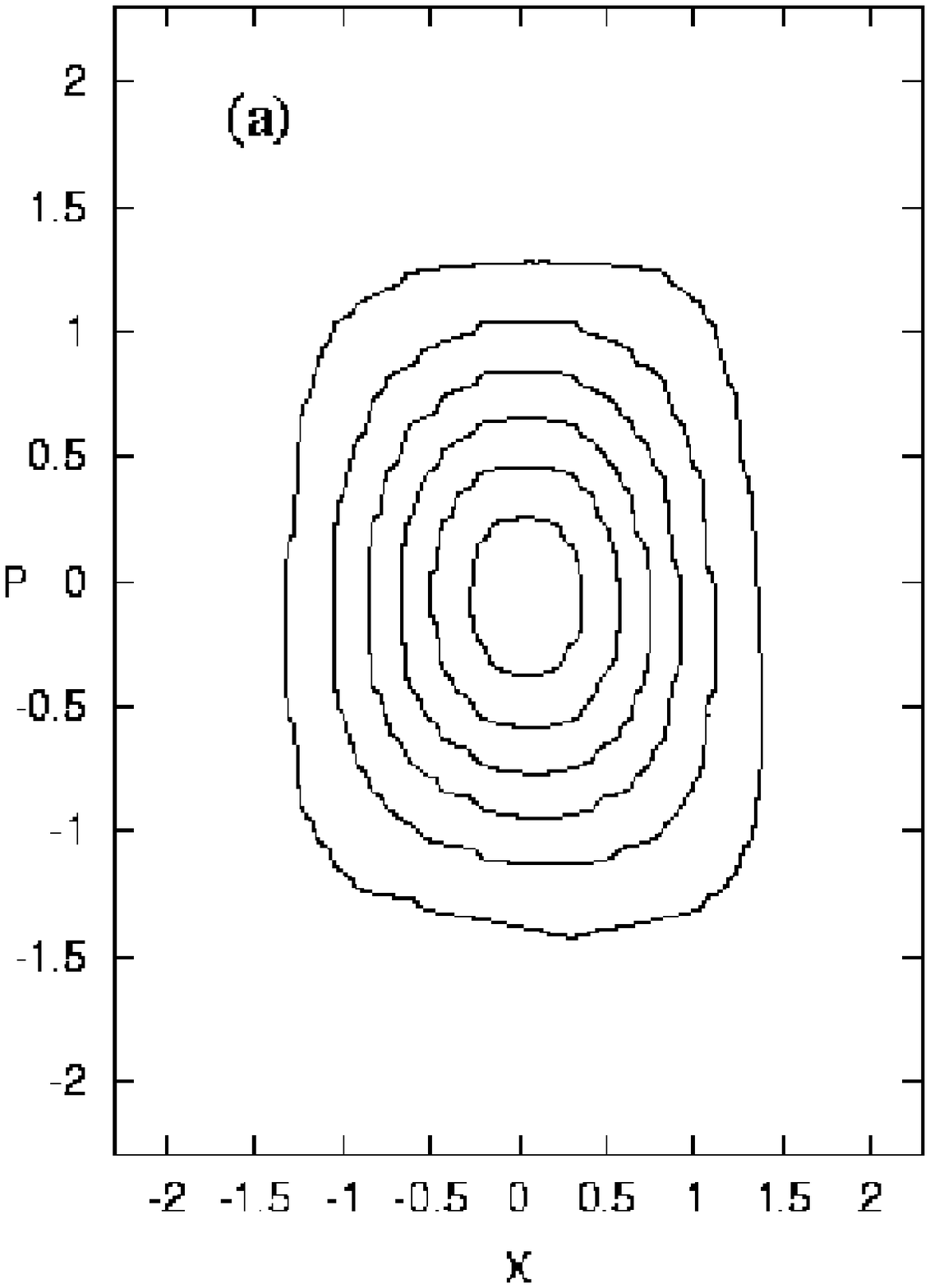,width=8cm,height=8cm}
       \psfig{file=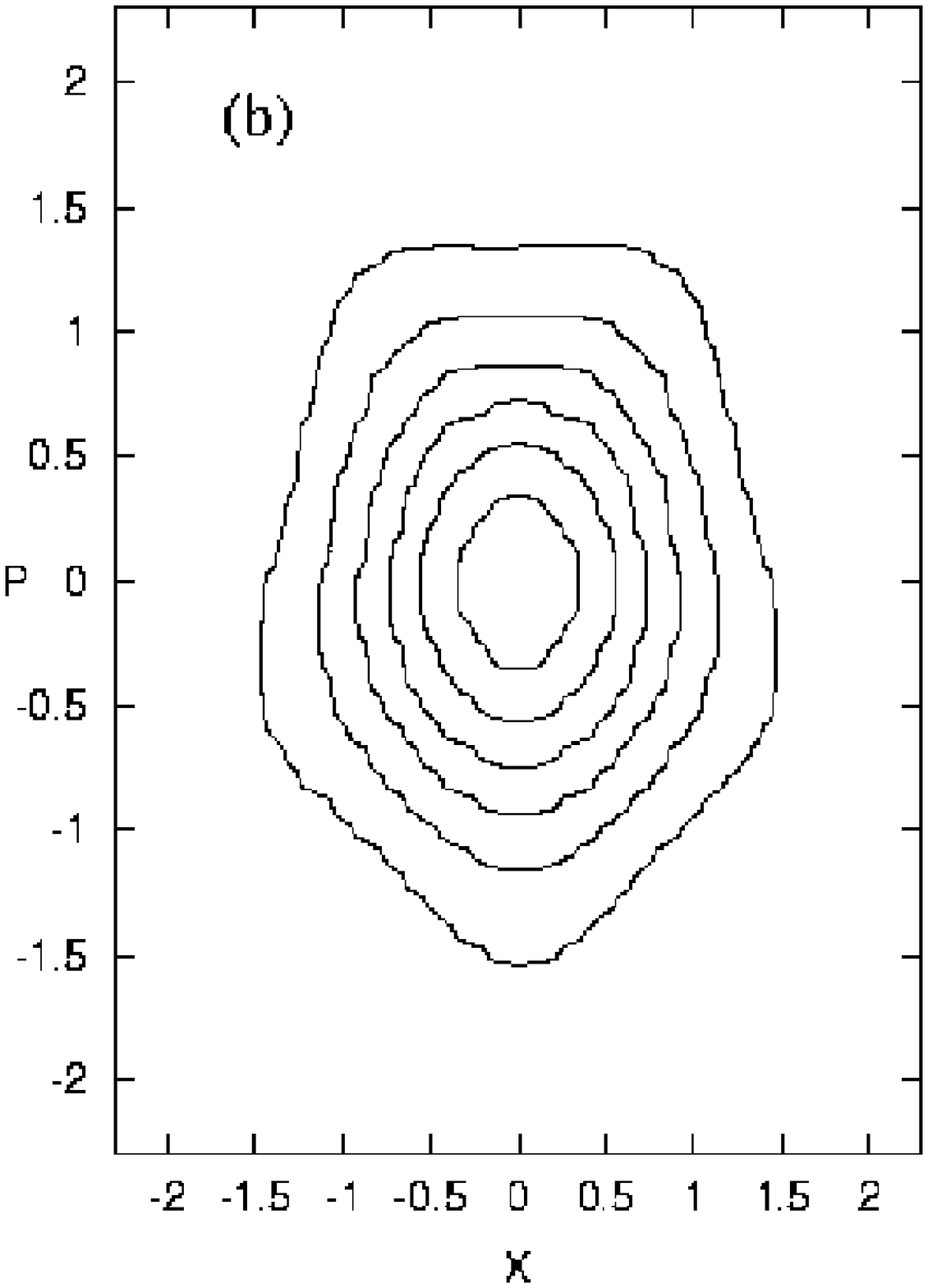,width=8cm,height=8cm}}
 \vspace{0.2cm}
 \caption{(a) The Husimi functions of the QGS QE state shown in Fig. 12 (a).
 (b) The Husimi functions of the QGS QE state shown in Fig. 12 (b).
 The cross-sections are plotted
 from the level $0.047$ with the increment $0.042$, $h=0.5$, $\epsilon=0.05$.} 
 \end{center}
 \label{fig:14}
 \end{figure}

\noindent
The matrix elements for $n\gg 1$ can be approximated by the 
Bessel functions $J_m$,\footnote{More precise form of matrix elements 
see in Ref.\cite{1}}  
\begin{mathletters}
\begin{equation}
\label{m.elements1}
V_{n,n+2m+1}=\frac{1}2
\frac{(-1)^m n^{m+1/2} e^{-\frac{h}4}}
{\sqrt{(n+1)\dots (n+2m+1)}}J_{2m+1}(\sqrt{2nh}) ,
\end{equation}
\begin{equation}
\label{m.elements2}
V_{n,n+2m}=\frac{1}2\frac{(-1)^m n^m e^{-\frac{h}4}}
{\sqrt{(n+1)\dots (n+2m)}}J_{2m}(\sqrt{2nh}).
\end{equation}
\end{mathletters}

As one can see from Eq. (\ref{al_eq}), in the resonance approximation 
the QE functions have the form $C_n^q=C_{\ell m}^q$ 
with $m=1,\,2,\dots$. In this case, the particle is allowed to move only 
between the states 
with the numbers $n=\ell m$. As shown in Ref. \cite{4}, 
a particular form 
of the QE function, $C_n^q=C_{\ell m}^q$, makes the 
Husimi functions, illustrated in Figs. 14 (a) and 14 (b), 
symmetric with the
axial symmetry of the order $\ell$.  
One can see that the width of the Husimi distribution 
in Figs. 14 (a) and 14 (b), 
is, $\Delta P\sim\Delta X\sim\sqrt h\sim 0.7$.

   \begin{figure}[tb]
 \begin{center}
 \mbox{\psfig{file=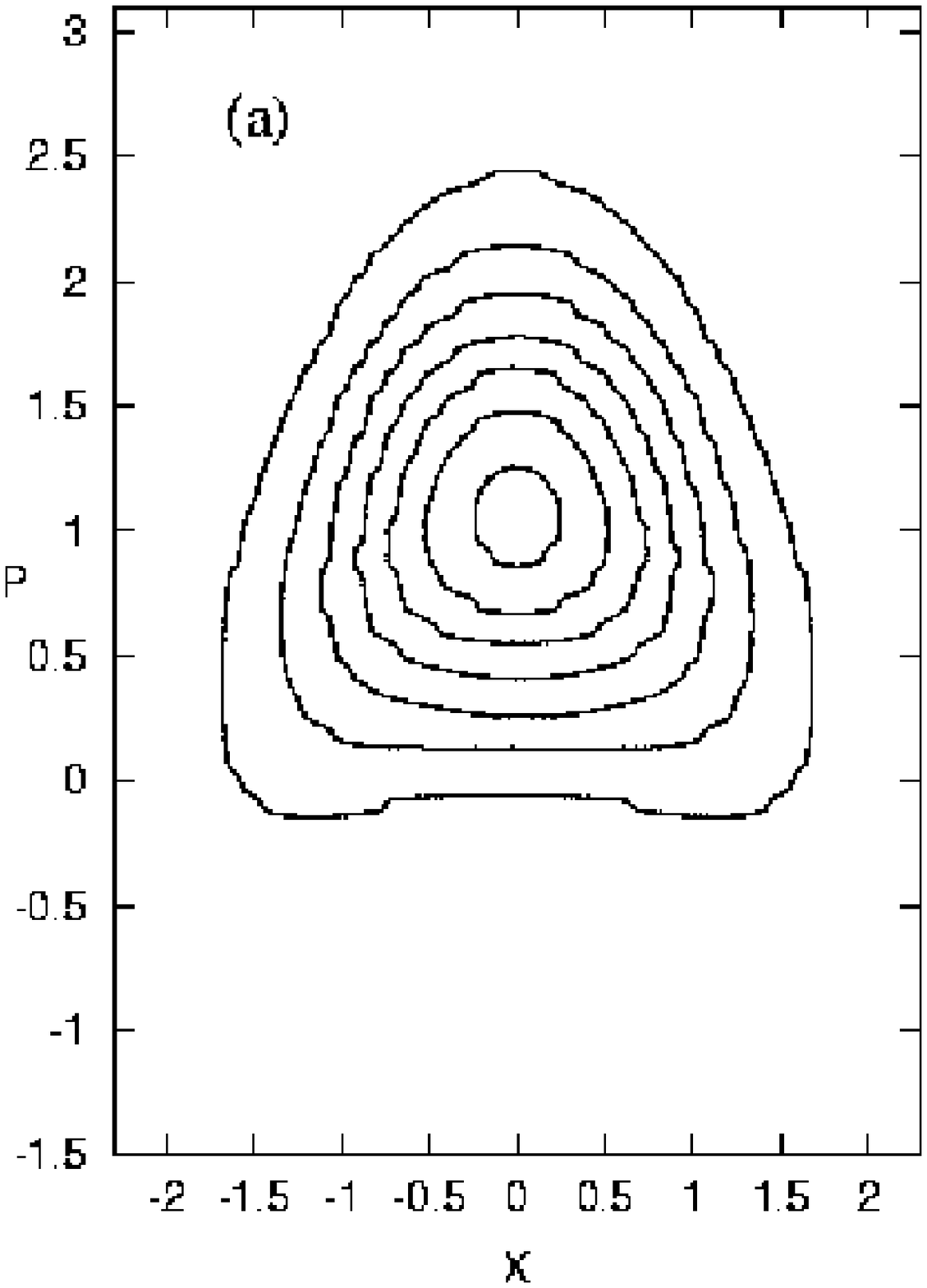,width=7.8cm,height=7.8cm}\hspace{0.2cm}
       \psfig{file=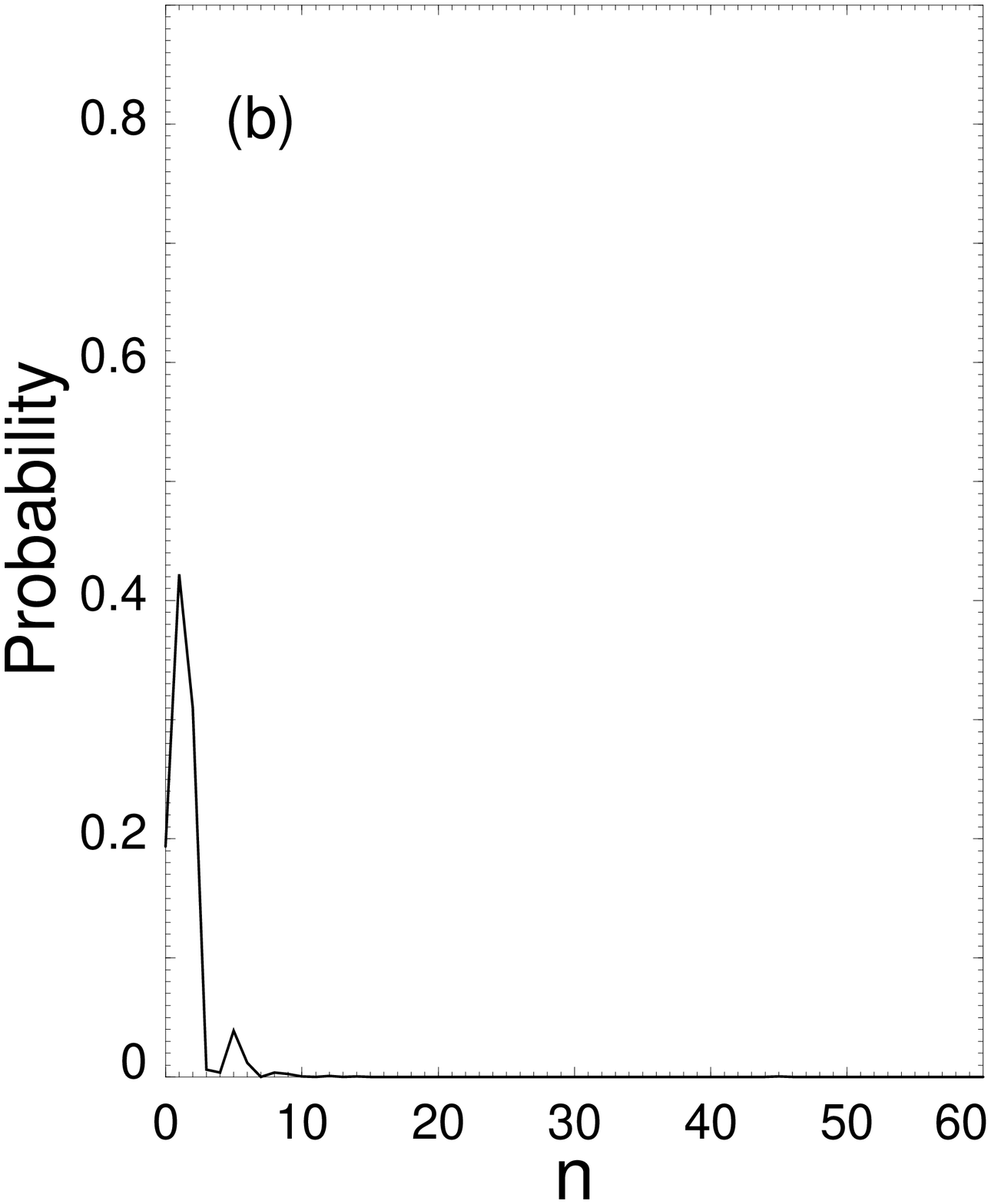,width=7.8cm,height=7.8cm}}
 \vspace{0.2cm}
 \caption{(a) The Husimi functions of the shifted from the 
 ground state QE function, shown in Fig. 15 (b).
 The cross-sections are plotted
 from the level $0.047$ with the increment $0.042$,
 $\epsilon=5.0$, $h=0.5$.} 
 \end{center}
 \label{fig:15}
 \end{figure}

Now, let us consider influence of dynamical chaos on the QE states,
when $\epsilon$ increases.
There are several QE states localized 
near the QGS of the harmonic oscillator. Some of them 
are shifted from the level with the number $n=0$. They can be associated 
with shifted up classical central resonance cell, when $\epsilon$ increases.  
In this case, the Husimi function in Fig. 15 (a) of the QE state with the 
probability 
distribution illustrated in Fig. 15 (b) (for $\ell=5$ and $\epsilon=5$)
has a similar form to the form of the trajectories in the classical phase space 
shown in Fig. 5 (a). 
At large enough values of the wave amplitude, 
($\epsilon=5$ in Fig. 15 (a)), there are no 
QE states localized in the nearest quantum cells, except for the 
QE states localized in the central cell. 
This corresponds to the chaotic classical dynamics  
shown in Figs. 7 (b), 7 (d) with the stable island in the center
of the phase space.  

\begin{figure}[tb]
\centerline{\psfig{file=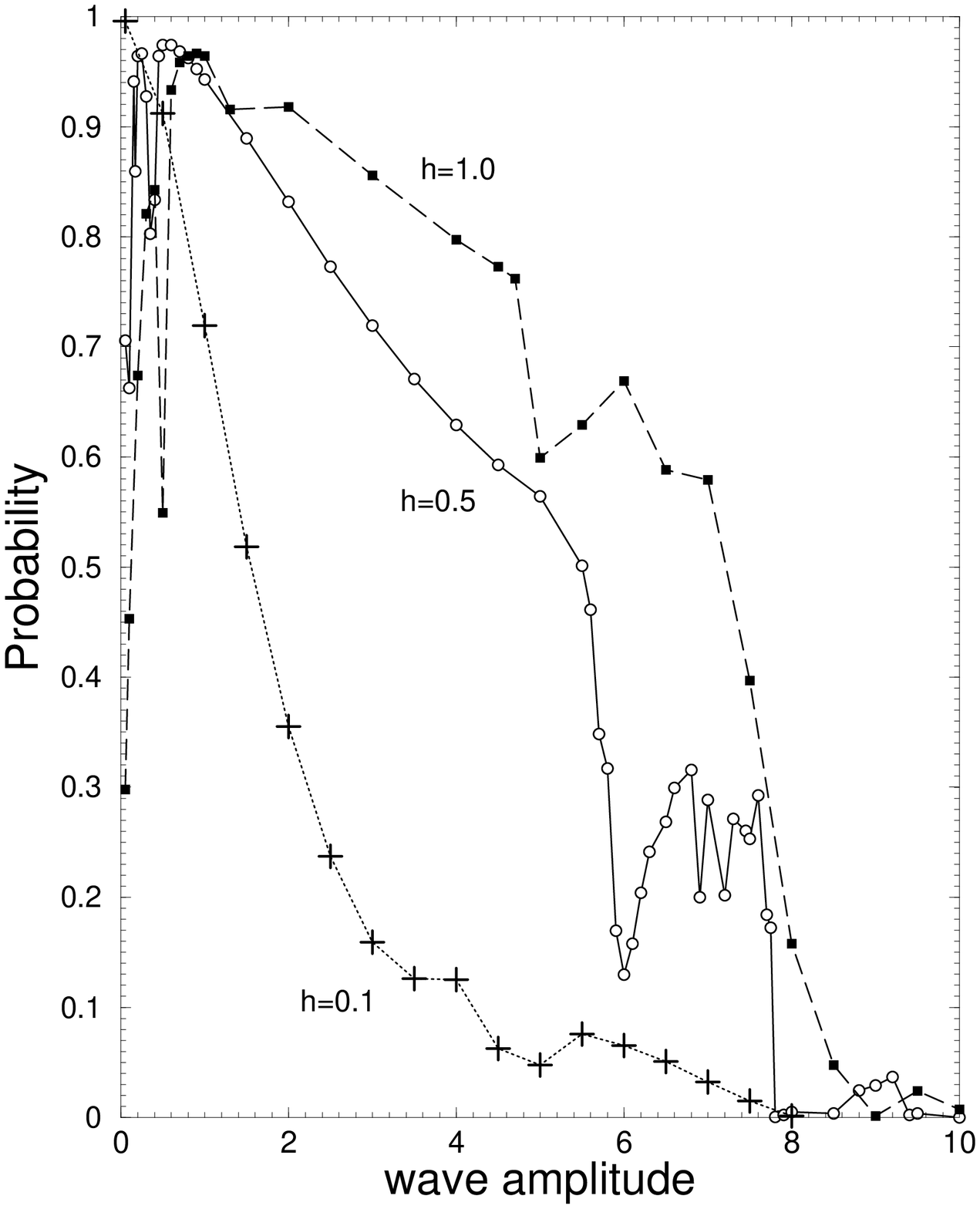,width=15cm,height=10cm}} 
\caption{The probability to find the system 
in the QGS of the 
harmonic oscillator, defined 
by the values $|C_{n=0}^{q'}|^4$ of the QE state $q'$ mostly localized at the 
state with $n=0$, versus the wave amplitude, $\epsilon$, for three values 
of the effective Planck constant, $h$; $\ell=5$.}
\label{fig:16}
\end{figure}

The QGS QE states, mostly localized at the CGS of the harmonic oscillator, 
are of the most interest for us, because they mainly define the dynamics of the 
quantum state initially located at the level with $n=0$. The 
time-evolution of the system with the initial 
state $C_n(0)=\delta_{n,n_0}$ is defined by the equation,  
\begin{equation}
\label{cnmt}
C_n(mT)=\sum_q C_n^{q*} C_{n_0}^q\exp(-iE_q mT/\hbar),
\end{equation}
where $m=0,\,1,\,2,\dots$.
The amplitude of probability to find the system in the initial state,  
$n_0$, is, 
\begin{equation}
\label{cn0mt}
C_{n_0}(mT)=\sum_q |C_{n_0}^{q}|^2 \exp(-iE_q mT/\hbar).
\end{equation}
Suppose that some QE state with the number $q'$ 
is mostly localized at the level $n=n_0$, i.e. 
$|C_{n_0}^{q'}|^2\gg |C_{n_0}^{q}|^2$ for all $q\ne q'$. Then, 
the term with $q=q'$ dominates in the sum in the right-hand side 
of Eq. (\ref{cn0mt}), and we can write, 
\begin{equation}
\label{cn0mt1}
C_{n_0}(mT)\approx |C_{n_0}^{q'}|^2 \exp(-iE_{q'} mT/\hbar).
\end{equation}
The probability, $P_{n_0}(mT)$, to find the system at the moments $t_m=mT$ 
in the state with $n=n_0$ is given by,
\begin{equation}
\label{cn0mt2}
P_{n_0}(mT)=|C_{n_0}(mT)|^2\approx |C_{n_0}^{q'}|^4.
\end{equation}
The value of $P_{n_0}(mT)$ in this approximation
is independent of the number of periods passed, 
$P_{n_0}(mT)\equiv P_{n_0}$. In the next approximation, 
the neglected terms in Eq. (\ref{cn0mt}) cause the probability $P_{n_0}$
slightly oscillate with time.  

In Fig. 16, we present a plot of the probability, $P_0=|C_{n_0=0}^q|^4$,
to find the system in the QGS
as a function of the wave amplitude, $\epsilon$,
if the initial state is the QGS of the harmonic oscillator. 
One can see that the dynamical chaos (the range of large enough $\epsilon$) decreases this probability. 
However, the process of
delocalization of the QGS QE state 
is extremely slow when $\epsilon$ increases,
in comparison 
with that in other nearest cells. For example, at $\epsilon=5$ all QE states
in the nearest cells are chaotic (delocalized) which corresponds 
to the chaotic classical 
dynamics in Fig. 7 (d), while the QE state located 
at the QGS remains localized with the probability 
$P_0\approx 0.56$ when $h=0.5$, and $P_0\approx 0.6$ when $h=1.0$.  
From comparison   

 \begin{figure}[tb]
 \begin{center}
 \mbox{\psfig{file=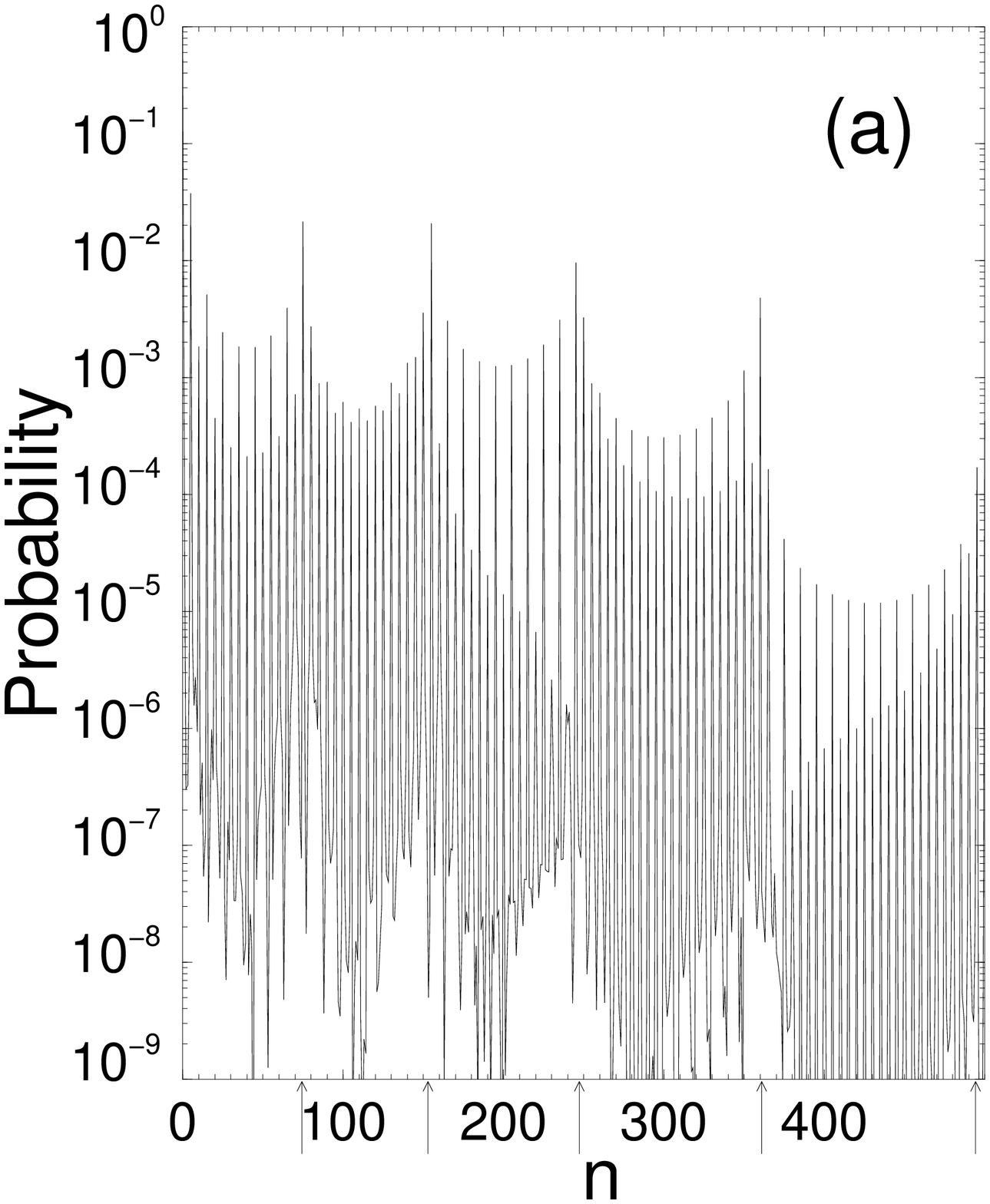,width=7.8cm,height=6.8cm}\hspace{0.5cm}
       \psfig{file=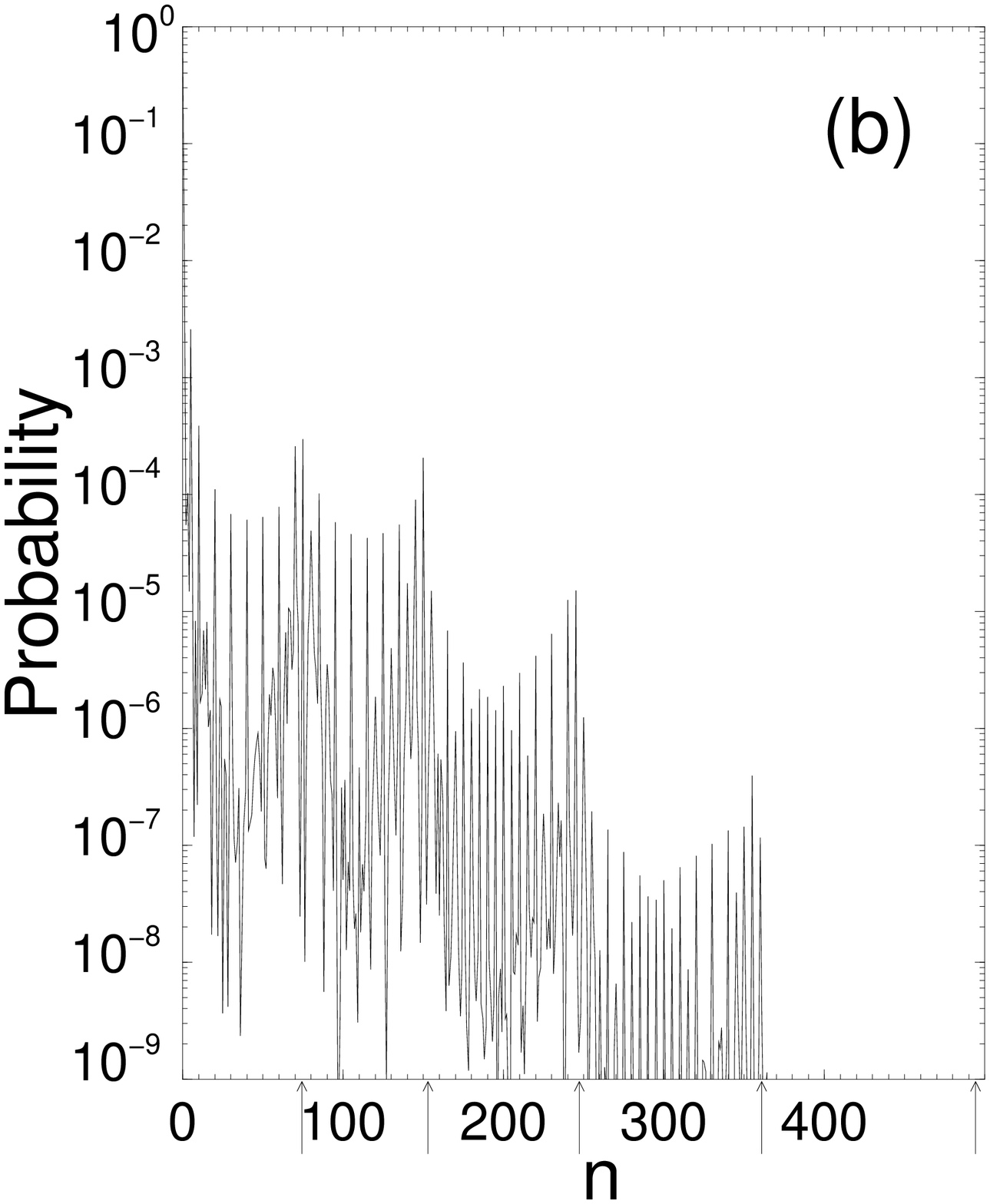,width=7.8cm,height=6.8cm}}
 \mbox{\psfig{file=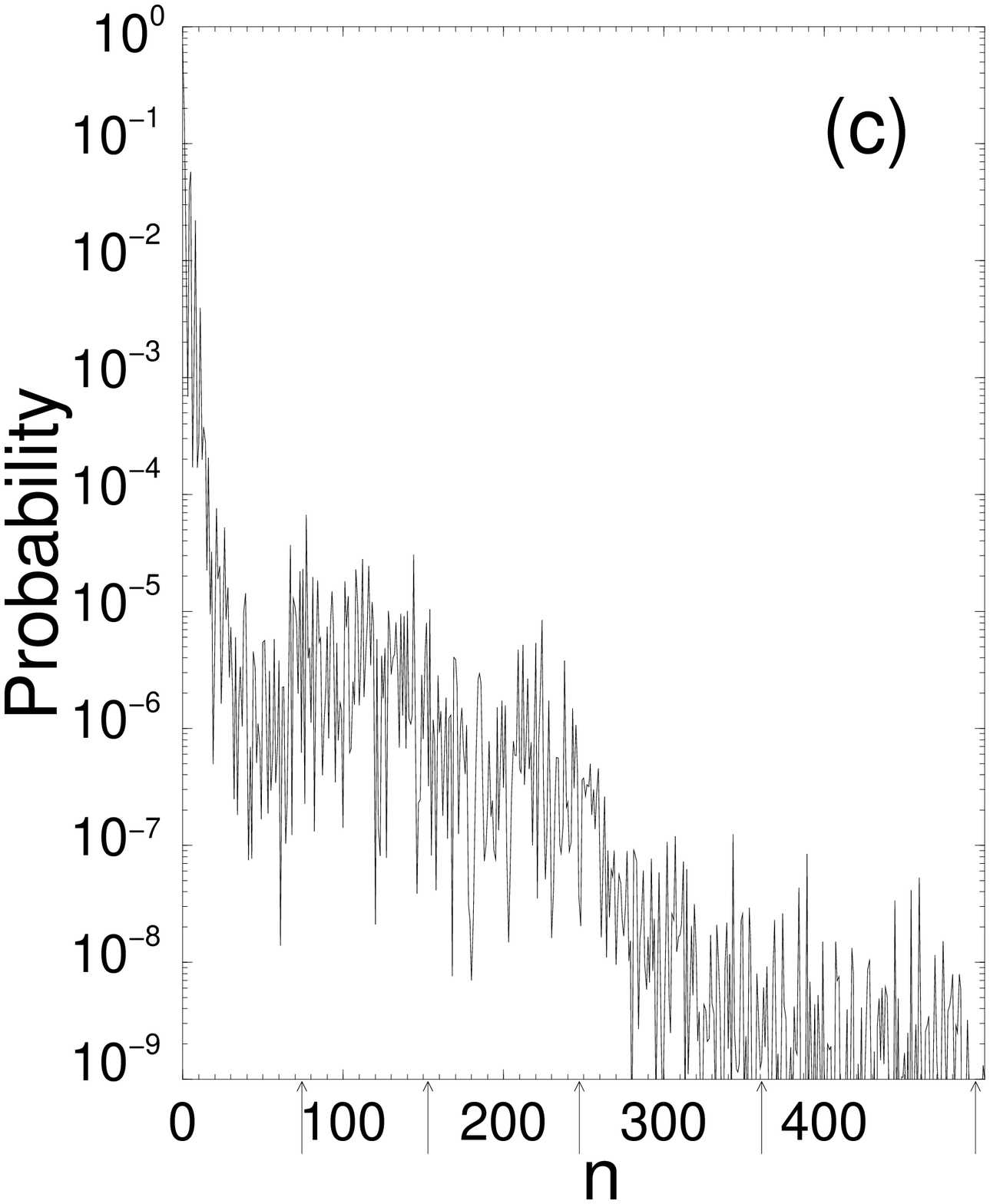,width=7.8cm,height=6.8cm}\hspace{0.5cm}
       \psfig{file=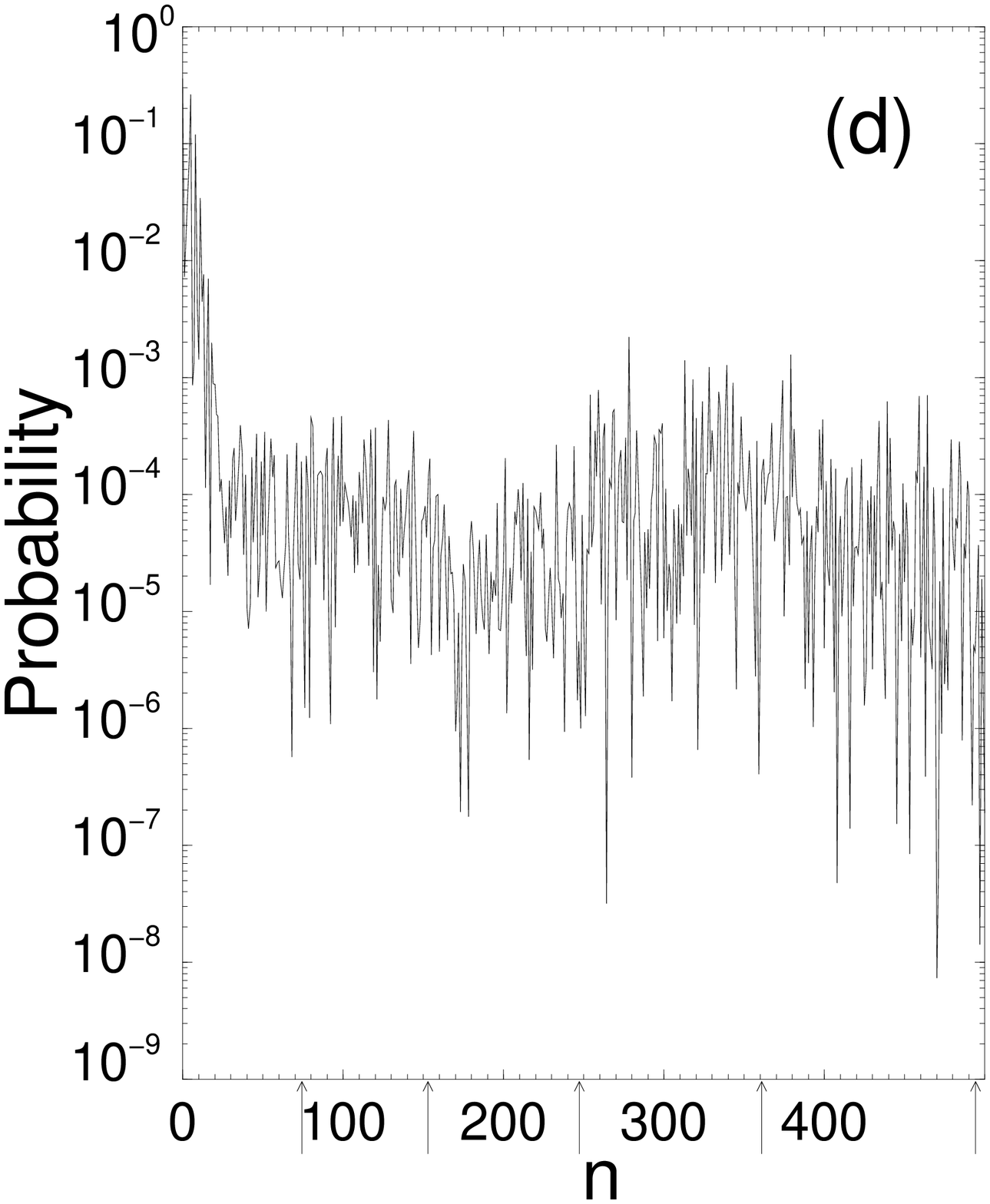,width=7.8cm,height=6.8cm}}
 \mbox{\psfig{file=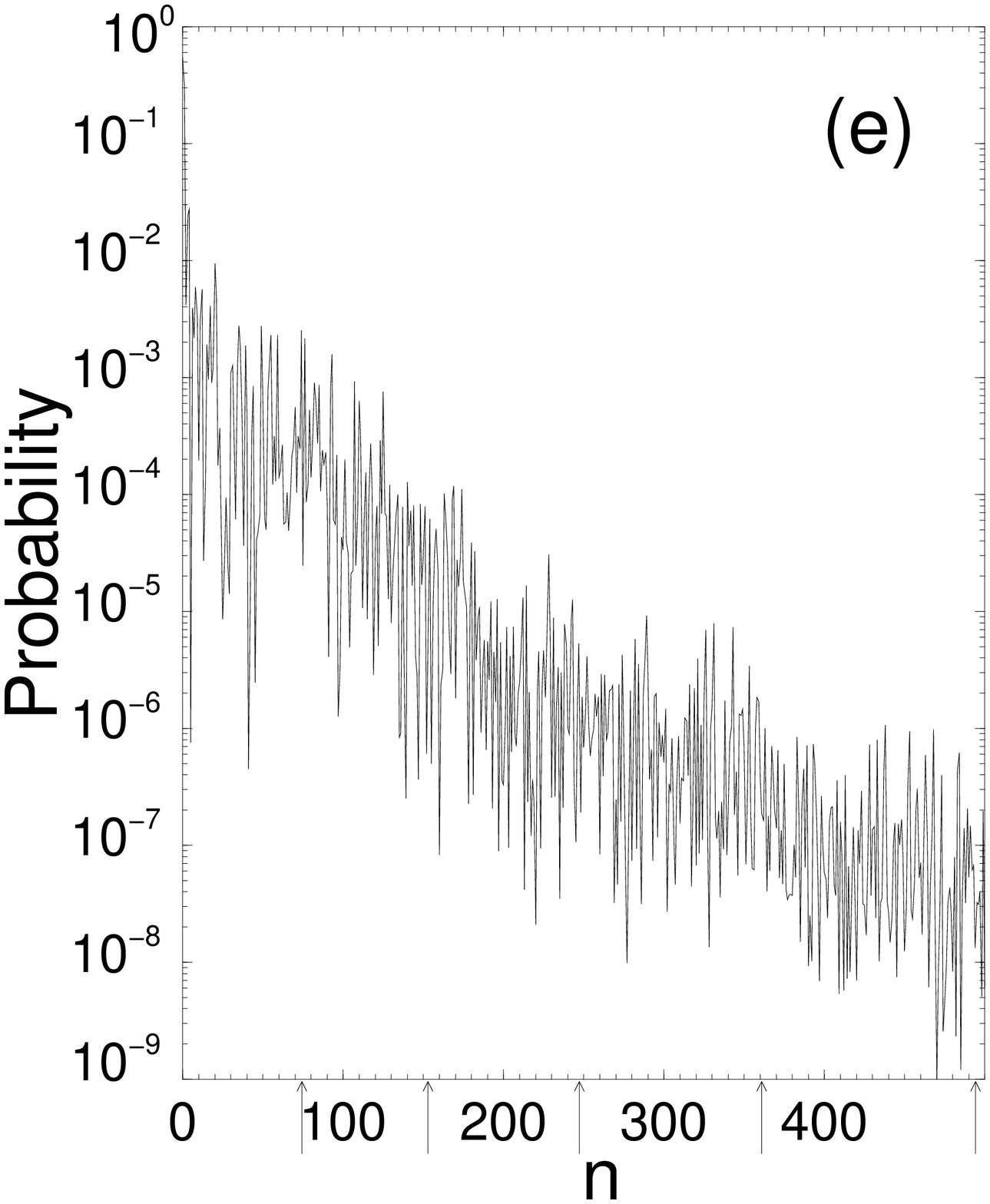,width=7.8cm,height=6.8cm}\hspace{0.5cm}
       \psfig{file=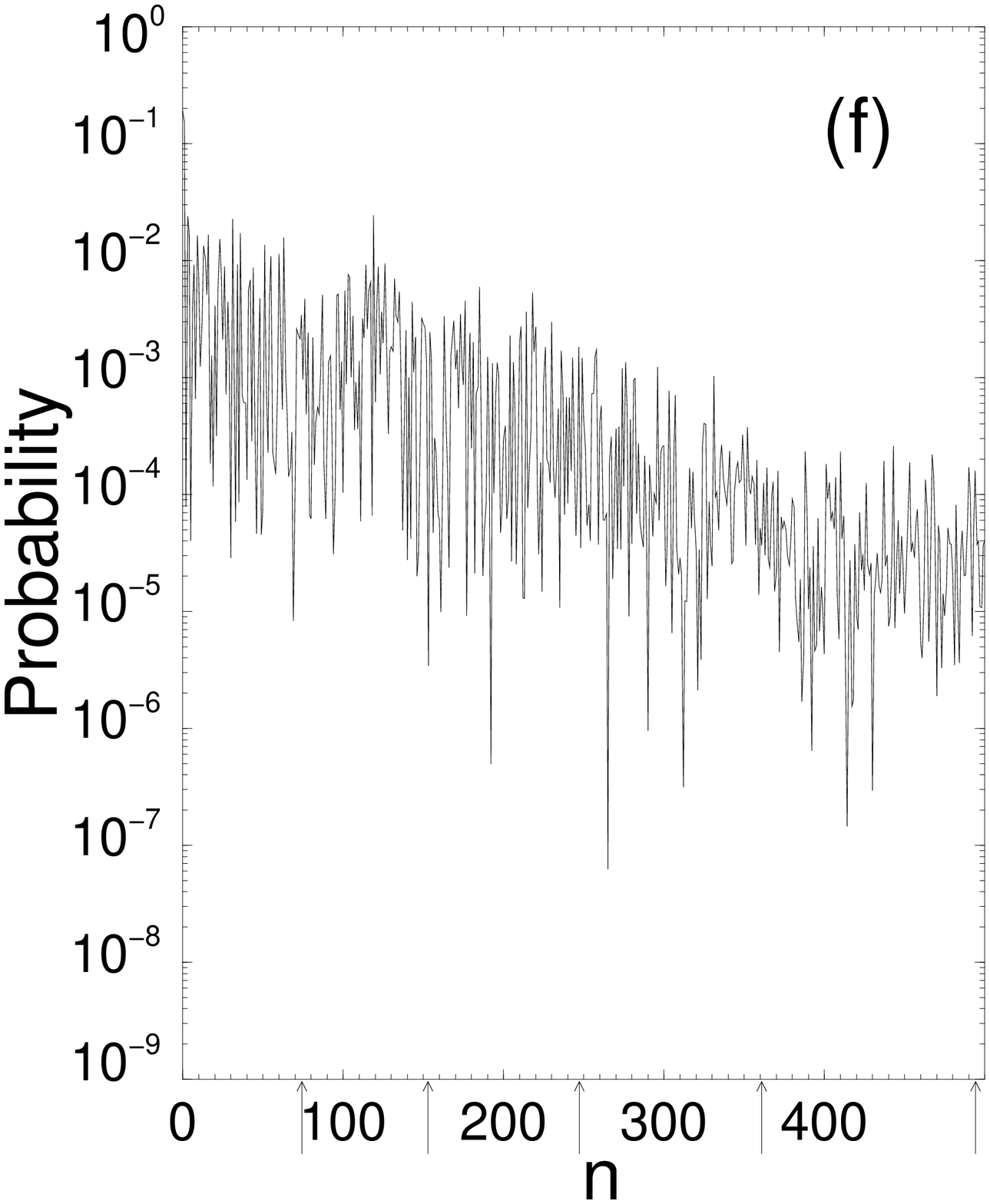,width=7.8cm,height=6.8cm}}
 \vspace{0.2cm}
 \caption{(a) The QE states mostly localized at the
 QGS of the harmonic oscillator (QGS QE states), $h=0.5$, 
 (a) $\epsilon=0.05$, (b) $\epsilon=0.5$, (c) $\epsilon=5.5$
 (d) $\epsilon=6.0$, (e) $\epsilon=7.6$, (f) $\epsilon=9.2$.
 The boundaries of the quantum cell are marked by arrows.}   
 \end{center}
 \label{fig:17}
 \end{figure}

\noindent
of different curves, for h=0.1,
$h=0.5$ and $h=1.0$, one can note the following features. 
i) At small values of $\epsilon$,  
increase of $h$ leads, on average, to decrease of stability
of the QGS. ii) The QGS, at large values of $h$ ($h=1$), 
is more stable
under the influence of chaos (the range of large enough $\epsilon$) 
than that at small values of $h$ ($h=0.1$).  
We should note, that oscillations of $P_0(mT)$ in time should increase when  
$P_0(mT)$ decreases. This happens in the region of 
large enough $\epsilon$ in Fig. 16, 
because in this case the influence of 
neglected terms in Eq. (\ref{cn0mt}) becomes more significant.   

More information about the stability of the QGS can be 
extracted from the analysis of the 
structure of QE states located at the QGS 
at different values of the wave amplitude, $\epsilon$,
shown in Fig. 17 (a) - 17 (f) (for $h=0.5$). The QGS QE state 
shown in Fig. 17 (a) for small value of $\epsilon$   
($\epsilon=0.05$) is similar
to the separatrix QE states, \cite{3} because it has the regular
structure (compare, for example, with Fig. 17 (c)),
and its maxima are located 
near the quantum separatrices, indicated in Fig. 17~(a) by arrows. 
Thus, the QGS QE state also possesses the properties of the 
``separatrix'' QE states, considered in Ref. \cite{3}. 

The separatrix QE states are of the quantum nature \cite{3} because
they are delocalized over several resonance cells. These QE states
provide the tunneling between the cells 
when chaotic regions in the phase space are negligible small, and 
the classical particle can not practically penetrate from one resonance cell to
another. The separatrix QE function mostly localized in the QGS
is a particular one, and it is different from other separatrix states 
studied before.\cite{3} On the one hand, it is delocalized over several
resonance cells (see Fig. 17 (a)) as other separatrix QE functions. 
On the other hand, this particular QE function is mostly concentrated 
on the QGS of the harmonic oscillator, unlike the other separatrix QE 
functions. These ``contradictory'' features of the QGS QE state
define the dynamics: on the one hand, the system mainly
remains localized in the QGS, but on the other hand, a small part 
of the probability distribution can tunnel to the states with large $n$, 
located in the other resonant cells.
When we increase the parameter $h$,
the separatrix
QE states become more delocalized, and the probability to tunnel to
other cells increases, which explain decrease of $P_0$ with increasing $h$
in Fig. 16 (compare the different curves in the region of small $\epsilon$).

At intermediate values of $\epsilon$, when
$1<\epsilon<7$, a stability of the QGS increases with increasing $h$.
However, we can not increase $h$ indefinitely because when
$h$ becomes larger than the resonance area in the phase space, the 
QGS becomes unstable. Thus, at $h=5$, $\ell=4$, $\epsilon=2$ 
(see the classical phase space in Fig. 4 (b)) and at
$h=5$, $\ell=5$, $\epsilon=5$ (Fig. 5 (a)) no localized QGS was
found. 

An increase of $P_0$ with increasing the 
wave amplitude, $\epsilon$, when $\epsilon$ is small 
($\epsilon<0.5$ for $h=0.5$ and $\epsilon<1.0$ for $h=1.0$),
shown in Fig. 16, is a consequence of a partial localization of the separatrix
QE state (see Fig. 17 (b)) under the influence of chaos   
explored in Ref. \cite{3} In this case, the QGS QE function, 
shown in Fig. 17 (b), loses its ``separatrix'' features. 
Further increase of $\epsilon$, makes 
the QGS QE state more delocalized. However, as one can see from 
Figs. 17 (c) ($\epsilon=5.5,\,h=0.5$)
and 17 (d) ($\epsilon=6.0,\,h=0.5$)
delocalization takes place mainly over the nearest oscillator states
with small numbers, $n$. At $\epsilon>5.5$ ($h=0.5$) the  
oscillations appear in the dependence $P_0=P_0(\epsilon)$, in Fig. 16. 
Thus, the QE function 
at $\epsilon=6$ in Fig. 17 (c) is less localized than the 
QE function at $\epsilon=7.6$ shown in Fig. 17 (d). 
At large values of $\epsilon$ ($\epsilon>7.6$), practically 
all QE states are delocalized, which
is the quantum 
manifestation of chaotization of the classical central cell
in the phase space. 

 \begin{figure}[tb]
 \begin{center}
 \mbox{\psfig{file=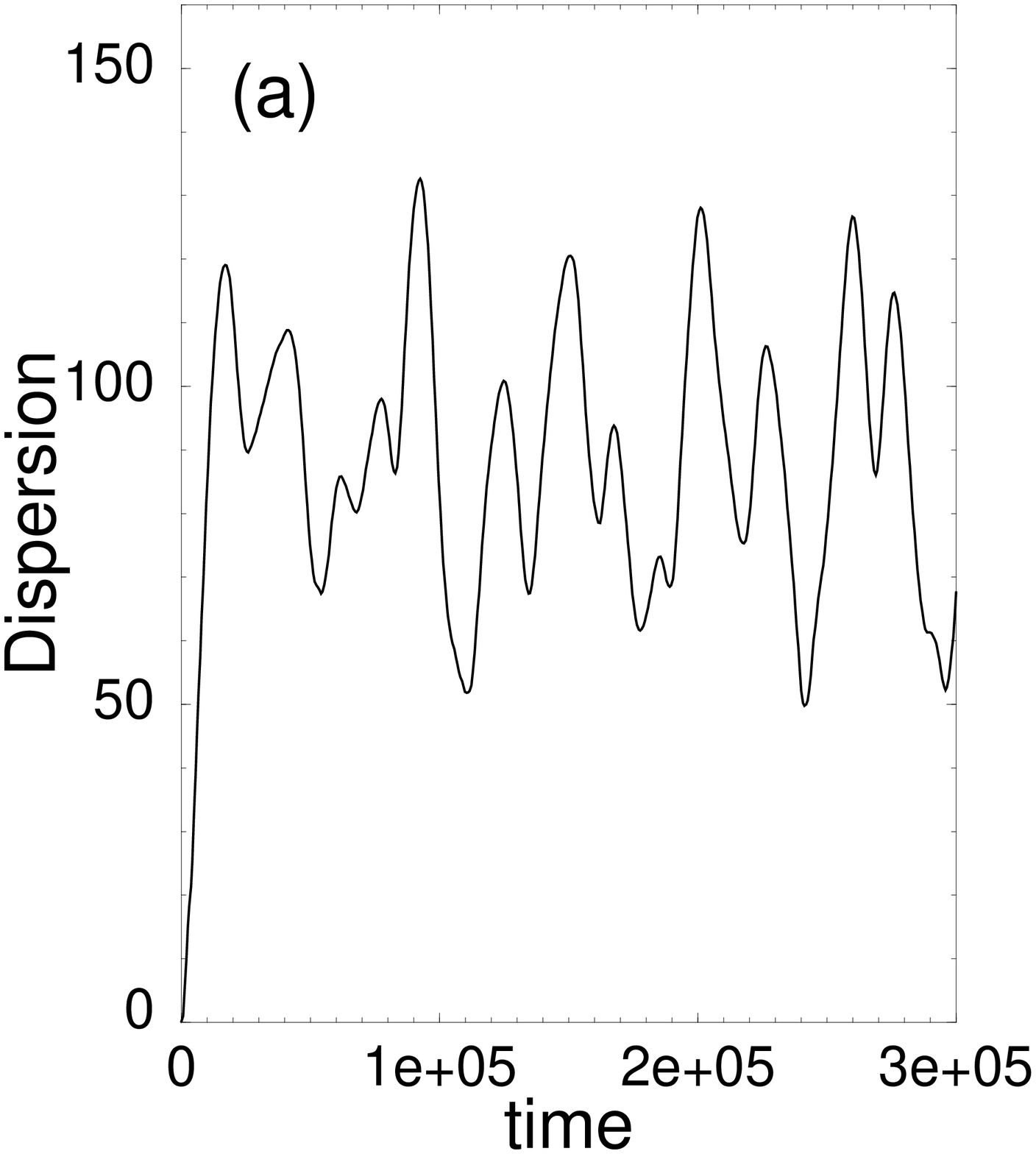,width=5.4cm,height=5.4cm}
       \psfig{file=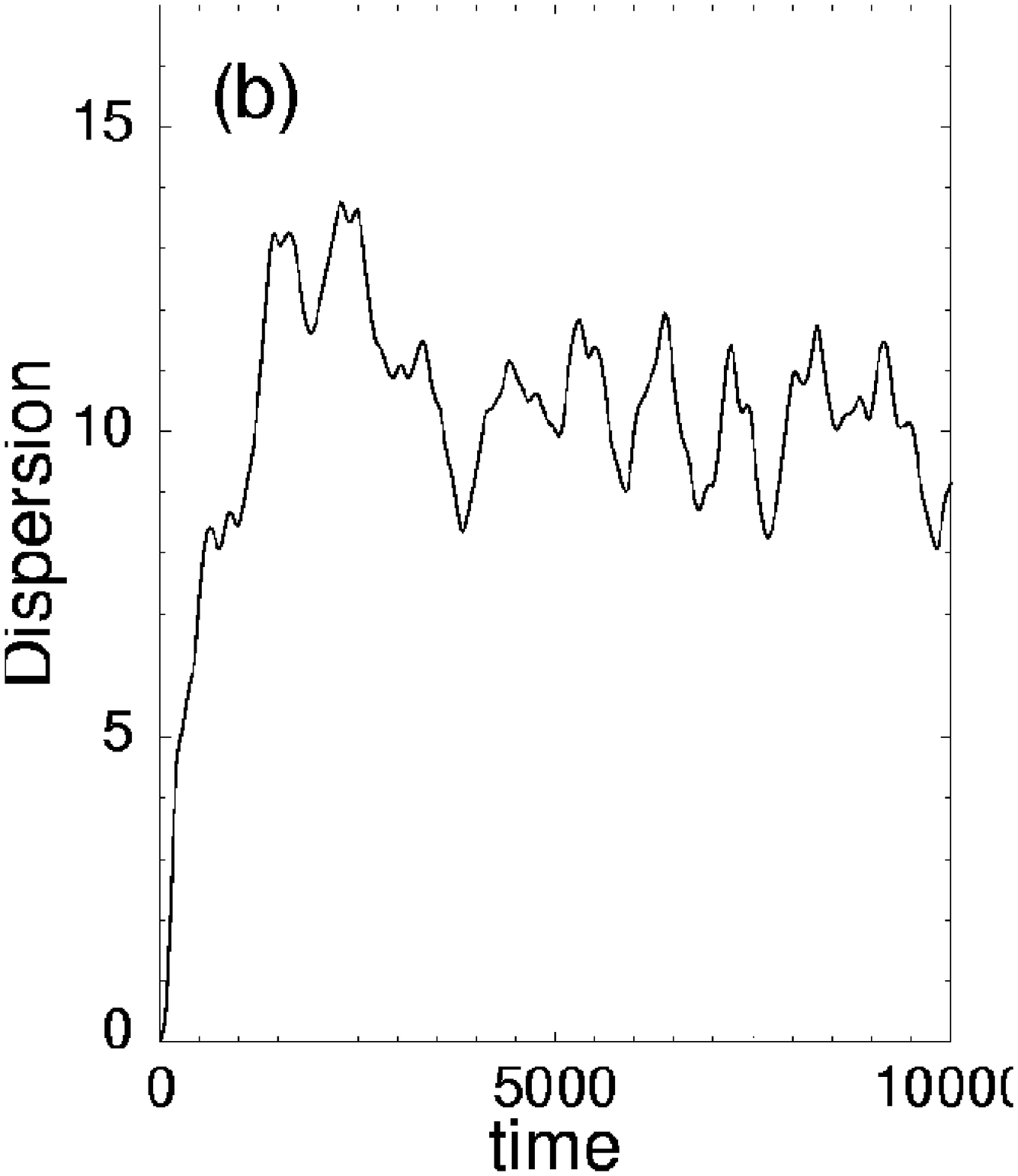,width=5.4cm,height=5.4cm}
       \psfig{file=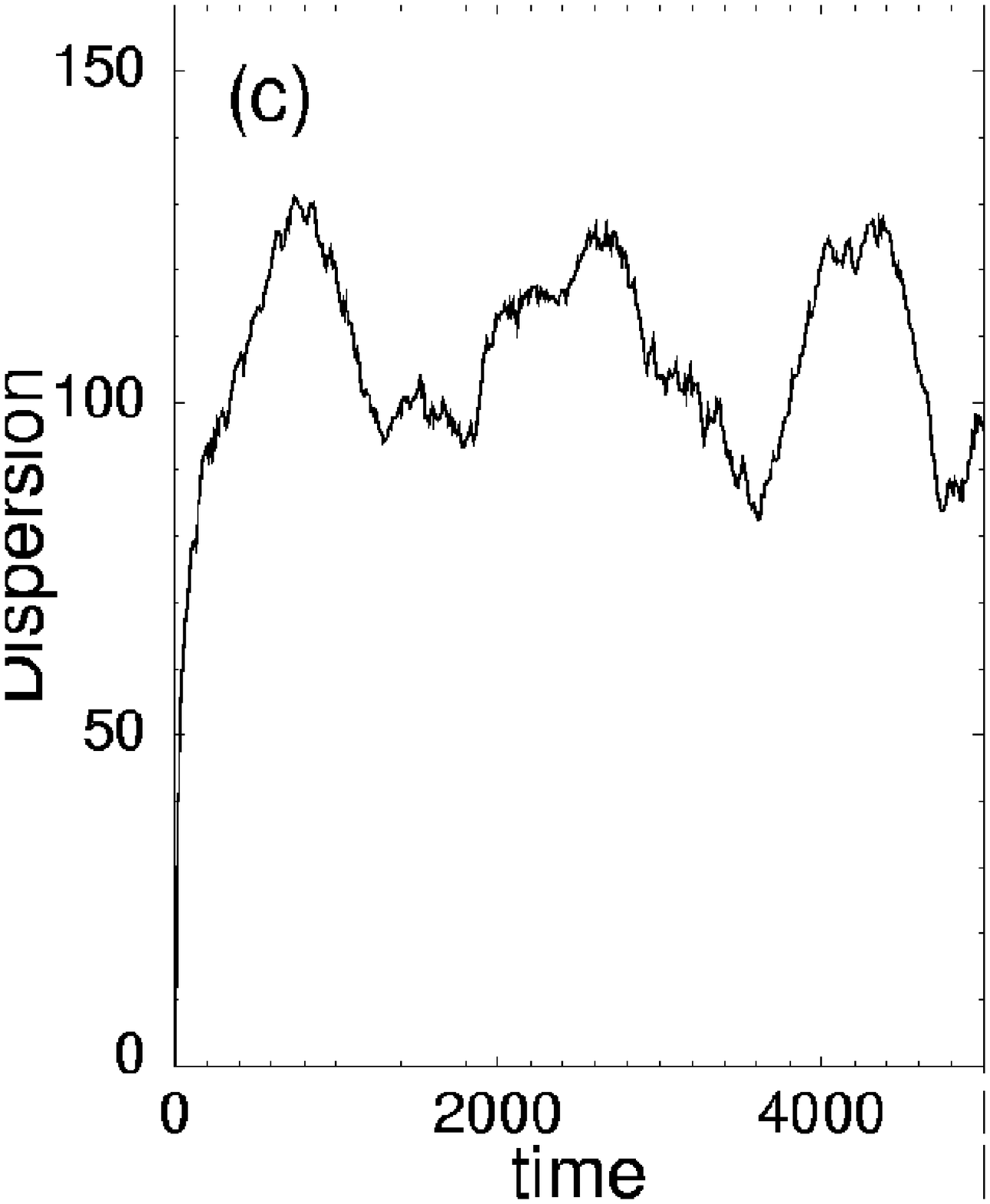,width=5.4cm,height=5.4cm}}
 \caption{Time-evolution of the dispersion $\sigma=\sigma(m)$,
 where $m=t/T$, for three values of the wave amplitudes:
 (a) $\epsilon=0.05$,
 (b) $\epsilon=0.5$, (c) $\epsilon=7.6$, and for $h=0.5$, $\ell=5$,
 $\delta=0$.}
 \end{center}
 \label{fig:18}
 \end{figure}

In order to illustrate  a non-monotonic character of 
localization of the QGS as a function of the wave amplitude at small $\epsilon$, 
we computed the dynamics of the quantum state initially concentrated 
on the ground state of the harmonic oscillator, $C_n(0)=\delta_{n,0}$,
using Eq (\ref{cnmt}). Time-evolution of the dispersion, 
\begin{equation}
\label{sigma}
\sigma(mT)=\sqrt{\sum_n|C_n(mT)|^2(n-\bar n(mT))^2}, 
\end{equation}
where $\bar n(mT)=\sum_n|C_n(mT)|^2 n$ is the average, 
is presented in Figs. 18 (a) - 18 (c), for three values of 
$\epsilon$. When the wave amplitude, $\epsilon$, is small (Fig. 18 (a)), 
a small 
part of the wave packet can propagate to large values of $n$ due to 
effect of diffusion via the separatrices as shown in Fig. 19 (a). 
Similar tunneling effect of  the wave packet between the resonance cells 
via the separatrices was explored in Ref. \cite{3} 
In spite of a small probability of tunneling to other cells,
contribution of this part to the dispersion, $\sigma$, is significant because it is
proportional to $(n-\bar n)^2$, where $n-\bar n\gg 1$.
At $\epsilon=0.5$ the separatrix QE 
states are destroyed by chaos, as shown in Fig. 17 (b), and QGS 
becomes more localized (see Fig. 19 (b)). This leads to a significant 
decrease of $\sigma$ in Fig. 18 (b) in comparison with the case of 
small $\epsilon$, shown in Fig. 18 (a). Further increase of $\epsilon$, 
up to the value $\epsilon=7.6$, results in delocalization of the 
QGS, as shown in Fig. 19 (c), and the dispersion, $\sigma=\sigma(mT)$,
in Fig. 18 (c) becomes large again. 

 \begin{figure}[tb]
 \mbox{\psfig{file=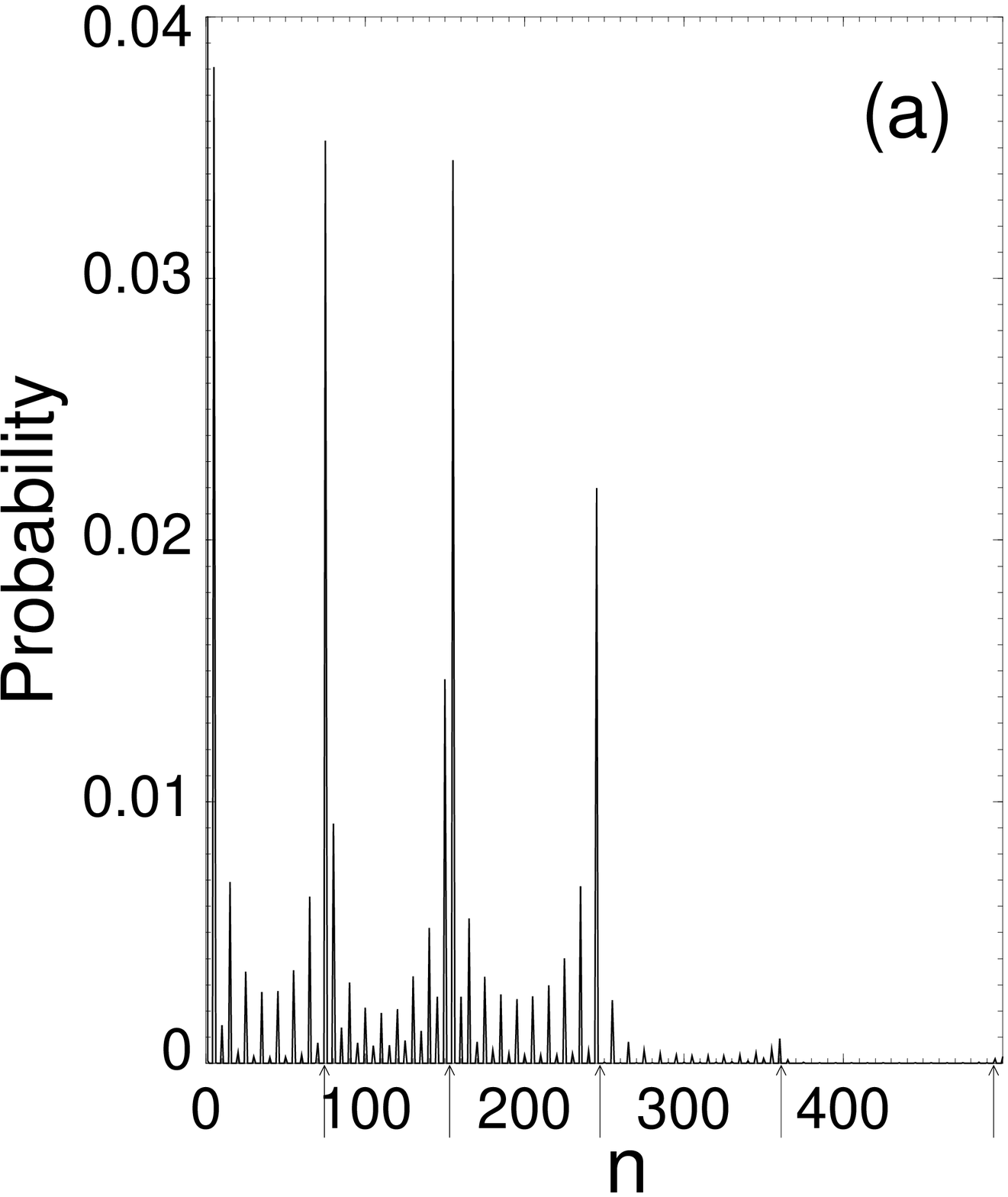,width=5cm,height=5cm}\hspace{0.2cm}
       \psfig{file=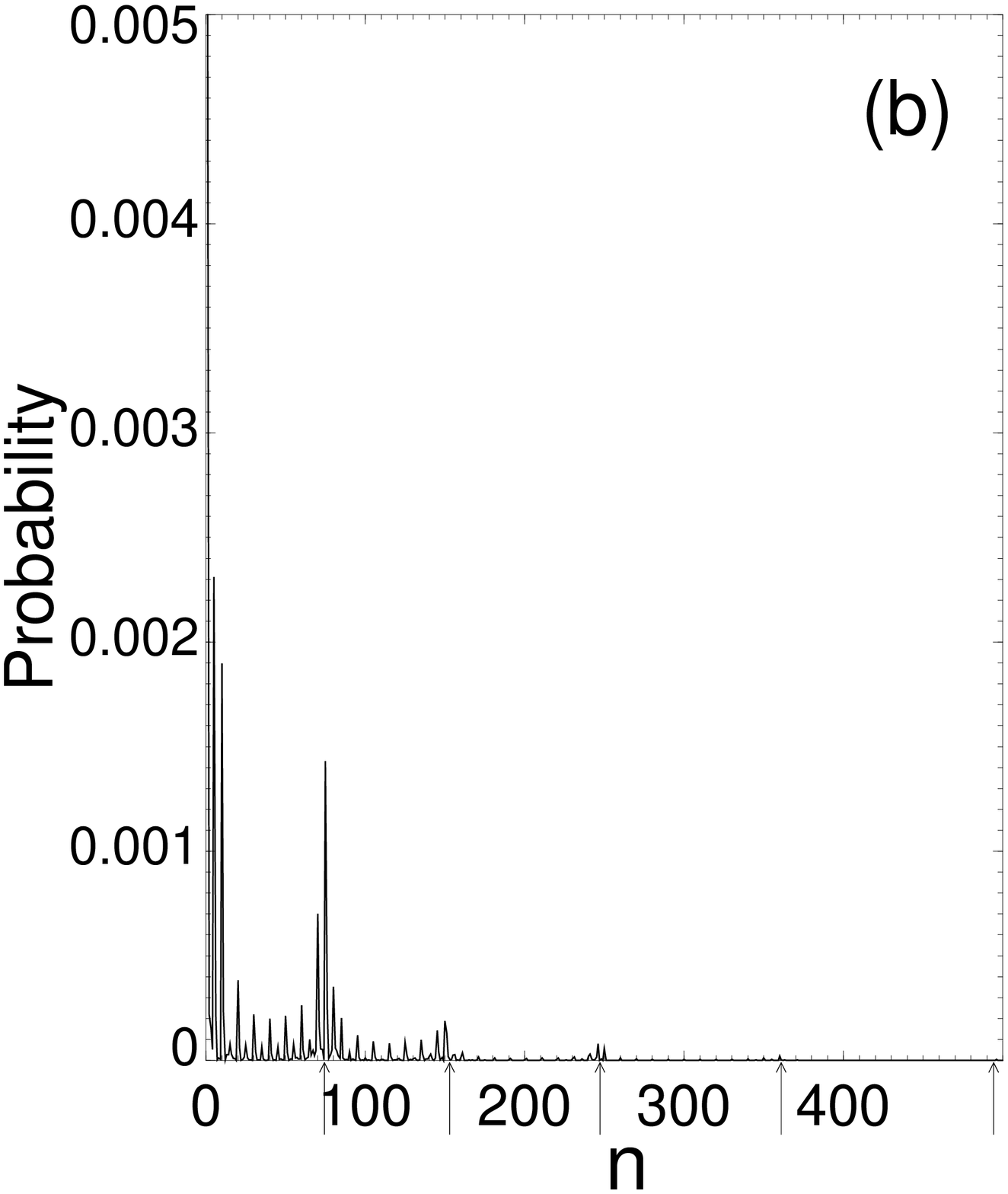,width=5cm,height=5cm}\hspace{0.2cm}
       \psfig{file=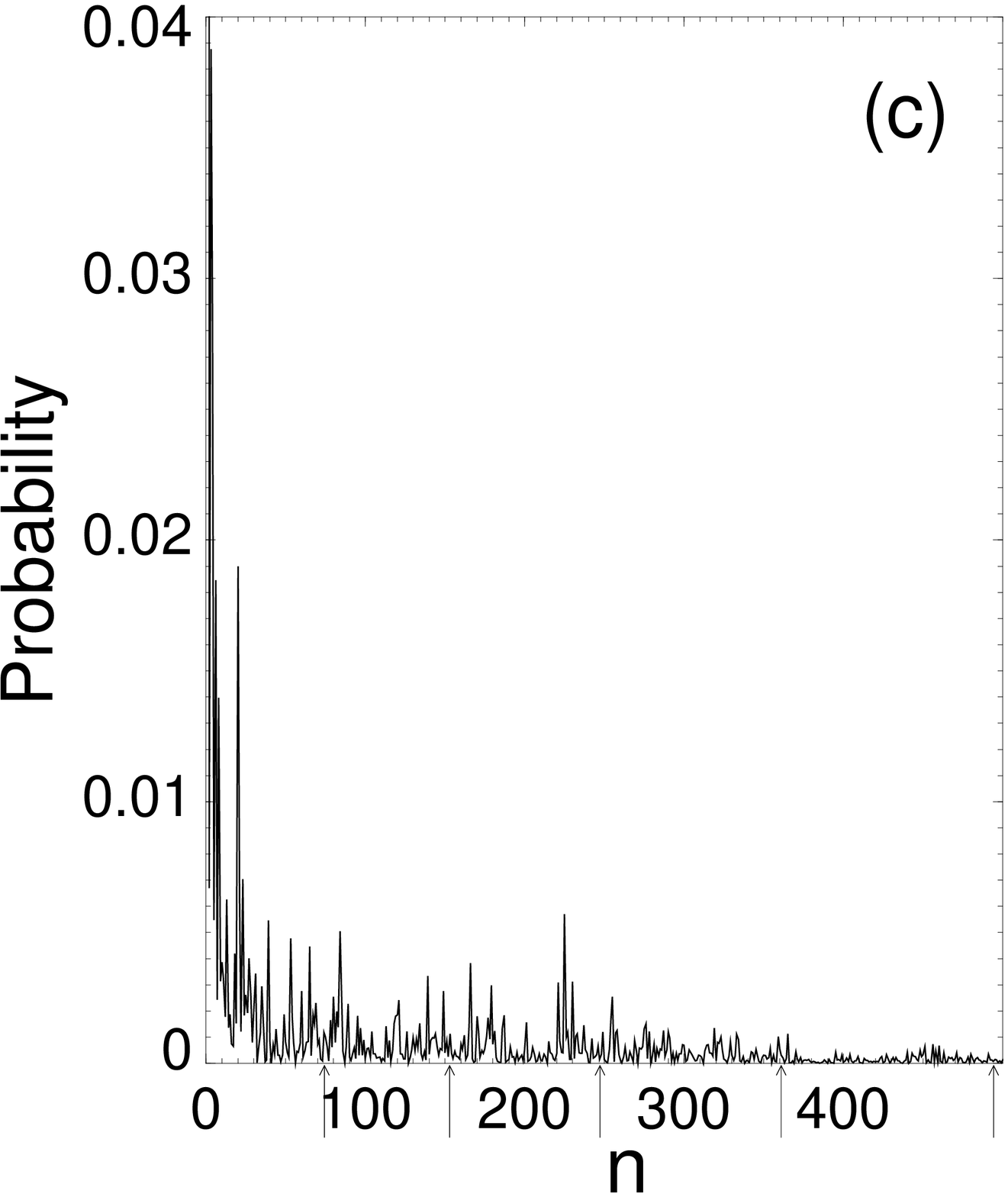,width=5cm,height=5cm}}
 \caption{Probability distribution in the system with
 initial condition $C_n(0)=\delta_{n,n_0}$ and
 (a) $\epsilon=0.05$, $mT=3\times 10^5T$; here $|C_0|^2=0.746$,
 (b) $\epsilon=0.5$, $mT=10^4T$; $|C_0|^2=0.982$,
 (c) $\epsilon=7.6$, $mT=5000T$; $|C_0|^2=0.137$, $|C_1|^2=0.423$.
 The boundaries of quantum cells (quantum separatrices)
 are marked by arrows; $h=0.5$, $\ell=5$, $\delta=0$.}
 \label{fig:19}
 \end{figure}

Comparison of Figs. 19 (a) and 
19 (c) allows us to conclude that delocalization of the QGS at very small, 
and at large values of $\epsilon$ is of a different nature. In the former 
case, the diffusion is caused by the separatrix QE states. These states 
are the quantum objects because they are delocalized,
and provide tunneling between the resonant cells. \cite{3}
As a consequence of the quantum nature of the separatrix QE states, 
increase of the dimensionless Planck constant, $h$, leads to  
delocalization of the separatrix QE states 
and decrease of localization of QGS at small $\epsilon$,
shown in Fig. 16. On the other hand, when $\epsilon$ is large we 
observe delocalization 
caused by chaos, which is manifested in the irregular form of the 
probability distribution, shown in Fig. 19 (c).

As follows from Fig. 16, in order to make the QGS more stable at 
small $\epsilon$ one should decrease the Planck constant, $h$.
A plot of the dispersion, $\sigma$, versus time for $h=0.1$ is presented in 
Fig. 20 (a). 
The system remains in the ground state with the probability 
$P_0=0.996$. However the dispersion is still large enough because the particle
can propagate with small probability to the levels with large $n\gg 1$, due to the 
diffusion via the separatrices which can be seen from the plot 
of the probability distribution presented in Fig. 20 (b).  

Another way to increase the stability of the ground state at small 
$\epsilon$, is to destroy the separatrix QE functions by choosing
the non-resonant value of the wave frequency, so that
$\delta=\ell-\Omega/\omega\ne 0$. 
For a non-resonant case ($\delta=0.01$),
the plot of the dispersion as a function of time is presented 
in Fig. 21 (a), and
the probability distribution at time $m=t/T=3\times 10^6$ is illustrated
in Fig. 21 (b). 
One can see from Fig 21 (a)
that introducing a detuning,
$\delta\ne 0$, results in considerable improvement of the stability of the 
QGS in comparison with  
the case of the exact resonance, illustrated in Fig. 18 (a). 
Thus, in order to make the ground state more stable
at small values of $\epsilon$ one must
detune the system from the exact resonance.  

The minimal value of detuning, $\delta$, required to 
destroy the separatrix structure, and
to make the ground state more 
stable, can be estimated from the quantum equations of motions 
in the resonance approximation (\ref{al_eq}). 
The term, proportional to $\delta n$ destroys the 
separatrix QE states as shown in Ref.\cite{1}  
This term becomes significant when it becomes of the order 
of the relation, $\epsilon/h$.
On the other hand, the separatrix 
QE function must, at least, occupy the two separatrices. Thus, we can 
estimate the number, $n=n_\delta$, of the oscillator's state at which the 
separatrix structure is destroyed, namely $n_\delta=\epsilon/\delta h$.
The separatrix QE functions decay exponentially
 with increasing $n$
in the region $n>n_\delta$.\cite{1}  
For the parameters 
in Fig. 21, $n_\delta=50$, which is much less than the position of the
first separatrix $n_1=385$ shown in Fig. 20 (b) by arrow. 
So, the separatrix QE states at these parameters do not exist, and we 
do not observe tunneling from the QGS to other resonance cells. 

 \begin{figure}[tb]
\centerline{\mbox{\psfig{file=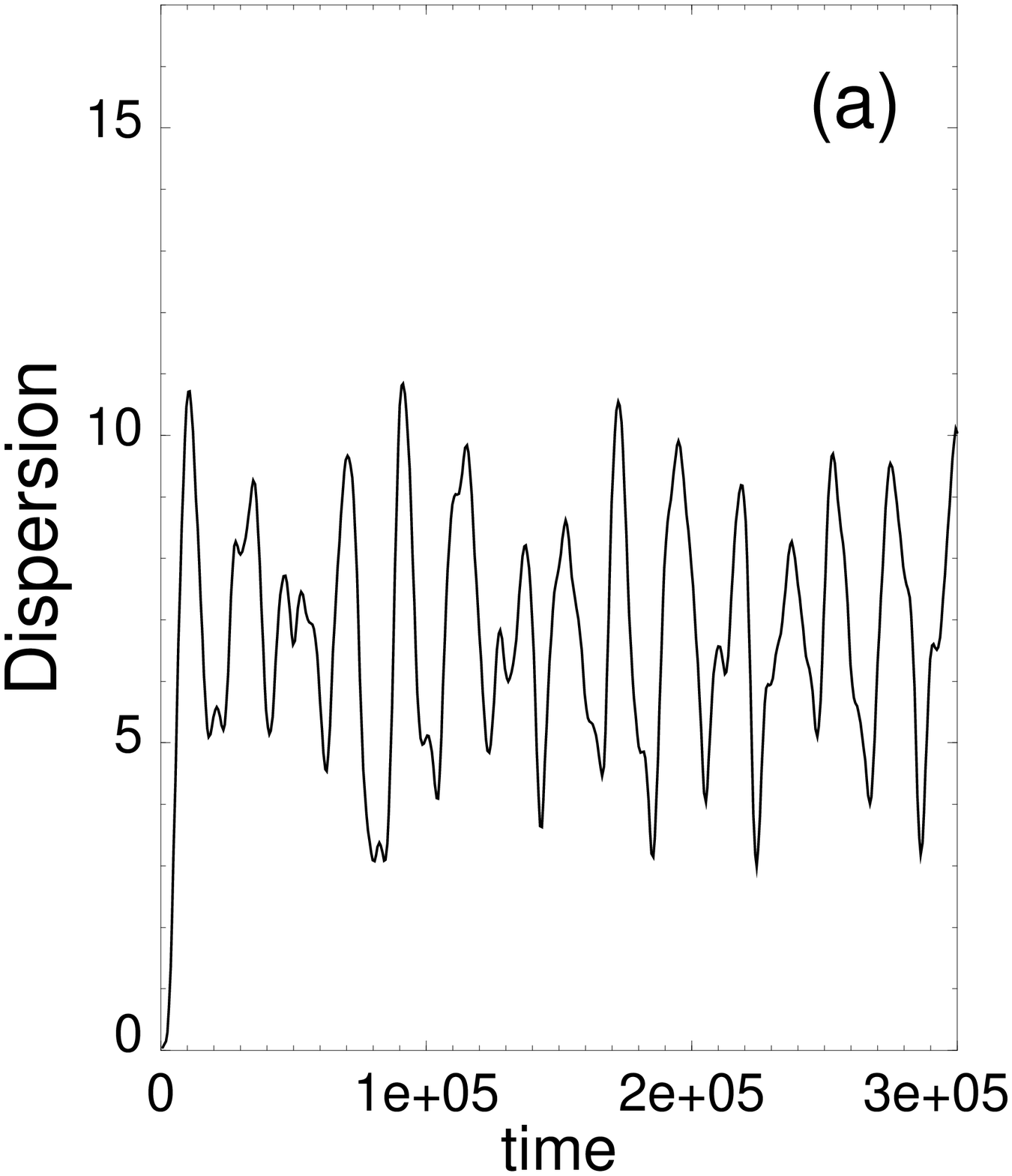,width=8cm,height=8cm}\hspace{0.4cm}
       \psfig{file=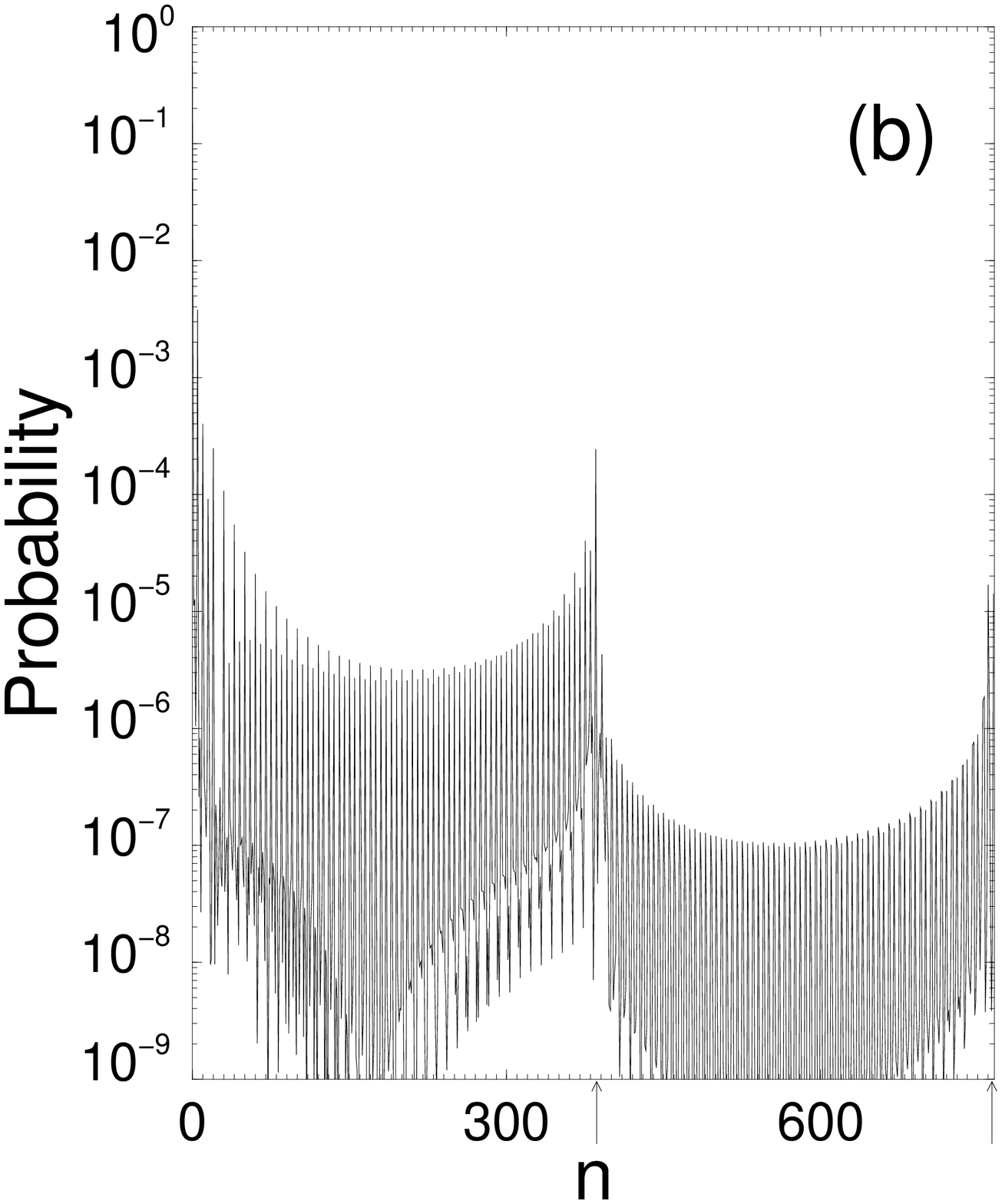,width=8cm,height=8cm}}}
 \vspace{0.4cm}
 \caption{(a) Time-evolution of the dispersion, $\sigma=\sigma(m)$,
 where $m=t/T$, and (b) probability distribution at time 
 $m=3\times 10^6$ for small value of $h$, $\delta=0$, $h=0.1$, $\epsilon=0.05$}
 \label{fig:20}
 \end{figure}
\begin{figure}[tb]
\centerline{\mbox{\psfig{file=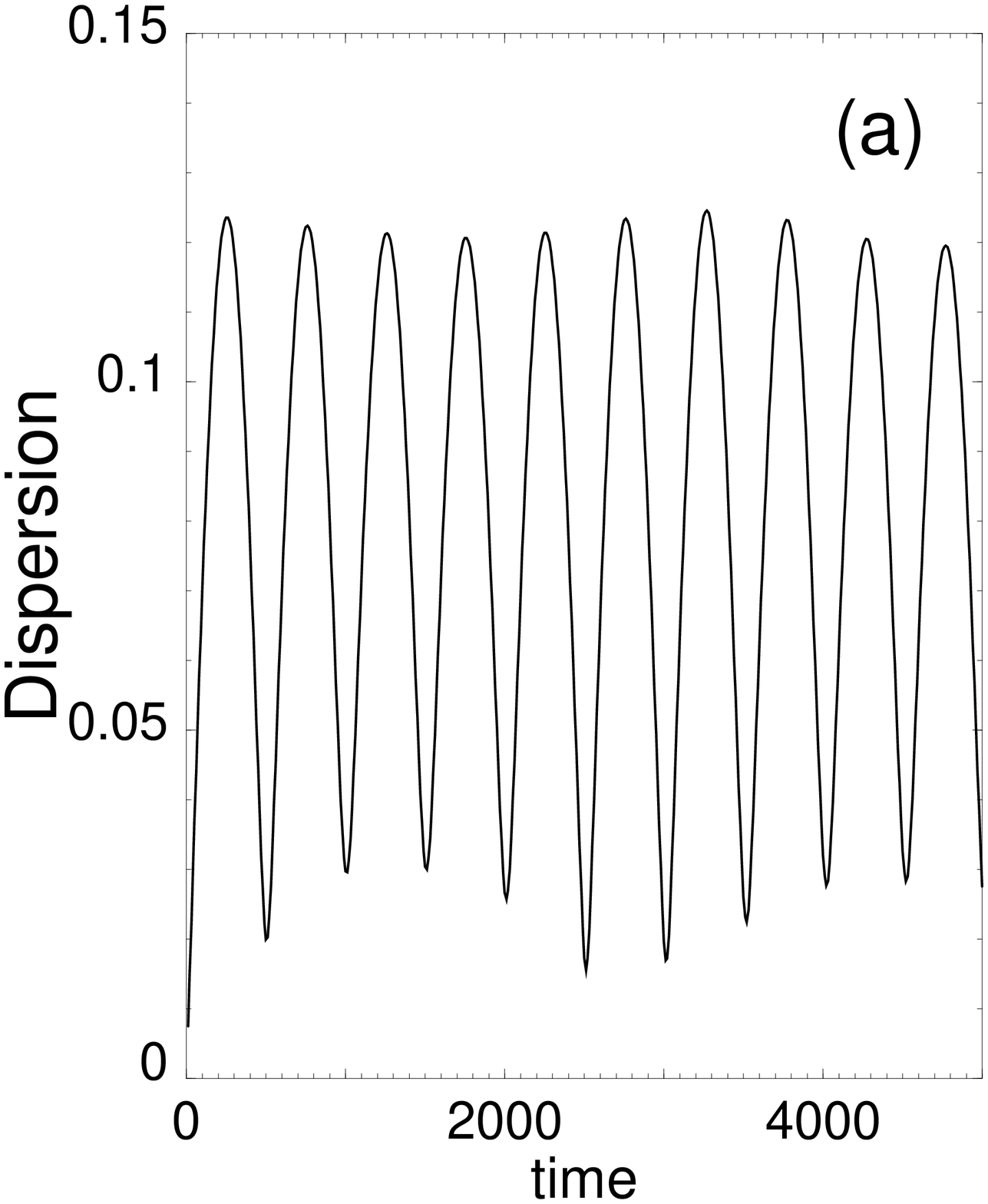,width=8cm,height=8cm}\hspace{0.4cm}
       \psfig{file=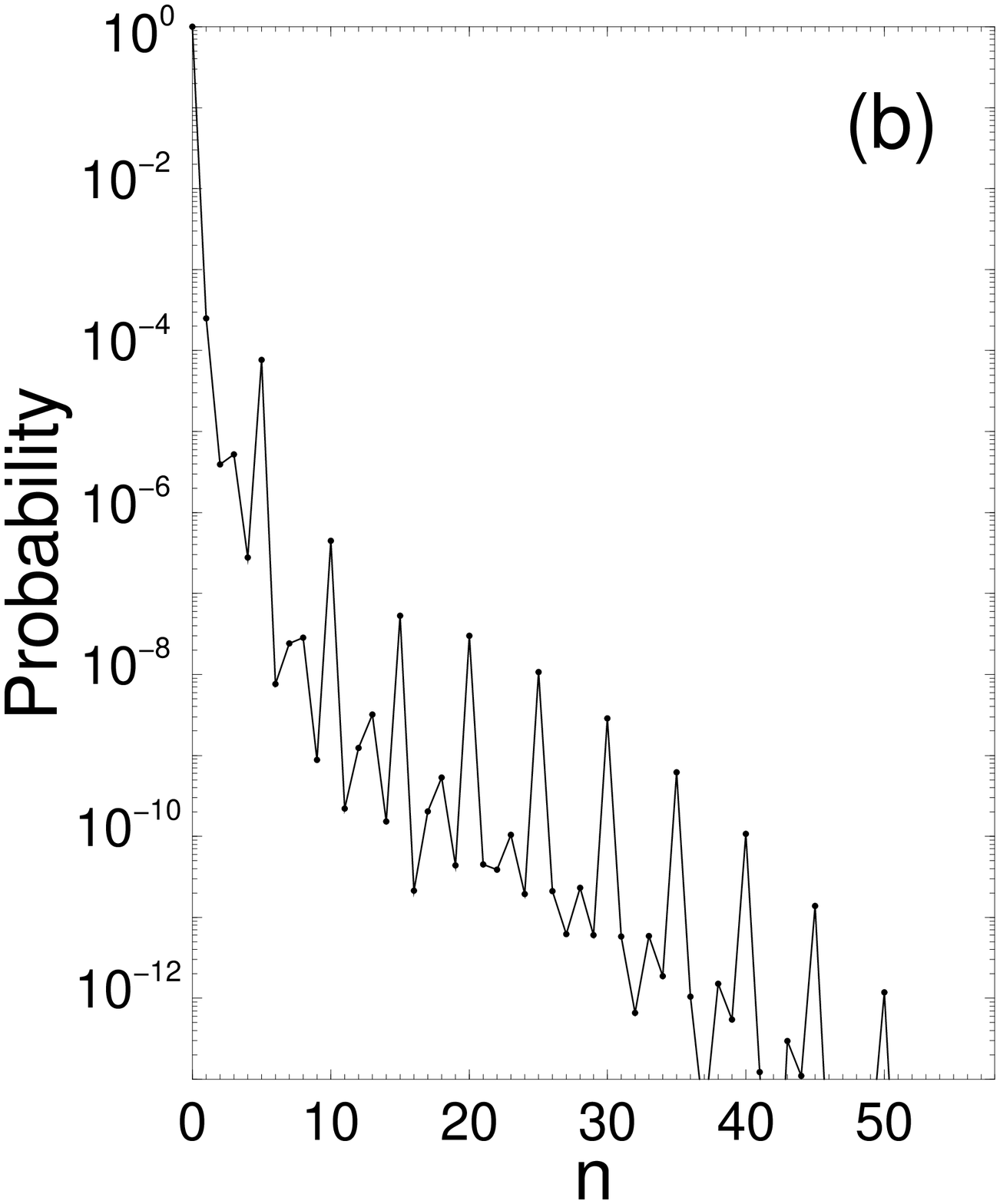,width=8cm,height=8cm}}}
 \vspace{0.4cm}
 \caption{(a) Time-evolution of the dispersion $\sigma=\sigma(m)$,
 where $m=t/T$, and (b) probability distribution at time 
 $m=3\times 10^6$ for the near resonance case, $\delta=0.01$; $h=0.1$, $\epsilon=0.05$}
 \label{fig:21}
 \end{figure}
Decreasing the value of $\epsilon$, in the near resonance case, 
makes the QGS more stable. 
Unlike the near resonance case, in the case of the exact resonance,
a stability of the QGS at $\epsilon\ll 1$ is independent 
of the wave amplitude, because in this case the localization 
properties of the QGS are defined by the structure of separatrix 
QE function, which
in the resonance approximation (when $\epsilon$ is small) is 
independent of $\epsilon$.\cite{1} 
One can see from a comparison of Fig. 21 (a) with Fig. 20 (a) 
($h=0.1$, $\epsilon=0.05$)
 that the dispersion in the near resonance case (Fig. 21 (a))
is much less than that in the exact resonance case (Fig. 20 (a)), in spite of
the probability to stay in the ground state, $P_0$, for these two cases 
does not differ significantly ($P_0=0.99897$ at $\delta=0.01$ and 
$P_0=0.996$ at $\delta=0$). The reason is that the dynamics in the exact 
resonance case is mainly determined by the separatrix QE function, and 
in this case there is a small probability for particle to tunnel 
to the high oscillator levels with $n\gg 1$, as shown in Fig. 20 (b).  
\begin{figure}[tb]
\centerline{\psfig{file=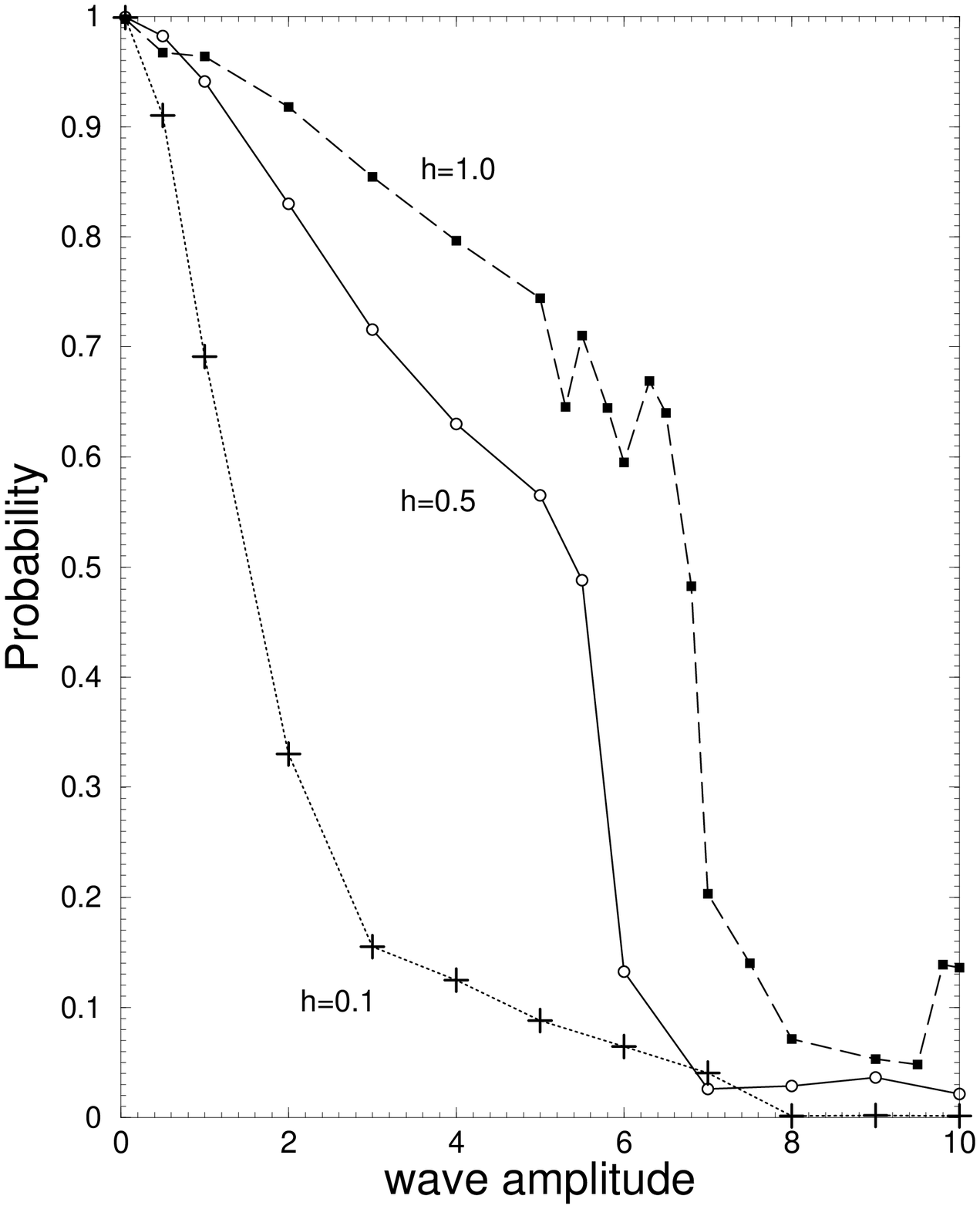,width=15cm,height=10cm}} 
\caption{The same as in Fig. 16, but for the near resonance case,
$\delta=0.01$.}
\label{fig:22}
\end{figure}
When the wave amplitude $\epsilon$ increases, the condition
$\delta\ne 0$ becomes less significant. This can be explained by 
less influence of the term proportional to $\delta$ on the
dynamics in comparison with influence of the wave with the amplitude
$\epsilon$ in the region where the value of $kr$ is relatively small
(see the classical Hamiltonian in Eqs. (\ref{cl_H0}),
(\ref{wave_decomposition}), (\ref{cl_H3}) and
quantum Eq. (\ref{al_eq})).
In Fig. 22 we plot the function
$P_0=P_0(\epsilon)$ for the near resonance case.
In Figs. 23 (a) - 23 (c) we compare the results for
the exact resonance case
($\delta=0$) with those for the near resonance case,
when $\delta=0.01$ and $\delta=0.1$.
One can see from  these figures that the dynamics in the vicinity
of the QGS in the near resonance
case is similar to that in the exact resonance case,
except for the region of small $\epsilon$, which was discussed above.
 \vspace{0.4cm}
 \begin{figure}[tb]
\centerline{\mbox{\psfig{file=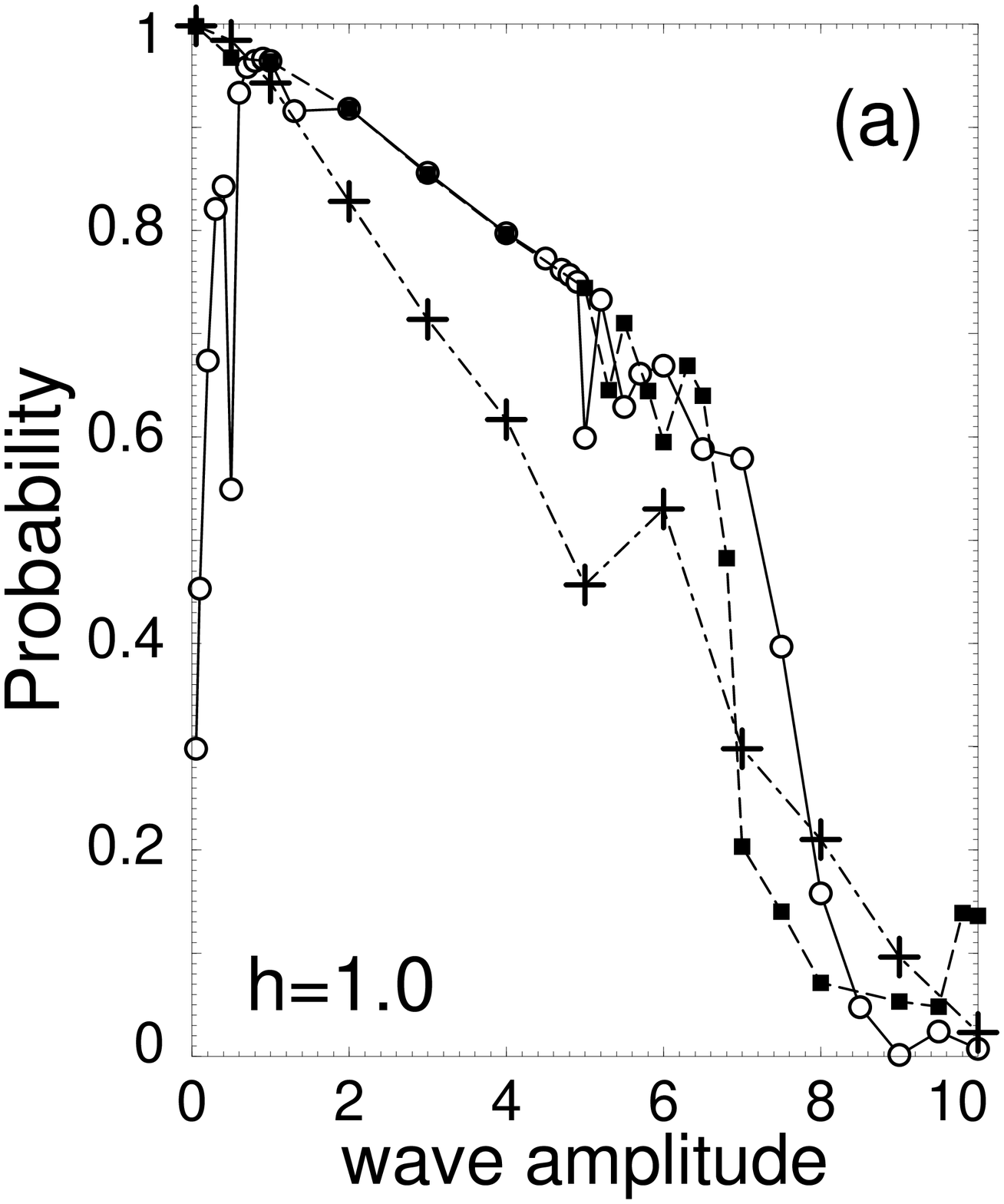,width=5.4cm,height=5.4cm}
       \psfig{file=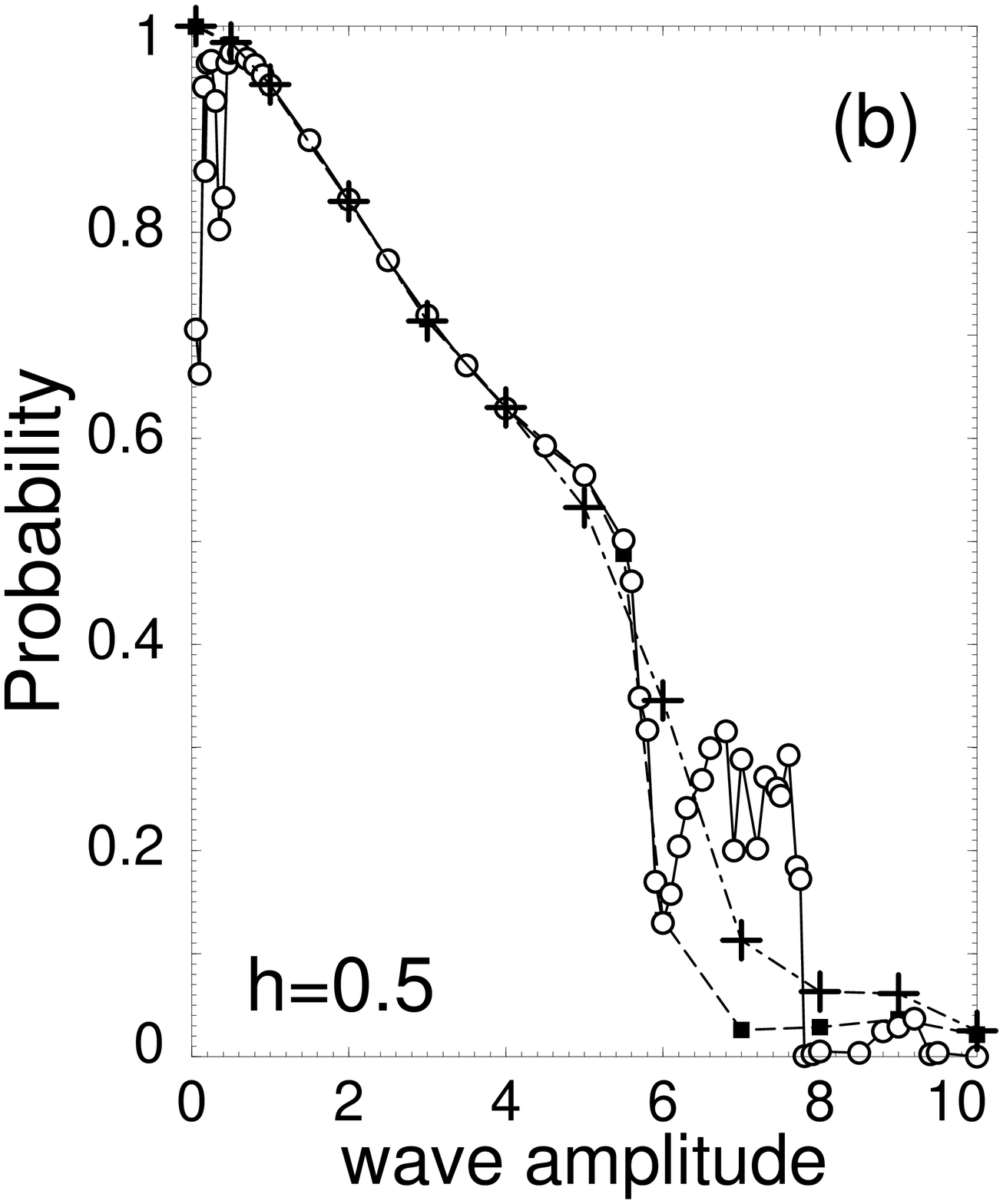,width=5.4cm,height=5.4cm}
       \psfig{file=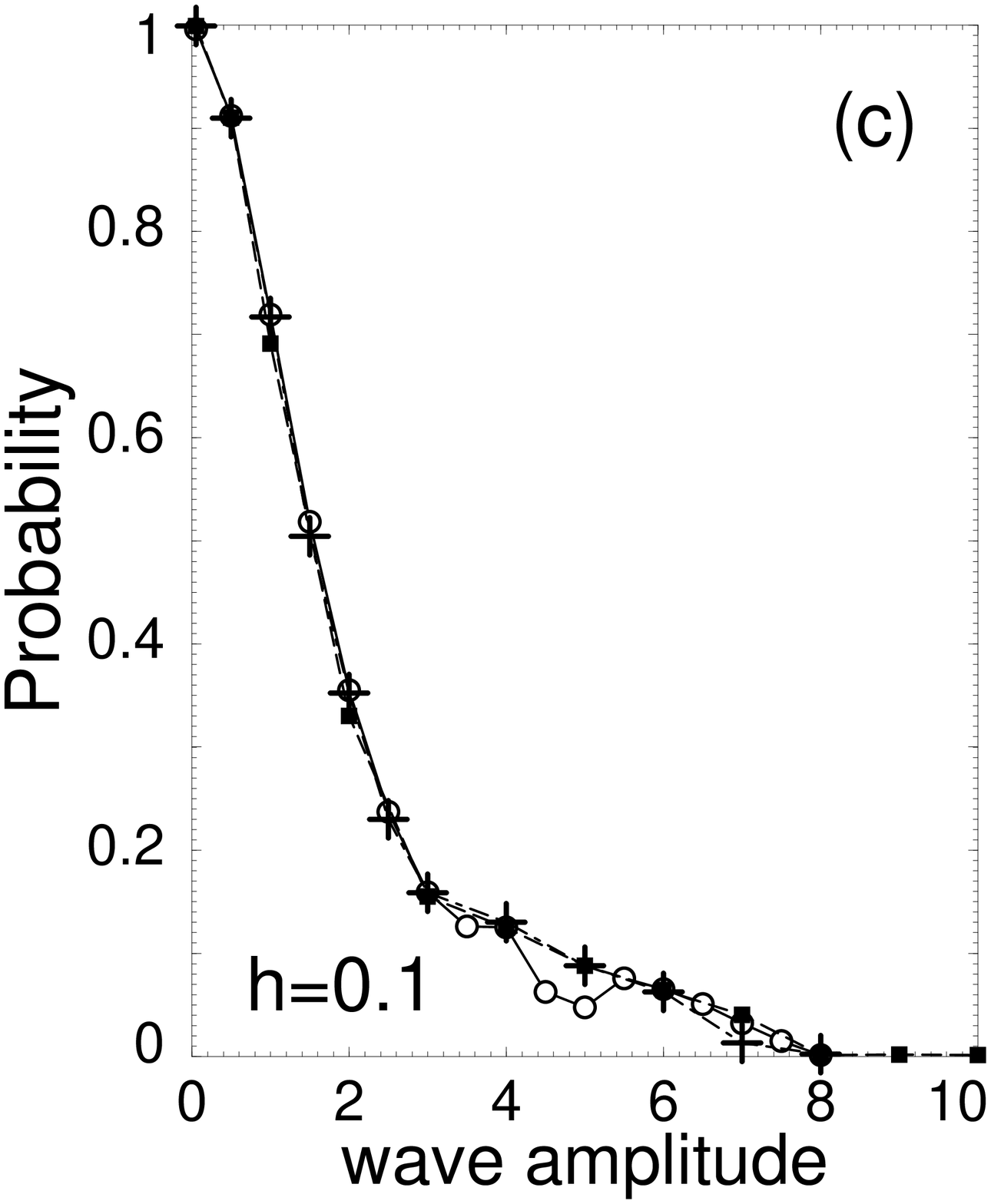,width=5.4cm,height=5.4cm}}}
 \vspace{0.5cm}
 \caption{Plot of $P_0$ versus $\epsilon$ for 
 the exact resonance case,  when $\delta=0$ (solid line and open circles),
 and two curves for the near resonance cases:
 $\delta=0.01$ (dashed line and filled squares), 
 $\delta=0.1$ (dot-dashed line and crosses);   
 $\ell=5$, (a) $h=1$, (b) $h=0.5$,
 (c) $h=0.1$.}   
 \label{fig:23}
 \end{figure}      
In the quantum case, similar to the classical one, at very small $\epsilon$
and finite $\delta$, there always exists the QGS QE state at any $\ell$
including the cases $\ell=1$ and $\ell=2$. In Fig. 24 we present a plot
$P_0=P_0(\epsilon)$ for the cases $\ell=1$ and $\ell=2$ when
$\delta=0.1$ and $h=0.5$. As one can see from Fig. 24 the QGS QE state
in these two cases becomes unstable at considerably less
values of $\epsilon$ than that in the case $\ell=5$ in Fig. 23, which
corresponds to the classical dynamics in Figs. 8 - 10. In the case 
$\ell=1$ the stable point shifts down from the region $X=0,\,P=0$ in 
Figs. 8 (a)- 8 (c), and in the case $\ell=2$ the stable point 
at  $\epsilon\approx 0.2$ becomes unstable (see Figs. 9 (a), 9 (b)) and 10).
The quantum manifestation of this process is a rapid decay of the value
of $P_0$ in the region $\epsilon\ge 0.2$ in Fig. 24.

In conclusion, the classical dynamics in the vicinity of the point ($x=0,\,p=0$) in
the classical phase space and quantum dynamics in the vicinity of the ground state of the
harmonic oscillator in a field of a monochromatic wave, is explored.
Both resonance and near resonance cases are analyzed.
It is shown that at small $\epsilon$ and finite detuning from the resonance,
$\delta$, the quantum ground state is always stable.
In the case $\delta=0$, the dynamics is
unstable for the resonance
numbers $\ell=1,2$.
Stability of the classical dynamics in the central cell 
and stability of the quantum dynamics near the ground state of the harmonic 
oscillator under the influence of chaos is analyzed. It is shown, 
that under certain conditions 
($\ell>2,\,\delta=0,\,\epsilon\ll 1,\,h\sim 1$)
the presence of chaos makes the quantum ground state more localized. 
Increase of the quantum parameter, $h$,  
at intermediate values of $\epsilon$, enhances localization of the 
QGS considerably.   
Experimental observation of discussed results in the system of an 
ion trapped in a linear ion trap and interacting with laser
fields represents a significant interest for understanding
the stability properties of this system.

\begin{figure}[tb]
\centerline{\psfig{file=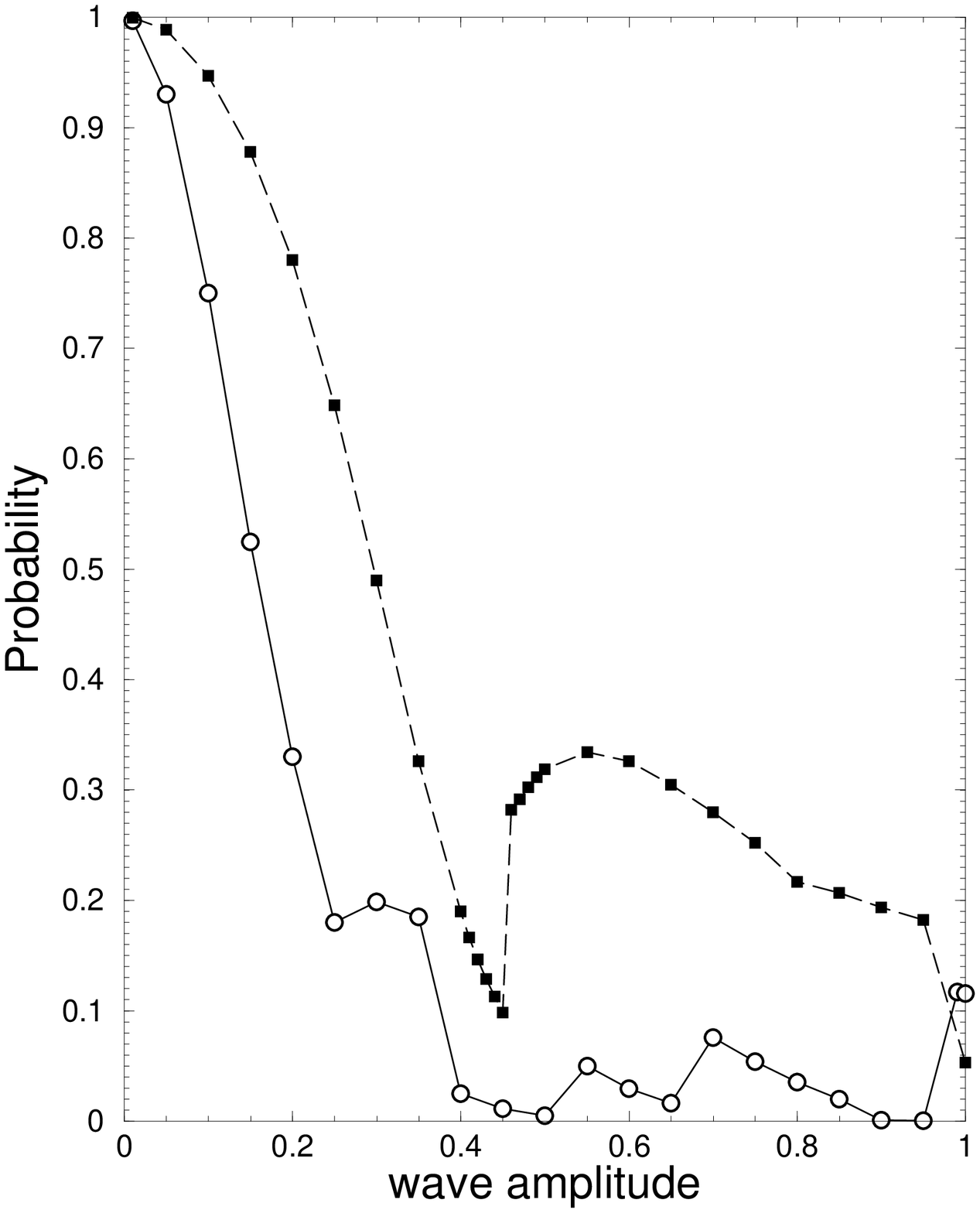,width=15cm,height=10cm}} 
\caption{Plot $P_0=P_0(\epsilon)$ for the resonance numbers 
$\ell=1$ (open circles and solid line) and 
$\ell=2$ (filled squares and dashed line) in the near resonance case, 
$\delta=0.1$; $h=0.5$.}
\label{fig:24}
\end{figure}

\section{Acknowledgments}
We are thankful to R.J. Hughes and G.D. Doolen for useful discussions.
This work was  supported by the National Security Agency, and by the
Department of Energy under the contract W-7405-ENG-36.
Work of D.I.K. was partly supported by Russian Foundation for Basic Research
(Grants No. 98-02-16412 and 98-02-16237).
%
{}

\end{document}